\documentclass[11pt,a4paper]{report}

\usepackage{amsmath, amssymb, amsthm}
\usepackage{graphicx,color}
\usepackage[left=1.5in, right=1in, top=1in, bottom=1in, includefoot, headheight=13.6pt]{geometry}
\usepackage[square, comma, numbers, sort&compress]{natbib}
\usepackage[T1]{fontenc}

\usepackage{feynmf}

\usepackage{mathpazo}
\usepackage[hang, small, bf, margin=20pt, tableposition=top]{caption} 
\numberwithin{equation}{section}
\renewcommand{\bibname}{References}

\usepackage{fancyhdr}
\makeatletter
\renewcommand{\chaptermark}[1]{\markboth{\textsc{\@chapapp}\ \thechapter:\ #1}{}}
\makeatother

\usepackage{sectsty}
\chapterfont{\large\sc\centering}
\chaptertitlefont{\centering}
\subsubsectionfont{\centering}

\usepackage[colorlinks=false]{hyperref}
\usepackage[figure,table]{hypcap}
\hypersetup{
	bookmarksnumbered,
	pdfstartview={FitH},
	citecolor={blue},
	linkcolor={red},
	urlcolor={black},
	pdfpagemode={UseOutlines}
}
\makeatletter
\newcommand\org@hypertarget{}
\let\org@hypertarget\hypertarget
\renewcommand\hypertarget[2]{%
  \Hy@raisedlink{\org@hypertarget{#1}{}}#2%
} 
\makeatother

\newcommand{\be}{\begin{eqnarray}}
\newcommand{\ee}{\end{eqnarray}}
\newcommand{\im}{{\mathrm i}}
\newcommand{\f}[2]{\frac{#1}{#2} }
\newcommand{\myref}[1]{(\ref{#1})}

\def\tl{\tilde}
\def\d{\delta}

\def\a{\alpha}
\def\b{\beta}

\def\t{\theta}
\def\s{\sigma}
\def\l{\lambda}
\def\m{\mu}
\def\n{\nu}
\def\r{\rho}

\def\k{\kappa}

\def\i{\iota}  
\def\e{\epsilon}  
\def\S{\Sigma}

\def\p{\partial}

\def\H{\mathcal{H}}
\def\L{\mathcal{L}}

\def\<{\langle}
\def\>{\rangle}

\newcommand{\Tr}[1]{ \mathrm{Tr}  \left( {#1}\right)}
\newcommand{\nn}{\nonumber}

\newcommand{\ket}[2]{ {\langle #1\,#2\rangle} }
\newcommand{\bra}[2]{ {[#1\,#2]}}

\begin{document}
	
	% Titlepage
	\title{
		\huge{\textbf{Graviton scattering amplitudes and Pure Connection Formulation of GR}} \\[1.2cm]
		\Large{	Gianluca Delfino} \\[1.2cm]
		\Large{Thesis submitted to The University of Nottingham \\ 
		for the degree of Doctor of Philosophy} \\ \vspace{1cm}
		\Large{March 2013} 
	} 
	\author{} \date{}
	\pdfbookmark[0]{Titlepage}{title}
	\maketitle
	
	% Dedication
 	\newpage \vspace*{8cm} 
	\pdfbookmark[0]{Dedication}{dedication}
	\begin{center} 
		\large Dedicated to my niece Arianna,\\
		\emph{may you always see the world through the curious eyes of a child.}
 	\end{center}

	% Abstract
	\newpage 
	\pdfbookmark[0]{Abstract}{abstract}
	\chapter*{Abstract}
	We show how the recently introduced ``Pure Connection Formulation'' of gravity provides a natural framework for approaching the problem of computing graviton scattering amplitudes. 
	In particular, we show that the interaction vertices are greatly simplified in this formalism as compared to the Einstein-Hilbert perturbation theory. This, in turns, leads to very simple Feynman rules that we employ for the direct computations. Furthermore, this framework naturally extends to wider class of gravitational theories, which encompasses General Relativity as a special case. 
	We compute all the possible tree-level graviton-graviton scattering amplitudes for a general theory from this class. In the GR case the results are in complete accordance with the known expressions in the literature. Moreover, for the general theory distinct from GR, we find new tree-level parity-violating amplitudes. The presence of this new amplitudes is a direct consequence of the fact that the general theory does not exhibit explicit parity invariance.

	% Acknowledgements
 	\chapter*{Acknowledgements}
	\pdfbookmark[0]{Acknowledgements}{acknowledgements}
	I would like to thank my Supervisor K.Krasnov for his help and guidance during my PhD, and without whom this thesis would have not been possible. I am also deeply thankful to have met so many wonderful friends in the past four years, who have been my brothers and sisters during this journey that I was lucky to share with them. I am especially thankful to my girlfriend for putting up with me all this time.
	Finally, for their support and for always being there for me, \textit{grazie Mamma e Pap\'a}. 
	
	\tableofcontents 
	\newpage	
	\pagenumbering{arabic}
 	
	% Include all chapter files
\chapter{Introduction}

General Relativity (GR) is known for being particularly burdensome to be treated perturbatively: DeWitt in a 1967 paper \cite{DeWitt:1967uc} said himself 

\textit{ ``The tediousness of the algebra involved in obtaining the graviton-graviton cross section may be inferred from the complexity of the vertex function, but the fact that the final result are ridiculously simple leads one to believe that there must be an easier way.''} 

Indeed, if one expands perturbatively the Einstein-Hilbert (EH) action $\int\sqrt{g}R$ around a fixed background obtains something that is far from similar to any other physical theory we know. For instance, even in the simplest case expanding around Minkowski, we see a proliferation of terms that greatly complicates the Feynman rules of the theory.
%First indication that the theory cannot be straightforwardly cast in the usual Yang-Mills terms is that the force carries in the latter are represented by spin one gauge connection, while the hypothetical carrier bosons of gravity ($gravitons$) are supposed to be spin two particles. Moreover gravity differs from ``usual'' gauge theories also in that the interaction is always attractive. Finally
Indeed, the complexity of the perturbative expansion of the EH action, even at orders as low as third or fourth, makes it an arduous task to derive scattering amplitudes by combining Feynman diagrams. As a matter of fact, already the fourth order variation of the Lagrangian is composed by, at least, half a page of terms (see \cite{Goroff:1985th} formula A.6).

The difficulty of the theory was reason for the birth of a number of distinct approaches to address the issue of graviton scattering. For instance, in the past two decades, many authors (see \cite{Choi:1994ax} \cite{Cachazo:2004kj} and references therein) have started employing the so called ``spinor-helicity methods'' to alleviate the complexity of the perturbation theory. 
The key point of this methods is to cleverly define the external states' polarisation tensors in order to simplify the computations. Indeed, the polarisation tensors can always be written in terms of the momentum of the external particle and in terms of an $arbitrary$ reference momentum. The arbitrariness of the latter is due to gauge invariance of the theory. Choosing these reference momenta appropriately leads to a number of simplifications. For instance, if ones aims to compute the 2-gravitons to 2-gravitons scattering amplitudes in GR, it is easy to show that using this technique one can ignore entirely the contribution coming from the fourth-order variation. This particular computation can be done then only combining a small number of vertices coming from the third-order expansion. 
(See Appendix \myref{EH_Computation})

Graviton scattering amplitudes can also be computed with string theory technology, recalling that the low energy limit of string theory is indeed GR plus quantum fields \cite{Green:1987sp}. In particular, the authors Kawai, Lewellen and Tye(KLT) \cite{Kawai:1985xq} derived a set of formulas expressing closed string amplitudes in terms of products of open string amplitudes. Then, in the low-energy limit of string theory, KLT formulas imply that similar relations must exist between amplitudes in gravity and gauge field theories. This means that, at tree-level, graviton scattering must be expressible as a sum of products of well defined pieces of non-abelian Yang-Mills scattering amplitudes. With this methods, graviton-graviton amplitudes were obtained in a form in which the integrands appearing in the expressions were given as products of integrands appearing in gauge theory \cite{Bern:1993wt} \cite{Dunbar:1994bn}. This suggested a much stronger relationship between gravity and gauge theories, which can be represented as: gravity $\sim$ (gauge  theory)$\times$ (gauge  theory) (see \cite{Bern:2010yg} or \cite{Bern:2002kj} for review material).

One can also derive graviton scattering amplitudes, without the need of Feynman diagrams, using recursion relations. Indeed, from the seminal Witten paper \cite{Witten:2003nn}, Britto $et$ $al.$ have derived recursion relations for both gauge theory and gravity \cite{Britto:2004ap} \cite{Britto:2005fq}. These BCFW recursion relations proved to be a very efficient technique for calculating scattering amplitudes both in QCD \cite{Berger:2006ci} \cite{Bern:2005hh} and in gravity \cite{Bedford:2005yy} \cite{Cachazo:2005ca}(for a detailed review on the subject we refer to \cite{Brandhuber:2007up}).

%This methods were first introduced in string theory, using open/closed string duality (see  \cite{Witten:2003nn} for review material), and at first were adopted to extract gluon-gluon scattering amplitudes from the low energy limit of the theory \cite{Bern:1993sx} \cite{Bern:1993wt} \cite{Green:1999pv}. Roughly, the spinor-helicity method is an expedient that consists in setting some of the fields to be $on$-$shell$ and using the properties of the graviton's helicity tensor to simplify the action.\footnote{The fields to set on shell have to be carefully chosen depending on the order of the action and the amplitude one wants to obtain.} In such way the perturbative expansion, at low orders, becomes more manageable.  Employing this methods, in fact, recursion relation have been obtained for computing general one-loop N-graviton scattering amplitudes \cite{Bern:1998sv} and (up to) 4-loops 4-graviton scattering amplitudes in N=8 supergravity \cite{Bern:2009kd}. Most importantly, it was observed that the resulting graviton scattering amplitudes exhibit a striking resemblance with the ``square'' of the respective QCD gluon scattering amplitudes (see \cite{Berends:1988zp} \cite{Bern:2010yg} and references therein). This resemblance however stood only in the final result, as it was not evident at the level of the interaction vertices.

This thesis is centred on computing graviton scattering amplitudes in a novel formulation of General Relativity. In fact, we will employ the  spinor-helicity methods in a formalism for gravity that exhibits a close resemblance, at the Lagrangian level, to Yang-Mills theory. % vertex\footnote{ Albeit in the GR sector of the theory and in its spinorial form, see section \myref{parity_preserving_cubic_vertex_section} }. 
In this formulation of General Relativity, introduced in the papers \cite{Krasnov:2011pp} \cite{Krasnov:2011up}, the only dynamical field is a complexified SO(3) connection $A^i$. Hence the name Pure Connection Formulation (PCF). In this approach GR is presented in a particularly elegant and simple manner. Indeed, this formalism lends itself to a simpler perturbative treatment of the theory than it is possible in the metric case.
 
 The aim of this thesis is threefold: 
\begin{itemize}
	\item To quantise the PCF and introduce a framework for computing its scattering amplitudes. Remarkably, we will see that the Feynman rules for this formulation of gravity are much simpler than the EH ones.
	\item To check that the amplitudes in the PCF coincide with the known expression for the graviton scattering amplitudes. We will compare all graviton-graviton amplitudes with any combination of incoming/outgoing helicities.
	\item To enlarge the discussion to a wider class of gravitational theories that can be described with the same formalism. Most importantly, we will see that the general theory from this class does not exhibit parity invariance, which is restored solely in the special case of GR. In fact, for the general theory different from GR, we find some $new$ scattering amplitudes that directly violate parity invariance.
\end{itemize}

 The PCF stems from the work previously done by Capovilla et al. in \cite{Capovilla:1989ac}, where the authors start from the Plebanski action \cite{Plebanski:1977zz} (see \cite{Krasnov:2009pu} for review material)

\be\nonumber
S_{\rm Pleb} (A, B,\eta)=\frac{1}{8\pi \im G}\int \left[B^i\wedge F^i (A)- \frac{1}{2} \eta_{ij} B^i\wedge B^j\right],
\ee

 where $\eta_{ij}$ is an appropriate Lagrange multiplier and $F^i$ the curvature of a (self-dual) SO(3) connection. Then one integrates out the two-form field  $B^i$ leaving the connection as the only dynamical field. However the resulting action in  \cite{Capovilla:1989ac} still depends on the Lagrange multiplier. The novelty of the work in \cite{Krasnov:2011pp} stands in the fact that $also$ the latter is integrated out and the action depends $only$ on the connection. Then, the theory takes the following simple form:
 
\be\label{Kirills_action}
S_{PCF}(A)= \im  \int_M f(F^i\wedge F^j),
\ee

where the $f$ is a properly and uniquely defined function for the case of General Relativity. However, the function $f$ above is purposely left unspecified because this framework can actually encompass a wider class of theories that propagate $two$ degrees of freedom as two polarization states of the graviton. In fact, the only constraints we require from the function $f$ are that it is a gauge-invariant function, that it is homogeneous of degree one and that it maps symmetric $3 \ \times \ 3$ matrices to complex numbers (see section \myref{General_action_section} for details). Then all the theories that can be written in such way can be regarded as gravitational, GR being a special case between them. Here, in fact, lies the power of the new formalism: the generic action \myref{Kirills_action} can describe in a compact form  a whole range of modifications of gravity (including the higher-derivative order terms of EH). We refer to this general theory represented by the action \myref{Kirills_action} as ``deformation of GR.'' One important remark about GR formulated in this terms, is that it makes sense (at the level of the action) only in the presence of a cosmological constant $\Lambda \neq 0$, as the action diverges in the $\Lambda \rightarrow 0$ limit \cite{Krasnov:2011pp}. As we will see, in fact, this theory is naturally expanded around a de Sitter (or anti-de Sitter) background.

As mentioned, one of the aims of this thesis is to introduce a new framework that allows to compute graviton scattering amplitudes in gravitational theories of the like of \myref{Kirills_action}.  In particular we are going to introduce the method to compute the amplitudes for all possible graviton-graviton scattering processes, both in the GR case and in the general theory. We will see that the actual computations, once defined the Feynman rules, % in ``spinor'' form,
are rather simple. This is  a direct consequence of the brevity and compactness of the variations of the action \myref{Kirills_action}. However, we first need to deal with some technical issues that are not present in the EH expansion. For instance, as we mentioned, the theory is naturally expanded around a de Sitter background, whereas quantum field theory(QFT) scattering amplitudes are usually defined in Minkowski. Therefore care will have to be taken when dealing with the limit to Minkowski. Further, we are working in a formulation of gravity based on a complex connection field, but we know that the graviton field must be ``real,'' hence we need to introduce also a proper reality condition\footnote{This problem exists in all formulations based on the Plebanski approach, where one trades six real components of the spin connection $\omega^{IJ}$ with indices in $SO(3,1)$, for three complex components of the self-dual connection $A^{i}=\omega^{0i}-1/2\im \e^{j}{}_{kl}\omega^{kl}$ with indices in $SO(3)_{\mathbb C}$. See \cite{Rovelli:1991zi} for review material.}. All this technicalities will be resolved in chapter \myref{General_PCF_chapter}.

We start chapter \myref{General_PCF_chapter} reviewing the classical PCF theory from \cite{Krasnov:2011up} \cite{Krasnov:2011pp}. In particular, in the sections \myref{Section_From_PB_to_GR}-\myref{section_free_theory}, we give a formal introduction to the formalism by deriving the general action \myref{Kirills_action} in details and studying its variations. Subsequently, in section \myref{Hamiltonian_analysis_section}, we deal with the Hamiltonian analysis of the theory and we show that, indeed, any of the theories we consider propagates two degrees of freedom.
In section \myref{Reality_condition_section}, instead, we discuss the issue of the reality condition. In section \myref{Canonical_transformation_section} we show, with a canonical transformation, what is the relation between this formulation of GR and the usual metric one. Then, in \myref{Mode_expansion_section}, we carry out the mode expansion of the connection field and we second quantise the theory. Finally, in section \myref{Discrete_symmetries_section}, we analyse its $CPT$ symmetries. An important conclusion of this chapter is that the mode decomposition of the connection field reveals that there is a certain asymmetry between the polarisation states. %Also, as a consequence of working in de Sitter space,
%Furthermore we will see in this chapter that the two polarisations are to be treated very differently. 
In fact, for all intents and purposes, the positive helicity graviton should be considered massive. All these sections follow the paper by this author $et$ $al.$ \cite{Delfino:2012zy}.

In the following chapter \myref{Amplitudes_chapter}, we compute the actual amplitudes. We start, in section \myref{LSZ_section}, by applying the usual QFT technology to the quantised theory obtained in the previous chapter to derive a prescription for how the scattering amplitudes can be computed. We also deal with the complication of working in de Sitter and taking the due Minkowski limit. After writing down the propagator in section \myref{Gauge_fixing_propagator_section} and the interaction vertices in \myref{Interactions_section}, we introduce the spinor formalism in sections \myref{spinor_technology_section} and \myref{Feynmans_rules_section} that we employ in the subsequent computations. Finally in section \myref{graviton_graviton_scattering_section} we find the expressions for all the possible tree-level graviton-graviton scattering amplitudes. These, in the GR limit, are only the processes that preserve helicities combinations, i.e. $++\rightarrow ++$, $+-\rightarrow +-$ and $--\rightarrow --$. In fact, in the special case of GR,  the amplitude for any other plus and minus combination is automatically zero. This is normal in GR and it is a consequence of parity invariance together with the particular structure of the interaction vertex.  We get, with our formalism, the same expressions for the GR amplitudes derived with other methods \cite{Bern:2002kj}. Most importantly, for the general action (\ref{Kirills_action}), we also find some new non-zero amplitudes which are not present in the GR case, like $++\rightarrow +-$ and $++\rightarrow --$. These new amplitudes derive from the new interaction vertices that disappear in the GR limit. This chapter follows the paper by this author  $et$ $al.$ \cite{Delfino:2012aj}.

We conclude with some remarks on the new amplitudes and on how the pure connection formulation, which correctly reproduced the known results for GR, can be further explored and studied.

%%%%%%%%%%%%%%%%%%%%%%%%%%%%%%%%%%%%%%%%%%%%%%%%%%%%%%%%%%%
%%%%%%%%%%%%%%%%%%%%%%%%%%%%%%%%%%%%%%%%%%%%%%%%%%%%%%%%%%%
\chapter{General Pure Connection Formulation}\label{General_PCF_chapter}
%%%%%%%%%%%%%%%%%%%%%%%%%%%%%%%%%%%%%%%%%%%%%%%%%%%%%%%%%%%
%%%%%%%%%%%%%%%%%%%%%%%%%%%%%%%%%%%%%%%%%%%%%%%%%%%%%%%%%%%

In this chapter we describe in depth a class of ``pure connection'' diffeomorphism invariant gauge theories. We will see how a whole class of gravitational theories, of which GR is a special case, can be described by a general formalism. Further, we provide a second quantisation of the general theory and discuss its discrete symmetries.

 We first follow \cite{Krasnov:2011up} and derive from Plebanski the Pure Connection Formulation of GR. 
Then, in section \myref{General_action_section}, we introduce the more general action that will encompass the GR case, derived in the previous section, as a member of a class of theories that propagate two degrees of freedom. This action is central in all our work.
After stating the general action we will perform the Hamiltonian analysis of the corresponding linearised theory in section \myref{Hamiltonian_analysis_section}. Further, in section \myref{Reality_condition_section}, we deal with the needed reality condition.
In section \myref{Canonical_transformation_section} we investigate the relation between this formulation of GR and the usual metric one.
We conclude, in section \myref{Mode_expansion_section}, with the mode expansion of the connection field and, in section \myref{Discrete_symmetries_section}, with a discussion on its discrete symmetries $CPT$.

 The chapter follows the paper by this author $et$ $al.$ \cite{Delfino:2012zy}, extending on the work done in the paper   \cite{Krasnov:2011up} by K. Krasnov. In particular the Hamiltonian analysis of the classical theory in section \myref{Hamiltonian_analysis_section} is done differently from the paper \cite{Krasnov:2011up} as we perform it by introducing two new differential operators $D$ and $\bar{D}$. This operators also facilitate the formulation of the
reality condition (section \myref{Reality_condition_section}), which was first introduced in the above mention paper of this author. The section deriving the quantisation of the connection field (\ref{Mode_expansion_section}) and the section dealing with the discrete symmetries of the quantum version of the theory (\ref{Discrete_symmetries_section}) were also first introduced in the paper \cite{Delfino:2012zy}.

%%%%%%%%%%%%%%%%%%%%%%%%%%%%%%%%%%%%%%%%%%%%%%%%%%%%%%%%%%%
\section{From Plebanski to Pure Connection GR}\label{Section_From_PB_to_GR}
%%%%%%%%%%%%%%%%%%%%%%%%%%%%%%%%%%%%%%%%%%%%%%%%%%%%%%%%%%%

We report here how to obtain the pure connection action that is central in all our work. The method we employ in this section follows the one used in the appendix of \cite{Krasnov:2011up}. 
We start from Plebanski formulation of gravity \cite{Plebanski:1977zz}:

\be
S_{\rm Pleb} (A, B,\Psi)=\frac{\im}{8\pi  G}\int \left[B^i\wedge F^i (A)- \frac{1}{2} \left( \Psi^{ij} + \f{\Lambda}{3}\d^{ij} \right) B^i\wedge B^j\right],
\ee

where $\im$ is the complex unit, $G$ is the Newton constant, $\Lambda$ is the cosmological constant, $B$ is an $\mathfrak{su}(2)$-valued two form field, $F^i(A) = dA^i+(1/2)\e^{ijk}A^j\wedge A^k$ is the curvature of a complex $SO(3)$ connection $A^i_\m$ and $\Psi^{ij}$ is a symmetric traceless Lagrange Multiplier (for a comprehensive review on the subject we refer to \cite{Krasnov:2009pu}). 
 %In particular we note that the conditions imposed on the Lagrange multiplier imply that it can be phrased in terms of a symmetric tensor $\tl{\Psi}$ whose trace-part has been projected away: $\Psi^{ij} = \tilde{\Psi}^{ij} - \d^{kl}(\tilde{\Psi}^{kl}/3) \ \d^{ij}$.

Varying with respect to $B$, $A$ and $\Psi$ we obtain the following equations of motion:

\begin{align}
F^i (A) &=  \left( \Psi^{ij} + \f{\Lambda}{3}\d^{ij} \right) B^j, \label{EOM_B}
\\
DB^{ij} &= 0,
\\
 B^i\wedge B^j &= \f{\d^{ij}}{3}\d^{kl} B^k\wedge B^l.
\end{align}

If we assume that the matrix $(\Psi+\Lambda/3\d)^{ij}$ is invertible, we can plug back \myref{EOM_B} into the action to obtain:

\be\label{Pure_connection_with_LM}
S (A, \Psi)=\frac{\im}{16\pi  G}\int \left( \Psi^{ij} + \f{\Lambda}{3}\d^{ij} \right)^{-1} F^i\wedge F^j.
\ee

The one thing left to do is to integrate out the Lagrange Multiplier. Before we perform this last step, let us introduce the following notation:

\be\label{Def_X}
\tilde{X}^{ij} = \f{1}{4}\tilde{\e}^{\m\n\r\s} F^i_{\m\n} F^j_{\r\s}. %= *\left( F^i\wedge F^j\right),
\ee

Here $\tilde{e}^{\m\n\r\s}$ is the completely anti-symmetric densitised tensor and $\tilde{X}^{ij}$ is a densitised matrix valued in the second symmetric power of the Lie algebra, i.e. $\tilde{X}^{ij}\in \mathfrak{su}(2)\otimes_s \mathfrak{su} (2).$ We have also used the Levi-Civita symbol $\tilde{\e}$ that can be defined on any orientable manifold without the use of a metric.\footnote{ The conventions we use here are $\tilde{\e}^{0123} = 1$ which implies $dx^\m\wedge dx^\n\wedge dx^\r\wedge dx^\s = \tilde{\e}^{\m\n\r\s}d^4x$.} Furthermore, it is convenient to rescale the Lagrange multiplier field $\Psi$ to absorb the constant in front of the action. Finally we define the (very small) parameter

\be\label{Alpha_def}
\a = \frac{16\pi \Lambda  G}{3} =  \f{M^2}{M_p^2} \sim 10^{-120}, \quad M^2 = \f{\Lambda}{3},\ M_P^2=\f{1}{16 G \pi};
\ee
where we have introduced the $M$ and $M_p$ constants that will frequently appear throughout this thesis.

Thus the \myref{Pure_connection_with_LM} with the rescaled fields becomes
 
\be\label{Pure_connection_with_LM3}
S (A, \Psi)=\im \int dx^4 \ \left( \tilde{\Psi}^{ij} + \a \d^{ij} \right)^{-1} \tilde{X}^{ij}. 
\ee

We now want to integrate out the rescaled field $\tl{\Psi}$. In the following we will assume that solution for $\tl{\Psi}$ can be written as a function of the matrix $X$ that admits representation as a series in powers of $\a$. 
To simplify the computation, we observe that we can apply an $SO(3)$ rotation to $X$ to diagonalise  it ( which is always possible at least locally) and look for solutions to $\tl{\Psi}$ which are also diagonal. Therefore we parametrise $\tl{\Psi}$ as a traceless diagonal matrix with elements $a$, $b$ as in $\tilde{\Psi}=diag(a,b,-(a+b))$ and, analogously, $X = diag(\lambda_1, \lambda_2, \lambda_3)$.

We are left with the following functional to vary with respect to $a$ and $b$ and substitute the solutions back into the action:
\be
F[a,b] = \f{\lambda_1}{\a+a} +\f{\lambda_2}{\a+b} +\f{\lambda_3}{\a-(a+b)}. 
\ee

Assuming none of the denominators are null, we can solve $\d F = 0$, which yields the two equations:

\be
(\a+a)^2\lambda_3 = (\a-(a+b))^2\lambda_1, \quad (\a+b)^2\lambda_3 = (\a-(a+b))^2\lambda_2.
\ee

Taking the positive branch of the square roots of either equations  we obtain expressions for $(\a+a)$ and $(\a+b)$, then summing the two and applying some simple algebra manipulations we obtain

\begin{align}\label{Solutions_of_a1_b1}
\nn   \a+a  &= 3\a \f{ \sqrt{ \lambda_1} }{\sqrt{ \lambda_1 } +\sqrt{ \lambda_2 }+\sqrt{ \lambda_3 } }
\\
\nn  \a+b  &= 3\a \f{ \sqrt{ \lambda_2} }{\sqrt{ \lambda_1 } +\sqrt{ \lambda_2 }+\sqrt{ \lambda_3 } }
\\
\a - (a+b) &= 3\a \f{ \sqrt{ \lambda_3} }{\sqrt{ \lambda_1 } +\sqrt{ \lambda_2 }+\sqrt{ \lambda_3 } }.
\end{align}

Therefore we see that substituting we have 
\be\label{GR_solved_functional}
F[\lambda] = \f{1}{3\a} \left( \sqrt{ \lambda_1 } +\sqrt{ \lambda_2 }+\sqrt{ \lambda_3 } \right)^2, 
\ee
which in the action translates to

\be\label{GR_Action_appendix}
S_{GR} (A)=\frac{\im}{ 16   \pi \Lambda  G}\int dx^4 \ \Tr{\sqrt{ \tl{X}} }^2. 
\ee

An alternative derivation of the same action is provided in section \myref{Appendix_AltDerivationOfAction}. Also, a similar analysis for theories that involve a $small$ modification of GR is presented in section \myref{Appendix_Mod_GR}.

Some considerations on the action in \myref{GR_Action_appendix} are in order.  The first thing is that we notice that there is not a well defined limit $\Lambda \rightarrow 0$, which was instead trivial in the Plebanski formulation. This is in fact an important characteristic of this kind of theories and special care will need to be taken when computing the scattering amplitudes (which are defined in Minkowski) in chapter \myref{Amplitudes_chapter}. Will will see in \myref{Background_section} that, indeed, the perturbative expansion of the theory is carried out on a de Sitter background and that, upon quantisation, the mode decomposition of the connection field in section \myref{Mode_expansion_section} will also reflect the presence of the constant $\Lambda$.

 Furthermore, due to the imaginary unit as a factor in front of the action, it is not obvious that this action describes a theory with unitary dynamics. Still, as we shall see in particular from the graviton scattering results, it describes the usual general relativity. The issue will be resolved in section \myref{Reality_condition_section} with the introduction of an appropriate reality condition.

%%%%%%%%%%%%%%%%%%%%%%%%%%%%%%%%%%%%%%%%%%%%%%%%%%%%%%%%%%%
\section{The action}\label{General_action_section}
%%%%%%%%%%%%%%%%%%%%%%%%%%%%%%%%%%%%%%%%%%%%%%%%%%%%%%%%%%%

After deriving the action in \myref{GR_Action_appendix}, we observe that one can generalise the theory to describe a class of diffeomorphism invariant gauge theories of the kind

\be\label{GEN_action}
S_f (A) = \im \int  \ f\left( F \wedge F\right),
\ee
where $F$ is again the curvature of a complex $SO(3)$ connection $A^i_\m$. Here however, differently from what we have in formula \myref{GR_Action_appendix}, we have introduced a function $f$ which is referred to as the ``defining function'' of the theory (for a thorough review on this action we refer to \cite{Krasnov:2012pd}). The only constraints we require on the function $f$ are that it is a gauge-invariant, homogeneous of degree one and that it maps symmetric $3 \ \times \ 3$ matrices to complex numbers, i.e. we have the following conditions:

\begin{itemize}
\item for $g \in SO(3)$ we have $f(Ad_g X) = f(X)$;
\item given $X\in \mathfrak{g} \otimes_s \mathfrak{g}$, then $f:X\rightarrow  \mathbb{C}$;
\item for $\a \in \mathbb{C}$ then $f(\a X)= \a f(X).$
\end{itemize}

 Then $f\left( F \wedge F\right)$ is a well defined 4-form and can be integrated. In fact, with our conventions $dx^\m\wedge dx^\n \wedge dx^\r \wedge dx^\s = \tl{\e}^{\m\n\r\s} d^4 x$, we have

\be 
\nn S_f (A) = \im \int \ f\left( \f{1}{4}F^i_{\m\n} F_{\r\s}d x^\m \wedge d x^\n \wedge d x^\r \wedge d x^\s  \right)
\\
\nn = \im \int \ d^4 x f\left( \f{1}{4}\tl{\e}^{\m\n\r\s}F^i_{\m\n} F_{\r\s}  \right)
\\ \label{gen_action_with_dx}
= \im \int \ d^4 x f\left( \tl{X}^{ij}\right), 
\ee
where we used \myref{Def_X} and 

\be\label{Def_X_with_dx}
dx^4 \tilde{X}^{ij} =  F^i\wedge F^j.
\ee

As we have seen, in the case of GR with non vanishing cosmological constant $\Lambda$ the defining function corresponds simply to the square of the trace of the square root, i.e. $f(X)\propto \Tr{\sqrt{X}}^2$. The first time this action was proposed was in \cite{Krasnov:2011pp}.

One important remark about this class of theories is that they are dynamically non-trivial, i.e. they describe propagating degrees of freedom as we will see below. The only exception is when the Hessian of $f$ is degenerate like for $f(X)=\Tr{X}$ which yields a topological theory without any propagating modes. However, as we will see in \myref{Matrices_for_gen_f} when we study the variations of the general action, this is a very special point in the theory space and that in general the Hessian is indeed non-zero.

Another important remark is that we note that there are no dimensionful constants involved in the definition of the theory \myref{GEN_action}. indeed, it is natural to take the dimensions of the connections to be those of $1/L$, $L$ being length (or, using the standard terminology, mass dimension one). The quantity $ F \wedge F$ is then of mass dimension 4, and due to the homogeneity of $f$, so is the Lagrangian. Thus, there are only dimensionless constants involved in the construction of the Lagrangian of our theory, and these are hidden as the parameters of the defining function $f$.

% We will also show later in \myref{Reality_condition_section} that the action in \myref{GEN_action} describes a self-interacting field with 2 propagating degrees of freedom.

%%%%%%%%%%%%%%%%%%%%%%%%%%%%%%%%%%%%%%%%%%%%%%%%%%%%%%%%%%%
\subsection{The equations of motion}
%%%%%%%%%%%%%%%%%%%%%%%%%%%%%%%%%%%%%%%%%%%%%%%%%%%%%%%%%%%

In this section we want to derive the equations of motion of the action \myref{GEN_action}. It is not hard to see the field equations are second order in derivatives (non-linear) partial differential equations for the connection components. 

We start from the action in (\ref{gen_action_with_dx}), 
% 
% by rewriting the action in \myref{GEN_action} ``extracting'' the volume form from the defining function $f$. Using  \myref{Def_X} and the conventions $dx^\m\wedge dx^\n \wedge dx^\r \wedge dx^\s = \tl{\e}^{\m\n\r\s} d^4 x$, we have
% 
% 
% 
% Then the action  \myref{GEN_action}  takes the form
% \be%\label{gen_action_with_dx}
% S(A)= \im \int d^4 x f(\tl{X}^{ij}).
% \ee
its first variation can be easily computed:

\be
\d S(A) = \im \int \ d^4 x \ \f{\partial f}{\partial \tilde{X}^{ij}} \ 2 F^i\wedge \d F^j.
\ee 

Then we recall

\be
\d F = \d\left( dA+\f{1}{2}[A,A] \right)= d \d A + \f{1}{2}\left([\d A,A]+[A,\d A] \right)= d \d A + [A,\d A],
\ee
which is equal, by definition, to $D_A \d A.$ Thus (dropping the constants) we have 
\be
\d S(A) \propto \int \f{\partial f}{\partial \tilde{X}^{ij}} F^i \wedge  D_A \d A^j. 
\ee 

Taking care of the by-parts integration involved, we can read off the equation of motion for the connection field:

\be\label{gen_theory_EOM}
 D_A \left( \f{\partial f}{\partial \tilde{X}^{ij}} F^i\right) = 0,
\ee

which is a set of second order partial differential equation. The fact that second order differential equations appear is reassuring, given that higher order field equations typically lead to instabilities in the theory.

%%%%%%%%%%%%%%%%%%%%%%%%%%%%%%%%%%%%%%%%%%%%%%%%%%%%%%%%%%%
\subsection{Background}\label{Background_section}
%%%%%%%%%%%%%%%%%%%%%%%%%%%%%%%%%%%%%%%%%%%%%%%%%%%%%%%%%%%
We are (eventually) interested in developing Feynman rules for the theories \myref{GEN_action}. One immediate difference with the case of metric-based GR is that we cannot directly expand around a background that corresponds to the Minkowski spacetime. Indeed, our action \myref{GEN_action}, strictly speaking, only describes the $\Lambda \ne 0$ situation, as it blows up if one sends $\Lambda \rightarrow 0.$ Thus, the best we can do (if we are after the Minkowski spacetime scattering amplitudes) is to expand around a constant curvature background and take the curvature scalar to zero at the end of the computation. This is the strategy that will be followed here. As we shall see below, the presence of the cosmological constant at intermediate stages of the computations will make available to us constructions that are simply impossible in the usual metric setting of zero $\Lambda.$

We shall consider perturbations around a fixed constant curvature background connection. To explain what constant curvature means in our setting let us start by describing a general homogeneous and isotropic in space $SO(3)$ connection. First, a general homogeneous in space connection is of the form

\be
A^i = a^{ij}(\eta)dx^j+b^i(\eta)d\eta,
\ee 
where we have indicated that the components can only be functions of the time coordinate $\eta$. It is obvious that we can kill the $b^i(\eta)$ components by a time-dependent gauge transformation. This leaves us with the first term only. We now require that the effect of and $SO(3)$ rotation of the coordinates $x^i$ (around an arbitrary centre) can be offset by an $SO(3)$ gauge transformation. This implies that $a^{ij}$ must be proportional to $\d^{ij}$ for all $\eta$\footnote{ The requirement of spherical symmetry implies that $a^{ij}$ is an $embedding$ of the Lie algebra $\mathfrak{so}(3)$ of the group of spatial rotations $SO(3)$ into the Lie algebra $\mathfrak{g}$ of the gauge group $G$, i.e. a map that sends  $\mathfrak{so}(3)$ commutators into $\mathfrak{g}$ commutators.  In case $G = SO(3)$ there is only one such (non-trivial) embedding where $a^{ij}\sim \d^{ij}$ \cite{Krasnov:2011hi}. }. Thus, we are led to consider the following connections:

\be\label{background_connection}
A^i= \f{c(\eta)}{\im} dx^i,
\ee 
where the function $c(\eta)$ is arbitrary, and we have introduced a factor $1/\im$ for future convenience. We now note that the curvature of this connection is given by

\be\label{Curvature_of_Background}
\nn F^i = d A^i+\f{1}{2}\e^{ijk}A^j \wedge A^k=
\\
\f{c'}{\im} d\eta \wedge dx^i - \f{c^2}{2}\e^{ijk} dx^j  \wedge  dx^k ,
\ee

where the prime denotes the derivative with respect to $\eta$. This means that we have 

\be
F^i\wedge F^j %= (\f{c'}{\im} d\eta \wedge dx^i - \f{c^2}{2}\e^{ikl} dx^k  \wedge  dx^l) \wedge (\f{c'}{\im} d\eta \wedge dx^j - \f{c^2}{2}\e^{j m n } dx^m  \wedge  dx^n ) 
\sim \d^{ij}.
\ee

Thus, for our chosen background \myref{background_connection} the matrix $\tilde{X}^{ij}$ is proportional to the identity matrix, which means the matrix of first derivatives of the function $f(X)$ is also proportional to the identity on the background. This implies that any connection \myref{background_connection} satisfies the field equation \myref{gen_theory_EOM}, as it reduces to 

\be\label{Bianchi_identity}
D_A F^i %= \nn  d F^i + \e^{ijk} A^j\wedge F^k = d \left( d A^i + \f{1}{2}\e^{ikl}A^k \wedge A^l \right) + \e^{ijk} A^j \wedge \left( d A^k + \f{1}{2} \e^{k m n} A^m\wedge A^n \right)
%\\ \nn = - \e^{ikl}  A^k \wedge d A^l   + \e^{ijk} A^j \wedge d A^k + \f{1}{2}  (\d^i_m \d^j_n -\d^i_n \d^j_m)  A^j \wedge A^m\wedge A^n
  = 0,
\ee

which is the Bianchi identity, therefore is satisfied by any connection. This happens for any $f$, i.e. for any of the theories in our theory space.

We now note that the curvature \myref{Curvature_of_Background} can be written as 

\be\label{Curvature_of_Background2}
\nn F^i = -c^2\left(
\f{\im c'}{c^2} d\eta \wedge dx^i + \f{1}{2}\e^{ijk} dx^j  \wedge  dx^k \right).
\ee
We can now chose the time coordinate conveniently so that

\be\label{c_equation}
\f{c'}{c^2}d \eta = dt,
\ee

and therefore we have

\be\label{Curvature_of_Background_final}
\nn F^i = -c^2\left(
\im  d t \wedge dx^i + \f{1}{2}\e^{ijk} dx^j  \wedge  dx^k \right) ,
\ee

where $c$ now should be thought as a function of $t$. In fact, solving \myref{c_equation} we have:

\be
c(t) = -\f{1}{t-t_0},
\ee

where $t_0$ is the integration constant. All in all, we see that, by an appropriate choice of the $t$ coordinate, we can rewrite the curvature of any connection \myref{background_connection} as 

\be\label{Curvature_of_Background_Sigma}
F^i = - M^2 \S^i ,
\ee

where 
\be\label{self_dual_forms}
\S^i  = a^2\left(
\im  d t \wedge dx^i + \f{1}{2}\e^{ijk} dx^j  \wedge  dx^k \right)
\ee

are self-dual two forms for the de Sitter metric

\be\label{de_sitter_metric}
ds^2 = a^2 \left( -dt^2 + \sum_i (dx^i)^2\right),
\ee
and 
\be\label{a_def_with_conformal_time}
a(t) = -\f{1}{M(t-t_0)}
\ee
is the usual de Sitter scale factor as a function of the (conformal) time $t$. % Analogously we define the Minkowski two form basis dropping the conformal factor in front $\S^i_M = \im  d t \wedge dx^i + \f{1}{2}\e^{ijk} dx^j  \wedge  dx^k$. 
 For more details on the algebra of the self dual forms in \myref{self_dual_forms} we refer to \myref{Self_dual_forms_Appendix}. 

Few remarks are in order. Note that we have introduced an arbitrary dimensionful parameter $M$ in \myref{Curvature_of_Background_Sigma}. This parameter is directly related to the radius of curvature of the de Sitter metric \myref{de_sitter_metric}. It is completely arbitrary, as we can always rescale both $M$ and $\S $ in \myref{Curvature_of_Background_Sigma} without changing the curvature. However, once introduced, it determines the metric and thus determines how all scales in the theory are measured. The condition \myref{Curvature_of_Background_Sigma}, which as we saw can be always achieved by choosing the time coordinate appropriately, is our constant curvature condition for the background connection. The essence of this condition is that introduces a (background) metric into our background-free (up to now) description, and fixes how all scales are measured. 

It is worth discussing the construction that introduced a metric into our so far metric-free theory in more details. This is a geometrical construction known for many years, and is in particular due to \cite{Urbantke:1984eb}. The idea is that when a triple of curvatures $F^i$ of the connection $A^i$ is linearly independent, the 3-dimensional space that it spans (in the space of all 2-forms) can be declared to be the space of self-dual 2-forms for some metric. This construction is the so-called Urbantke metric and can be defined modulo conformal transformations:

\be
\tilde{g}_{\m\n} \propto \tilde{\e}^{\a \b \gamma \d } \e^{ijk} F^i_{\m \a} F^j_{\n \b} F^k_{\gamma \d}.
\ee

  This is precisely how the metric \myref{de_sitter_metric} appeared from the background connection \myref{background_connection}. We have also made a further choice of the conformal factor so that the connection becomes one of constant curvature in the sense of equation \myref{Curvature_of_Background_Sigma}. Fixing $M$ in that equation to be constant eliminates the conformal freedom in the choice of the metric, up to constant rescalings. A choice of a particular constant $M^2$ in that equation is then equivalent to a choice of units in which all other quantities in our theory are measured. In this sense $M$ is not a parameter of the theory, it is rather a scale in terms of which all other scales in the theory get expressed. We shall see, in chapter \myref{Amplitudes_chapter}, how the gravitons' interaction strength (Newton constant) appears as constructed out of $M$ and the dimensionless coupling constants present in our theory.

\subsection{A convenient way to write the action}
Let us now consider the value of $\tl{X}^{ij}$ at the background. We have

\be
\tl{X}^{ij} \hat{=} \f{M^4}{4}\tl{\e}^{\m\n\r\s}\S^i_{\m\n}\S^j_{\r\s}= 2\im M^4\sqrt{-g}\d^{ij},
\ee

where our convention is that the hat means ``evaluated at the background''. Here we made use of the formulae in appendix \myref{Self_dual_forms_Appendix}. It is very convenient to rescale the $\tl{X}$ variable by $2\im M^4 \sqrt{-g}$ so that the result equals to the Kronecker delta on the background. Thus, we introduce

\be
\hat{X}^{ij} := \f{ \tl{X}^{ij} }{ 2\im M^4 \sqrt{-g} } \ \hat{=} \ \d^{ij}.
\ee
We want to stress the fact that, adopting this rescaling, we are not working with densitised tensors anymore.
We can now rewrite the general gravity action \myref{gen_action_with_dx} in terms of $\hat{X}$. We have

\be\label{gen_action_with_dx_rescaled}
S(A) = - 2M^4 \ \int \ d^4 x \ \sqrt{-g}f(\hat{X}^{ij}). 
\ee
For the GR action this becomes:

\be\label{GR_Action_Chapter1}
S_{GR}(A) = - \f{2}{3} M^2 M^2_p \ \int \ d^4 x \ \sqrt{-g} \left( \Tr{\hat{X}^{ij} } \right)^2, 
\ee

where we re-introduced the constants

\be
 M^2 = \f{\Lambda}{3}, \quad M_P^2=\f{1}{16 G \pi}.
\ee

It then becomes a simple exercise to compute the variations of the action, which we will do in section \myref{Variations_Section}.

%%%%%%%%%%%%%%%%%%%%%%%%%%%%%%%%%%%%%%%%%%%%%%%%%%%%%%%%%%%
\subsection{Evaluating the action at the background}
%%%%%%%%%%%%%%%%%%%%%%%%%%%%%%%%%%%%%%%%%%%%%%%%%%%%%%%%%%% 
Let us also discuss the value of the actions \myref{gen_action_with_dx_rescaled} and \myref{GR_Action_Chapter1} when evaluated on the background. We have, for the general actions

\be\label{gen_action_at_background}
S(A)\ \hat{=} \ -2 M^4 f(\d) \int \ d^4 x \sqrt{-g},
\ee 
and for GR

\be
S(A)_{GR}\ \hat{=} \ -6 M^2_P M^2 \int \ d^4 x \sqrt{-g}= -\f{\Lambda}{8\pi G}\int  \ d^4 x \sqrt{-g},
\ee 
which is the same as the value of the Einstein-Hilbert action

\be
S_{EH}(g)=-\f{1}{16\pi G}\int d^4 x \sqrt{-g}(R-2\Lambda)
\ee
evaluated on the de Sitter metric \myref{de_sitter_metric}.
%  In fact, reminding the Einstein equation of motions
% 
% \be
% R_{\m\n}-\f{1}{2}R g_{\m\n} + \Lambda g_{\m\n} = 8\pi G T_{\m\n}
% \ee
% in vacuum and contracting with the inverse metric $g^{\m\n}$
% 
% \be
% R-\f{1}{2}4 R + 4\Lambda = 0,
% \ee
% which implies 
% 
% \be
% R = 4 \Lambda.
% \ee
We see from \myref{gen_action_at_background} that for a general theory the dimensionless quantity $f(\d)$ plays the role of the combination $3M_P^2/M^2$ in the case of GR. We emphasise, however, that for a general theory there is no notion of Planck constant, at least not until graviton interactions are considered. In chapter \myref{Amplitudes_chapter} we compute the graviton interactions strength and will extract an appropriate dimensionful coupling constant this way. It is however, not guaranteed that the Planck mass obtained from this Newton constant will be related the the dimensionless parameter $f(\d)$ in front of the background-evaluated action in exactly the same way as in GR.

%%%%%%%%%%%%%%%%%%%%%%%%%%%%%%%%%%%%%%%%%%%%%%%%%%%%%%%%%%%
\section{Variations}\label{Variations_Section}
%%%%%%%%%%%%%%%%%%%%%%%%%%%%%%%%%%%%%%%%%%%%%%%%%%%%%%%%%%% 
%[Reminding $F^i_{\m\n}= 2 (D A)_{\m \n}$ ] 
We start by computing the variations of $\hat{X}$, as a function of the connection, evaluated at the background $\hat{X} \ \hat{=} \ \d^{ij}.$ We have

\be
\nn \d \hat{X}^{ij} \ \hat{=} \ - \f{1}{M^2}\S^{(i \m\n}D_\m \d A^{j)}_\n,
\\
\nn \d^2 \hat{X}^{ij} \ \hat{=} \ \f{1}{\im M^4}\e^{\m\n\r\s} D_\m \d A^i_\n D_\r \d A^j_\r - \f{1}{M^2}\S^{(i\m\n}\e^{j)kl}\d A^k_\m \d A^l_\n,
\\
 \d^3 \hat{X}^{ij} \ \hat{=} \ \f{3}{\im M^4}\e^{\m\n\r\s} D_\m \d A^{(i}_\n \e^{j)kl}\d A^k_\r \d A^l_\s.
\ee

Finally, the fourth variation is zero $\d^4 \hat{X} = 0$ even away from the background. In all expressions above $D_\m$ is the covariant derivative with respect to the background connection. Thus, it is important to keep in mind that $D$'s do not commute:

\be\label{curvature_from_derivative_commutator}
2 D_{[\m}D_{\n]}V^i = \e^{ijk} F^j_{\m\n} V^k,
\ee
for an arbitrary Lie algebra valued function $V^i$. here $F^i_{\m\n}$ is the background curvature \myref{Curvature_of_Background_Sigma}. thus, the commutator \myref{curvature_from_derivative_commutator} is of order $M^2.$ this has to be kept in mind when (in the limit $M \rightarrow 0$) replacing the covariant derivatives $D$ with the usual partial derivatives.

%%%%%%%%%%%%%%%%%%%%%%%%%%%%%%%%%%%%%%%%%%%%%%%%%%%%%%%%%%%
\subsection{Variations of the general action}
%%%%%%%%%%%%%%%%%%%%%%%%%%%%%%%%%%%%%%%%%%%%%%%%%%%%%%%%%%%
We will now explain a procedure that can be used for computing the perturbative expansion of the action \myref{gen_action_with_dx_rescaled}. It is completely algorithmic, and is not hard to implement to an arbitrary order. In this chapter we only need the second variation, but we will explain the general procedure already here since once the general principle is understood, it is not hard to implement it to get the interactions as well. First, let us define a convenient notation

\be
f^{(n)}_{ijkl\ldots} = \f{\partial^n f}{\partial \hat{X}^{ij} \partial \hat{X}^{kl}\ldots}\Bigg|_\d,
\ee
where the derivatives are all evaluated at the background $\hat{X} = \d^{ij}.$ The variations of the action are then given by (we have dropped the $\sqrt{-g}$ for brevity):

\be\label{variations_gen_action}
\nn \d S\ \hat{=} \ -2M^4\int \ f^{(1)}_{ij} \d \hat{X}^{ij} , \quad \nn \d^2 S\ \hat{=} \ -2M^4\int \bigg[ f^{(2)}_{ijkl} \d \hat{X}^{ij}\d \hat{X}^{kl}+ f^{(1)}_{ij}\d^2 \hat{X}^{ij}\bigg],
\\
\nn \d^3 S\ \hat{=} \ -2M^4\int \bigg[ f^{(3)}_{ijklmn}\d \hat{X}^{ij} \d \hat{X}^{kl}\d \hat{X}^{mn}+ 3 f^{(2)}_{ijkl} \d^2 \hat{X}^{ij}\d \hat{X}^{kl}+ f^{(1)}_{ij}\d^3 \hat{X}^{ij}\bigg],
\\
\nn \d^4 S\ \hat{=} \ -2M^4\int \bigg[f^{(4)}_{ijklmnpq}\d \hat{X}^{ij}\d \hat{X}^{kl}\d \hat{X}^{mn}\d \hat{X}^{pq}+6 f^{(3)}_{ijklmn}\d^2 \hat{X}^{ij} \d \hat{X}^{kl}\d \hat{X}^{mn}
\\
+ 4 f^{(2)}_{ijkl} \d^3 \hat{X}^{ij}\d \hat{X}^{kl}+ 3 f^{(2)}_{ijkl} \d^2 \hat{X}^{ij}\d^2 \hat{X}^{kl} \bigg].
\ee

Below we shall explain how the derivative matrices appearing here can be parametrised conveniently. However, let us first consider the special case of the GR action.

%%%%%%%%%%%%%%%%%%%%%%%%%%%%%%%%%%%%%%%%%%%%%%%%%%%%%%%%%%%
\subsection{Variations of the GR action}
%%%%%%%%%%%%%%%%%%%%%%%%%%%%%%%%%%%%%%%%%%%%%%%%%%%%%%%%%%%

For the case of GR we have
\be
f_{\rm GR}(\hat X)=\frac{M_p^2}{3M^2} \Tr{\sqrt{\hat X}}^2.
\ee
The variations are now easily obtained by defining $Y=\sqrt{\hat{X}}$, and writing
\be\label{GR'''}
S_{\rm GR}[A] =- \frac{2}{3} M^2_p M^2 \int \left( \mathrm{Tr} Y  \right)^2,
\ee
where we have dropped the integration measure $d^4x \sqrt{-g}$ for brevity. The variations are then easily computed:
\be
\nn \delta S_{\rm GR}[A]=-\frac{2}{3} M^2_p M^2 \int  2   \  \Tr{Y}   \Tr{  \delta Y} , 
\\
\nn \delta^2 S_{\rm GR}= -\frac{2}{3} M^2_p M^2 \int  2 \ \left[  \ \Tr{ \delta Y} \Tr{ \delta Y}    +\Tr{Y}     \  \Tr{ \delta^2 Y }\right], 
\\
\nn \delta^3 S_{\rm GR}= -\frac{2}{3} M^2_p M^2 \int  2 \ \left[  3 \  \Tr{ \delta Y} \Tr{ \delta^2 Y} 
  +\Tr{Y}     \  \Tr{ \delta^3 Y}\right], 
\\
\nn \delta^4 S_{\rm GR}=-\frac{2}{3} M^2_p M^2  \int  2 \ \left[  3  \Tr{\delta^2 Y}  \Tr{\delta^2 Y} +4 \Tr{\delta Y}  \Tr{\delta^3 Y}   +\Tr{Y}   \  \Tr{\delta^4 Y}\right].
\\
\ee

It thus remains to obtain a relation between the variations of $Y$ and those of $\hat{X}$. This is easily done by varying the relation $Y^2=\hat{X}$ (any required number of times), and then solving the resulting equations for $\delta^k Y$. We only need these variations on the background, where we have $Y^{ij}\hat{=}\delta^{ij}$. This procedure gives:
\begin{align}
 \nn & \delta Y\hat{=}  \frac{1}{2}\delta \hat{X},
\\
\nn  & \delta^2 Y \hat{=} \frac{1}{2} \delta^2 \hat{X} - \delta Y\delta Y =\frac{1}{2}\left( \delta^2 \hat{X} -\frac{1}{2}\delta \hat{X} \delta \hat{X}  \right),
\\
\nn  & \delta^3 Y  =  \frac{1}{2}\delta^3 \hat{X} -\frac{3}{2} \delta Y  \delta^2 Y- \frac{3}{2}  \delta^2 Y  \delta Y = \frac{1}{2}\delta^3 \hat{X} -\frac{3}{8}\left(   \delta^2 \hat{X}\delta \hat{X} +  \delta \hat{X}\delta^2 \hat{X}-\delta \hat{X}\delta \hat{X}\delta \hat{X}\right),
\\
  & \delta^4 Y  = -2 \delta Y  \delta^3 Y  - 2 \delta^3 Y \delta Y  - 6 \delta^2 Y  \delta^2 Y.
   \end{align}

The above results can be put into the general form \myref{variations_gen_action} by writing:

\be
 \nn (3M^2/M_p^2) f^{(1)}_{ij}=3\delta_{ij},\\  \label{f2-GR}
  (3M^2/M_p^2)f^{(2)}_{ijkl}=-\frac{3}{2}P_{ijkl},\\ 
 \nn (3M^2/M_p^2)f^{(3)}_{ijklmn} =\frac{9}{4} \sum_{\text perm} \frac{1}{3!} P_{ijab}P_{klbc}P_{mnca}+\frac{1}{2}\left(\delta_{ij} P_{klmn}+\delta_{kl} P_{ijmn}+\delta_{mn} P_{ijkl}\right)
 \\ \nonumber 
 (3M^2/M_p^2)f^{(4)}_{ijklmnpq}=-\frac{39}{16} \sum_{perm}\f{1}{3} \ P^{ij|kl}P^{mn|pq}  + \ldots,
 \ee
 where
 \be\label{P}
 P_{ijkl}:=\frac{1}{2}(\delta_{ik}\delta_{jl} +\delta_{il}\delta_{jk}) - \frac{1}{3}\delta_{ij}\delta_{kl}
 \ee
 is the projector on the symmetric tracefree matrices, and the dots in the last formula stand for terms containing at least one $\delta_{ij}$ in one of the 4 external ``legs''. The sum over permutations in the last two formulas is needed to make the result on the right-hand-side symmetric. Eventually we are going to contract $f^{(3)}, f^{(4)}$ with copies of the same matrix $\delta\hat{X}^{ij}$, and this sum over permutations (with the associated combinatorial factor) will disappear. Also, the reason why we don't write the remaining terms in the expression for $f^{(4)}$ is that (in chapter \ref{Amplitudes_chapter}) we shall see that these terms will not play any role (in the 4-vertex) as they will be killed on-shell by the external states, or killed by the symmetries of the propagator when the vertices are used in Feynman graphs. 

%%%%%%%%%%%%%%%%%%%%%%%%%%%%%%%%%%%%%%%%%%%%%%%%%%%%%%%%%%%
 \subsection{Matrices $f^{(n)}_{ijkl...}$ for a general
 $f$}\label{Matrices_for_gen_f}
%%%%%%%%%%%%%%%%%%%%%%%%%%%%%%%%%%%%%%%%%%%%%%%%%%%%%%%%%%%
For the case of a general theory we can, to a large extent, fix the derivatives of the function $f$ evaluated at the background $\hat{X}^{ij} = \d^{ij}$ from the properties of $f$ itself. The key point is that we know $f$ to be an $SO(3)$ invariant function. The background that we work with is also $SO(3)$ invariant. Thus, the same will be true for the matrices $f^{(n)}_{ijkl\ldots}$. This, in particular, implies that the matrix of the first derivatives must be proportional to $\d^{ij}$ %for the same reasons A was proportional to $\d^{ij}$.
 The proportionality coefficient can then be fixed from the homogeneity property of $f$ that implies

\be\label{f_and_homogeneity}
\f{\partial f}{\partial \hat{X}^{ij}}\hat{X}^{ij} = f.
\ee

Thus, we have

\be
f^{(1)}_{ij} = \f{f(\d)}{3}\d^{ij}.
\ee

We also know from \myref{gen_action_at_background} that $f(\d)$ is the analogue of the parameter $3M_p^2/M^2$ in GR for a general theory. We can now differentiate the equation \myref{f_and_homogeneity} once with respect to $\hat{X}^{ij}$ and obtain

\be
\f{\partial^2 f}{\partial \hat{X}^{ij} \partial \hat{X}^{kl} } \hat{X}^{ij} = 0.
\ee

In other words, the background itself is among the flat directions of the Hessian of $f$. This, together with the $SO(3)$ invariance of the matrix $f^{(2)}_{ijkl}$ implies that it is of the form

\be\label{f_second_variation}
f^{(2)}_{ijkl}=-\f{g^{(2)}}{2}P_{ijkl},
\ee
where $g^{(2)}$ is some parameter and $P_{ijkl}$ is the projector \myref{P} introduced above. This must be true for any $f$. Note that this is also true for the function $f(\hat{X})\sim \Tr{\hat{X}}$, i.e. for the topological theory, but in this case we have $g^{(2)} = 0$. We shall see that there are propagating degrees of freedom whenever $g^{(2)} \neq 0$. Finally, we note that we put a minus sign in \myref{f_second_variation} because there is one in the case of GR, see \myref{f2-GR}. It is natural to be interested in theories that are not too far from GR, and so it is natural to have the same sign  in \myref{f_second_variation} as in GR. For this reason we shall assume $g^{(2)}>0$ in what follows.
The higher derivatives $f^{(n)}_{ij\ldots}$ can be all determined in a similar fashion. This, one takes higher and higher derivatives of the equation \myref{f_and_homogeneity} and evaluates the result on $\hat{X}^{ij}=\d^{ij}$. One gets 

\be\label{recursive_relation_for_f_derivatives}
f^{(n)}_{i_1 j_1 i_2 j_2 \ldots i_n j_n}\d^{i_n j_n}+(n-2)f^{(n-1)}_{i_1 j_1 i_2 j_2 \ldots i_{n-1} j_{n-1} }=0,
\ee

which is a recursive relation for the matrices of derivatives. We see that the new independent term that appears at each order is always of the form of $n$ projectors \myref{P} contracted with each other in a loop, with a symmetrisation over index pairs $i j$ later taken to form a completely symmetric expression. There are also terms where the projectors are contracted in smaller groups. Thus, we can write

\be\label{f_nth_variation}
f^{(n)}_{i_1 j_1 i_2 j_2 \ldots i_n j_n} = (-1)^{n-1} g^{(n)}\sum_{perm} \f{1}{n!}P_{i_1 j_1 a_n a_1}P_{i_1 j_1 a_1 a_2}\ldots P_{i_n j_n a_{n-1} a_n}+\ldots
\ee
where the dots denote terms that contain smaller groups of $P$ contractions, as we have terms that do not vanish when contracted with $\d^{ij}$ in one of the channels. The coefficients in front of these latter terms are related to the lower $g^{(n)}$ via \myref{recursive_relation_for_f_derivatives}. For example, for $f^{(3)}$ we have

\be\label{f3}
f^{(3)}_{ijklmn} =  g^{(3)} P^{ij|ab}P^{kl|cb}P^{mn|ac} - \frac{1}{3}(-\frac{g^{(2)}}{2})\left( P^{ij|kl}\delta^{mn}+P^{kl|mn}\delta^{ij}+P^{mn|ij}\delta^{kl} \right).
\ee
Furthermore, the matrix of fourth derivative we have

\be\label{f_fourth_variation}
\nn f^{(4)}_{ijklmnpq} = - g^{(4)}\sum_{perm}\f{1}{3} \ P^{ij|kl}P^{mn|pq}  
\\
  - \frac{8}{3}g^{(3)} \sum_{perm}  \frac{1}{4} \delta^{ij}P^{pq|ab}P^{kl|cb}P^{mn|ac}  
-\frac{2}{3}g^{(2)}  \sum_{perm} \frac{1}{6} P^{pq|kl}\delta^{ij} \delta^{mn}.
\ee

The above relation \myref{f_nth_variation} for the derivatives of $f$ makes it clear that for a general theory there is an infinite number of independent coupling constants $g^{(n)}$, with a number of new couplings appearing at each order of the derivative of the defining function. In turn, we could have chosen to parametrise $f$ by its independent couplings $g^{(n)}$ . We also want to stress  that all these couplings are dimensionless. For GR we get

\be\label{GR_constants}
f_{GR}(\d) = \f{3M_p^2}{M^2}, \quad g^{(2)}_{GR}= \f{M_p^2}{M^2}, \quad g^{(3)}_{GR}= \f{3M_p^2}{4M^2}, \quad g^{(4)}_{GR}= \f{13 M_p^2}{16 M^2} ,
\ee
where, again, $M^2_p = 1/16\pi G$ and $M^2=\Lambda/3$.

We would like to emphasise that the procedure used to obtain the action variations is completely algorithmic and can be continued to arbitrary order without any difficulty. 

%%%%%%%%%%%%%%%%%%%%%%%%%%%%%%%%%%%%%%%%%%%%%%%%%%%%%%%%%%%
\section{Free theory}\label{section_free_theory}
%%%%%%%%%%%%%%%%%%%%%%%%%%%%%%%%%%%%%%%%%%%%%%%%%%%%%%%%%%%

The linearised action worked out below first appeared in \cite{Krasnov:2011up}, where also the Hamiltonian analysis (in the Minkowski limit) is contained. The novelty of this section is in the extension to the analysis to the more non-trivial de Sitter background. Also, the very compact form \myref{DaDa_action} of the completely symmetry reduced action was first introduced in the paper from the present author et al. \cite{Delfino:2012zy}. The most important new aspect of this section is in the realisation that the connection cannot be taken to be real. This is invisible in the Minkowski version of the linearised action analysed in the previous works. Thus, our treatment of the reality condition corrects and supersedes what appeared earlier in \cite{Krasnov:2011up} and \cite{Krasnov:2012pd}. 

%%%%%%%%%%%%%%%%%%%%%%%%%%%%%%%%%%%%%%%%%%%%%%%%%%%%%%%%%%%
\subsection{Linearised  Lagrangian}\label{linearised_lagrangian_section}
%%%%%%%%%%%%%%%%%%%%%%%%%%%%%%%%%%%%%%%%%%%%%%%%%%%%%%%%%%%
In this section we consider the linearised theory. The second order action (obtained as $1/2$ of the second variation) reads:

\be
\nn S^{(2)} = \int d^4 x \sqrt{-g} \bigg[\f{g^{(2)}}{2}P_{ijkl}\S^{i\m\n}D_\m \d A^j_\n \S^{k\r\s} D_\r \d A^l_\s 
\\
-\f{f(\d)}{3} \bigg( \f{1}{\im} \e^{\m\n\r\s} D_\m \d A^i_\n D_\r \d A^i_\s- M^2 \S^{i\m\n} \e^{ijk}\d A^j_\m \d A^k_\n \bigg) \bigg],
\ee
where $\sqrt{-g}$ is the square root of the determinant of the metric and $f(\d), g^{(2)}$ are the ``defining variables'' associated with the first and second variation of $f(X)$ respectively.
We first note that we can integrate by parts in the second term, with the result cancelling the last term precisely. One uses \myref{curvature_from_derivative_commutator} to verify this. The integration by parts is justified on the connection perturbations of compact support (in both space and time directions), and this is what we assume. Let us also absorb the prefactor $-g^{(2)}$ into the connection perturbation and define a new (canonically normalised as will be verified later) field

\be
a^i_\m = \f{ \sqrt{ g^{ (2) } } }{ \im } \d A^i_\m.
\ee

The free theory Lagrangian density takes the following simple form

\be\label{PDaDa_action}
\mathcal{L}^{(2)} = -\f{1}{2} P_{jikl}\S^{i\m\n}D_\m a^j_\n \S^{k\r\s} D_\r a^l_\s.
\ee

We emphasise that this is a Lagrangian for gravitons in de Sitter space. It should be compared with a significantly more complicated linearised Lagrangian in the metric-based description of GR (see the appendix in \cite{Goroff:1985th} to verify that the metric-based second variation is more involved even in the Minkowski case). It is worth stressing that the linearised theory is the same for any member of the class \myref{GEN_action}.

In this section we study this theory in some details. We start by listing the symmetries of the theory.
 
%%%%%%%%%%%%%%%%%%%%%%%%%%%%%%%%%%%%%%%%%%%%%%%%%%%%%%%%%%%
\subsection{Symmetries}\label{symmetries_section}
%%%%%%%%%%%%%%%%%%%%%%%%%%%%%%%%%%%%%%%%%%%%%%%%%%%%%%%%%%%
The free theory \myref{PDaDa_action} is invariant under the following local symmetries

\be\label{symmetries_on_action}
\d_\phi a^i_\m= D_\m \phi^i \quad (gauge), \quad \d_\xi a^i_\m = \xi^\a \S^i_{\m\a} \quad (diffeo). 
\ee

Note that the action of diffeomorphisms in this language is very simple, and corresponds to mere shifts of the connection in some direction. The first formula here is the usual action of the gauge symmetry. The second formula follows by writing the general transformation of the connection $a^i_\m$ under diffeomorphisms and modding out gauge transformation. The general (infinitesimal) action of diffeomorphisms on the connection is given by its Lie derivative along a (Killing) vector field $\xi^\a$  

\be
\d_{\xi}a^i = \mathcal{L}_{\xi}a^i = \i_\xi d a^i +d( \i_\xi a^i),
\ee
%  
%  \be
%  \d_{\xi}a^i_\m = \mathcal{L}_{\xi}a^i_\m = \xi^\a (d a^i)_{\a\m} +d_\m( \xi^\a a^i_\a),
%  \\substract \ a\ gauge\ transformation 
% \\ \nn= \xi^\a (d a^i)_{\a\m} +d_\m( \xi^\a a^i_\a) - D_\m( \xi^\a a^i_\a) = 
% \\ \nn= \xi^\a (d a^i)_{\a\m} +d_\m( \xi^\a a^i_\a) - d_\m( \xi^\a a^i_\a) + \e^{ijk} a^j\wedge ( \xi^\a a^k_\a) =
% \\ \nn= \xi^\a (d a^i)_{\a\m} +d_\m( \xi^\a a^i_\a) - d_\m( \xi^\a a^i_\a) + \f{1}{2} \xi^\a( \e^{ijk}  a^j\wedge  a^k_\a) = 
% \\ \nn= \xi^\a (d a^i)_{\a\m}  + \f{1}{2} \xi^\a( \e^{ijk}  a^j\wedge  a^k_\a) = \xi^\a F_\a\m
%  \ee
where $\i_\xi$ is the interior product with the vector field $\xi$. Then we notice that the second part is just a (normal) gauge transformation, which we can then mod out. The infinitesimal diffeomorphism of the connection is then simply

\be
\d_{\xi}a^i = \i_\xi F.
\ee
Using \myref{Curvature_of_Background_Sigma} we get \myref{symmetries_on_action}. 

The invariance under the usual gauge rotations is easy to see using the result for the commutator of two covariant derivative \myref{curvature_from_derivative_commutator} and then the algebra \myref{sigma_algebra} of $\S$-matrices. To verify the invariance under diffeomorphisms we use the Bianchi identity \myref{Bianchi_identity} together with  \myref{Curvature_of_Background_Sigma} to get $D_{[\m}\S^i_{\n\r]}=0$. Writing this identity as 
% 
% (given) 
% \be
% T_{[abc]}= \f{1}{3!} \bigg( T_{abc}-T_{acb}+T_{bca}-T_{bac}+T_{cab}-T_{cba} \bigg)
% \ee

\be
D_{[\r}\S^i_{\s]\a} = -\f{1}{2}D_\a \S^i_{\r\s}
\ee 
the variation of the Lagrangian \myref{PDaDa_action} becomes

\be\label{Variation_diffeo_PDaDa}
  \d_\xi  \mathcal{L}^{(2)} = % -  P_{jikl}\S^{i\m\n}D_\m a^j_\n \S^{k\r\s} D_\r (\xi^\a \S^l_{\s\a})
% \\
% = -  P_{jikl}\S^{i\m\n}D_\m a^j_\n \S^{k\r\s}   D_\r\xi^\a \S^l_{\s\a} -  P_{jikl}\S^{i\m\n}D_\m a^j_\n \S^{k\r\s}  \xi^\a  D_\r \S^l_{\s\a} 
% \\
 -  P_{jikl}\S^{i\m\n}D_\m a^j_\n  \bigg( -\f{1}{2} \S^{k\r\s}  \xi^\a D_\a \S^l_{\r\s} \bigg)
\ee

Here we have used the fact that in the term where the covariant derivative acts on the $\xi$ field and the $\S$ matrix is taken outside of the sign of the derivative, the algebra of the $\S$-matrices gives an expression that is either anti-symmetric in $\d^{kl}$ or a pure trace. Both are killed by the projector $P_{ijkl}$, and so only the term present in brackets in \myref{Variation_diffeo_PDaDa} remains. But now we note that the expression in the brackets can be replaced with

\be
-\f{1}{4}\xi^\a D_\a \bigg( \S^{k\r\s}\S^l_{\r\s}\bigg)
\ee
in view of the $kl$-symmetrisation implied by the projector $P_{ijkl}$. this expression, however, is proportional to the covariant derivative of the Kronecker $\d$ in view of the algebra satisfied by the $\S$'s, and this is zero. This establishes the invariance under diffeomorphisms as well.

%%%%%%%%%%%%%%%%%%%%%%%%%%%%%%%%%%%%%%%%%%%%%%%%%%%%%%%%%%%
\section{Hamiltonian Analysis of the general Theory}\label{Hamiltonian_analysis_section}
%%%%%%%%%%%%%%%%%%%%%%%%%%%%%%%%%%%%%%%%%%%%%%%%%%%%%%%%%%%
We now follow the textbook procedure of the Hamiltonian analysis of \myref{PDaDa_action}, to prepare the theory for the canonical quantisation. Unlike what was done in  \cite{Krasnov:2011up} we would like to remain in de Sitter background and not make the $M\rightarrow 0$ limit, at least not at this stage. We shall see that many subtleties, including those of the reality condition, can only be understood for a non-zero value of $M$. So, we live in the de Sitter space \myref{de_sitter_metric}, with the self-dual two-forms given by \myref{self_dual_forms}. We will also need a convenient expression for the background connection \myref{background_connection}, and this is given by
 
\be\label{background_connection_with_Hubble}
A^i_\m = \f{a'}{\im a}(dx^i)_\m := (\H/\im)(dx^i)_\m,
\ee

where the prime denotes the (conformal) time derivative and we have introduced $\H = a'/a$. Note that one should not confuse the conformal factor $a$ with the lower-case $a^{ij}$ which are the re-scaled components of the variation of the field $\d A^{i}_\m$. The equation \myref{a_def_with_conformal_time} then implies $\H'=\H^2=M^2 a^2.$
We can now compute the quantity $\S^{i \ \m\n}D_\m a^j_\n$ in terms of the temporal $a^j_0$ and spatial $a^j_i$ components of the connection. We get:

\be\label{SDa}
 a^2 \S^{i \ \m\n}D_\m a^j_\n = -\im \partial_t a^{ij} + \im D^i a_0^j + \e^{imn} D_m a_n^j,
\ee
%a^2 is the conformal parameter that sits in the de Sitter sigma $\S_{\m\n}$. With the metric going as $g^{\m\n}\sim a^{-2}, then $\S^{\m\n}\sim a^{-2}$
where $D_i$ is the covariant derivative with respect to the background connection \myref{background_connection_with_Hubble}. Explicitly 

\be
D_k a^i_l = \partial_k a^i_l - \im \H \e^{ikm} a^m_l,
\ee
where we have used \myref{background_connection_with_Hubble}. The convention in \myref{SDa} is that the first index in $a^{ij}$ is the spatial one.

We now decompose the spatial connection in its irreducible components 

\be
a^{ij} = \tl{a}^{ij} + \e^{ijk} c^k + \d^{ij}c,
\ee 
where $\tl{a}^{ij}$ is the symmetric tracefree component (i.e. spin 2). We substitute this into \myref{SDa} and immediately find that the spin zero component $c$ gets projected away by the projector $P_{ijkl}$ that multiplies this quantity in the Lagrangian. Keeping only the symmetric tracefree parts we get

\be\label{aPSDa}
a^2 P \S^{i \ \m\n} D_\m a^j_\n = -\im\partial_t \tl{a}^{ij}+ \im\partial^i (a_0^j+\im c^j) + \e^{ikl} \partial_k \tl{a}^j_l + \im \H \tl{a}^{ij}.
\ee

We see that the dependence on the anti-symmetric part $c^i$ can be reabsorbed into a shift of the temporal part. We therefore see that only the spin 2 part $\tl{a}^{ij}$ of the spatial connection is dynamical. We drop the tilde from now on. We can now rewrite the action in \myref{PDaDa_action} as

\be
S^{(2)} = -\f{1}{2}\int dt \int d^3 x  ( -\im\partial_t  a^{ij}+ \im\partial^i (a_0^j+\im c^j) + \e^{ikl} \partial_k a^j_l + \im \H a^{ij})^2. 
\ee
Here we simplified the conformal factors $1/a^2$ coming from \myref{aPSDa} with the $\sqrt{-g}=a^4$. The conjugate momentum to $a^{ij}$ is

\be\label{pi_def}
\pi^{ij} = \f{\partial \L}{\partial \dot{a}^{ij}} = \partial_t a^{ij} - P\partial^i (a_0^j + \im c^j) + \im B^{ij}-\H a^{ij},
\ee
where we have introduced the ``magnetic'' field $B^{ij}=P \e^{(ikl}\partial_k a^{j)}_l$, $P$ again being the symmetric tracefree projector. The action in the Hamiltonian form becomes:

\be
S^{(2)} = \f{1}{2}\int dt \int d^3 x \pi^{ij}\pi_{ij} = \int dt \int d^3 x (\pi_{ij}\partial_t a^{ij}-\H),
\ee

% (given)
% 
% \be
% \pi = \f{\partial \L}{\partial \dot{a}}
% \ee
% 
% and
% 
% \be
% H= \pi \dot{a}(\pi) - \L
% \ee

where in the second step we just substituted the definition (\ref{pi_def}), and in the last step the Hamiltonian density is 
\be\label{Hamiltonian_pi_a_with_lagrange_multipliers}
H = \f{1}{2} \pi_{ij}\pi^{ij}-\im \pi_{ij}B^{ij} + \H \pi_{ij}a^{ij} -(a_0^i+\im c^i)\partial^j \pi_{ij}.
\ee
We have integrated by parts in the Gauss constraint term and absorbed the projector $P$ in the $\pi_{ij}$ field.

%%%%%%%%%%%%%%%%%%%%%%%%%%%%%%%%%%%%%%%%%%%%%%%%%%%%%%%%%%%
\subsection{Gauge fixing}
%%%%%%%%%%%%%%%%%%%%%%%%%%%%%%%%%%%%%%%%%%%%%%%%%%%%%%%%%%%
It is convenient to fix the gauge at an early stage, and work with only the physical propagating modes. We see that the variation of the action with respect to the Lagrange multiplier $a^i_0$ gives the Gauss constraint

\be
\partial_i \pi^{ij} = 0.
\ee
This constraint generates gauge transformations
%first class primary constraints generate gauge transformations. Primary means that is zero off shell, first class that commutes with all other constraints..
\be
\delta a^{ij} 
% =\int d^3y  \xi_a(y) [Pa^{ij}(t,x),\f{\partial}{\partial y^m} \pi^{m a}(t,y)]_{PB}
% \\
% =\int d^3y P  \xi_a(y)\f{\partial}{\partial y^m}  [a^{ij}(t,x),\pi^{m a}(t,y)]_{PB}=
% \\
% =\int d^3y P \xi_a(y)\f{\partial}{\partial y^m} \d^{(im}\d^{j)a} \d^3(x-y)
% \\
% = P \xi_{(j} \partial_{i)}  =
% \ee
% or
% \be
% \delta a^{ij} 
% =\int d^3y  \xi_a(y) [\f{\partial}{\partial y^m} \pi^{m a}(t,y),Pa^{ij}(t,x)]_{PB}
% \\
% =-\int d^3y  \f{\partial}{\partial y^m} \xi_a(y) [\pi^{m a}(t,y),Pa^{ij}(t,x)]_{PB}
% \\
% =-\int d^3y P \f{\partial}{\partial y^m} \xi_a(y)\d^{(im}\d^{j)a}\d^3(x-y)(-)
% \\
% = P \f{\partial}{\partial x^m} \xi_a(y) \d^{(im}\d^{j)a} 
% \ee
% \be
 =P\partial^{(i}\xi^{j)}
\ee
where the projection is taken onto the tracefree part. This action can be used to set to zero the longitudinal part of $a^{ij}$

\be
\partial_i a^{ij}% = \partial_i a'^{ij} +\partial_i P\partial^{(i}\xi^{j)} 
= 0,
\ee
which is our gauge-fixing condition. Thus, our dynamical fields are a pair $(a_{ij},\pi^{ij})$ of symmetric traceless transverse tensors, as is appropriate for a spin 2 particle. We now note that the quantity $\e^{ijk}\partial_k a^j_l$ is automatically symmetric tracefree and transverse on $a^{ij}$ that are symmetric tracefree and transverse. Thus, the projector in the definition of $B^{ij}$ can be dropped.
% 
% Symmetric:
% \be
% \e^{iqj} \e^{ikl}\partial_k a^j_l= (\d_q^k \d_j^l-\d_q^l \d_j^k)\partial_k a^j_l = \partial_q a_j^j-\partial_j a^j_q = 0
% \ee
% transverse
% 
% \be
% \partial_i \e^{ikl}\partial_k a^j_l = 0,
% \\
% \partial_j \e^{ikl}\partial_k a^j_l = 0.
% \ee 
% 
% tracefree
% \be
% \e^{ikl}\partial_k a^i_l = 0.
% \ee

%%%%%%%%%%%%%%%%%%%%%%%%%%%%%%%%%%%%%%%%%%%%%%%%%%%%%%%%%%%
\subsection{Convenient notation}
%%%%%%%%%%%%%%%%%%%%%%%%%%%%%%%%%%%%%%%%%%%%%%%%%%%%%%%%%%%
The first-order differential operator $a^{ij}\rightarrow \e^{ikl}\partial_k a^j_l$ acts on the space of symmetric tracefree transverse tensors. It will appear on many occasions below, and so it is convenient to introduce a special notation for it

\be
(\e \partial a)^{ij} = \e^{ikl}\partial_k a^j_l.
\ee

It is then not hard to show that

\be
(\e \partial )^2 = -\Delta.
\ee
% \be
% (\e \partial )^2 a^{ij} = \e^{imn}\partial_m \e^{nkl}\partial_k a^j_l = (\d^i_k\d^m_l-\d^i_l\d^m_k)\partial_m \partial_k a^j_l =
% \partial_l \partial_i a^j_l - \partial_k \partial_k a^j_i
% \ee

It is also not hard to see that $\e \partial$ is self-adjoint with respect to the scalar product on the space of symmetric tracefree tensors $x^{ij},j^{ij}$:

\be
(x,y) = \int d^3 x \ x^{ij} y_{ij}.
\ee
%
% \be
% \int \pi^{ij} (\e\partial a)_{ij} =\int \pi^{ij} \e^{ikl}\partial_k a^{jl} = -\int a^{jl}  \e^{ikl}  \partial_k \pi^{ij} =
% = \int a^{jl}  \e^{lki}  \partial_k \pi^{ij} = \int a^{jl} (\e\partial \pi)_{jl}
% \ee

Overall, after gauge fixing, we note that we can rewrite the Hamiltonian \myref{Hamiltonian_pi_a_with_lagrange_multipliers} as

\be\label{gauge_fixed_Hamiltonia_pi_a}
H = \f{1}{2}\pi^2 -\im \pi(\e\partial a + \im \H a),
\ee
where we omitted the indices for brevity.

%%%%%%%%%%%%%%%%%%%%%%%%%%%%%%%%%%%%%%%%%%%%%%%%%%%%%%%%%%%
\subsection{Evolution equations}
%%%%%%%%%%%%%%%%%%%%%%%%%%%%%%%%%%%%%%%%%%%%%%%%%%%%%%%%%%%
Let us introduce two first order differential operators that are going to play an important role below.
We define

\be\label{DDbar_operators}
D:= -\im \partial_t +\e\partial +\im\H, \quad \bar{D}=\im \partial_t +\e\partial +\im\H,
\ee
where $\bar{D}$ is clearly the adjoint of $D$ with respect to the scalar product that also involves the time integration. We note that $Da$ is essentially the projected quantity $a^2 P \S^{i\m\n}D_\m a_\n^j,$ with the gauge-fixed spatial connection and its conjugated momentum satisfying the Gauss equation.
The Hamiltonian \myref{gauge_fixed_Hamiltonia_pi_a} then results in the following Hamilton equations

% remind
% \be
% \dot{\pi} = -\f{\partial H}{\partial a}, \quad \dot{a} = \f{\partial H}{\partial \pi} 
% \ee
% and using self-adjointness of $\e\partial$

\be
-\im \pi = D a, \quad \bar{D} \pi = 0,
\ee

which give immediately the following evolution equation

\be\label{evolution_equations_a}
 0 = \bar{D}D a %= (\im \partial_t +\e\partial +\im\H)(-\im \partial_t +\e\partial +\im\H) a =
% \\
% (\im \partial_t (-\im \partial_t a +\e\partial a  +\im\H a) 
% \\
%  +\e\partial  (-\im \partial_t a +\e\partial a  +\im\H a)  
% \\
% +\im\H (-\im \partial_t a +\e\partial a  +\im\H a) )=
% \\
%  + \partial_t^2  a +\im \partial_t\e\partial a  - \partial_t (\H a)
% \\
%     -\im \e\partial \partial_t a + \e\partial \e\partial a  +\im\H \e\partial a  
% \\
%  +\H \partial_t a +\im\H\e\partial a  - \H^2 a=
% \\
% \partial_t^2  a + \e\partial \e\partial a +2\im\H \e\partial a  - \partial_t (\H a) +\H \partial_t a - \H^2 a
%\\
=\partial_t^2  a - \Delta a +2\im\H \e\partial a   - 2 \H^2 a . 
\ee
Here I have used $\partial_t \H = \H' = \H^2.$ Because of the term with $\e \partial$ that has a factor of $\im$ in front, this equation is complex. It becomes a non-trivial problem that imposing $a^{ij}$ to be real would not be consistent with the evolution, because if one starts with a real $a^{ij}$, the evolution will generate an imaginary part. Thus, a more sophisticated strategy for dealing with this problem is needed.

%%%%%%%%%%%%%%%%%%%%%%%%%%%%%%%%%%%%%%%%%%%%%%%%%%%%%%%%%%%
\subsection{Second-order formulation}
%%%%%%%%%%%%%%%%%%%%%%%%%%%%%%%%%%%%%%%%%%%%%%%%%%%%%%%%%%%

Let us rewrite the original action \myref{PDaDa_action} as a functional on the space of the symmetric tracefree transverse tensors $a^{ij}$. This can be also obtained integrating out the momentum variable. Using the operators \myref{DDbar_operators} the corresponding second-order action can be written very compactly as 

\be\label{DaDa_action}
S^{(2)} = -\f{1}{2}\int d^4 x (D a)^2,
\ee
with \myref{evolution_equations_a} following immediately as the corresponding Euler-Lagrange equations.

%%%%%%%%%%%%%%%%%%%%%%%%%%%%%%%%%%%%%%%%%%%%%%%%%%%%%%%%%%%
\subsection{Degrees of freedom count}
%%%%%%%%%%%%%%%%%%%%%%%%%%%%%%%%%%%%%%%%%%%%%%%%%%%%%%%%%%%
We conclude this section by writing out explicitly the count of the degrees of freedom that propagate in the action \myref{DaDa_action}.
Before gauge fixing, we started with a complex $a^i_\m$ field, where the Latin index spans $i=1,2,3$ and the Greek index spans $\m=0,1,2,3$. This accounts for a total of $12$ complex components.

In the Hamiltonian analysis performed in previous sections we obtained the following results:

\begin{itemize}
	\item The 3 components $a^i_0$ of the field are Lagrange multiplier, as the Lagrangian does not depend on their time-derivatives.
	\item The 3 Lagrange multiplier impose 3 first class constraints $\p_i \pi^{ij}=0$.
	\item In the field $a^{ij}$ only the symmetric traceless part propagate. Eliminating 4 degrees of freedom. 
\end{itemize}
Total count is then
\be
DOF = 12-3-3-4=2 (complex).
\ee

To ensure that the theory describes gravity, we need to slash down the number of degrees of freedom to 2 real. In the next section, we deal with this introducing an appropriate reality condition.

%%%%%%%%%%%%%%%%%%%%%%%%%%%%%%%%%%%%%%%%%%%%%%%%%%%%%%%%%%%
\section{The reality condition}\label{Reality_condition_section}
%%%%%%%%%%%%%%%%%%%%%%%%%%%%%%%%%%%%%%%%%%%%%%%%%%%%%%%%%%%
The treatment of the connection field reality condition is this section was first introduced by this author et al. in \cite{Delfino:2012zy}. This section constitutes one of the most important new results of   \cite{Delfino:2012zy}.

%%%%%%%%%%%%%%%%%%%%%%%%%%%%%%%%%%%%%%%%%%%%%%%%%%%%%%%%%%%
\subsection{Evolution equation as an eigenfunction equation}
%%%%%%%%%%%%%%%%%%%%%%%%%%%%%%%%%%%%%%%%%%%%%%%%%%%%%%%%%%%
For our later purposes, it is very convenient to write the evolution equation \myref{evolution_equations_a} in a slightly different form. Thus, we use the fact that
\be
[D,\bar{D}] = 2 \H^2
\ee
which easily follows from $\H' = \H^2$, and write the evolution equation as an eigenfunction equation
\be\label{evolution_equations_a_with_E}
Ea=a,\quad where \quad  E=\f{1}{2\H^2}D\bar{D}.
\ee
This is the form that is going to be most useful below.

%%%%%%%%%%%%%%%%%%%%%%%%%%%%%%%%%%%%%%%%%%%%%%%%%%%%%%%%%%%
\subsection{An important identity}
%%%%%%%%%%%%%%%%%%%%%%%%%%%%%%%%%%%%%%%%%%%%%%%%%%%%%%%%%%%
We now prove an identity that lies at the root of the reality condition that is going to be imposed.
First, we note that
\be\label{identity_to_pull_2H_from_Dbar}
\bar{D}\f{1}{2\H^2} = \f{1}{2\H^2} D^*,
\ee
where $D^*=\im \partial_t +\e\p-\im\H$ is the complex conjugate to $D$. The above identity allows us to pull out a factor of $1/2\H^2$ from the derivative operator $\bar{D}$, at the expense of introducing the complex conjugate of $D$.
We now consider the square of the evolution equation operator $E$:
\be
E^2 =\f{1}{2\H^2}D\bar{D} \f{1}{2\H^2}D\bar{D}.
\ee
We use \myref{identity_to_pull_2H_from_Dbar} to convert $\bar{D}$ into $D^*$ and then use the fact that $D$ and $D^*$ commute $[D,D^*] = 0.$ We then use the complex conjugate of the identity \myref{identity_to_pull_2H_from_Dbar}. Overall, we get the following sequence of transformations 

\be\label{Esquare_identity}
E^2 = \f{1}{2\H^2}D\f{1}{2\H^2} D^* D \bar{D} = \f{1}{2\H^2}D\f{1}{2\H^2}DD^* \bar{D} = \f{1}{2\H^2}D\bar{D}^*\f{1}{2\H^2}D^* \bar{D} = RR^*,
\ee
where we have introduced 
\be
R:=\f{1}{2\H^2}D \bar{D}^*.
\ee
Note that $R$ is a dimensionless operator, since $\H$ carries the dimension of a mass.% The identity \myref{Esquare_identity} in particular implies that $E^2$ is a real operator, which is not at all obvious because $E$ is not real.

%%%%%%%%%%%%%%%%%%%%%%%%%%%%%%%%%%%%%%%%%%%%%%%%%%%%%%%%%%%
\subsection{The reality condition}
%%%%%%%%%%%%%%%%%%%%%%%%%%%%%%%%%%%%%%%%%%%%%%%%%%%%%%%%%%%
%Now, as our relation \myref{Esquare_identity} demonstrates, in spite of the fact that the evolution equation  \myref{evolution_equations_a_with_E} is complex, we see that its square $E^2a=a$, which is clearly implied by \myref{evolution_equations_a_with_E}, is a real equation. This fourth order equation 
We note that the relation $E^2a=a$, which is clearly implied by \myref{evolution_equations_a_with_E}, is not so interesting in itself but it introduces a new second-order differential operator $R$, such that $E^2=RR^*$. In other words, $R$ is a ``square root'' of the equation operator $E^2$, similar to the Dirac operator being a square root of the Klein-Gordon one. It is the clear that if we define

\be
\mathcal{R} = R\circ \dagger,
\ee
where $\dagger$ indicates Hermitian conjugation. Then the reality condition 
\be\label{reality_condition_with_R}
\mathcal{R}a = a
\ee
is compatible with the evolution equation $Ea=a$. Indeed, the compatibility is just a re-phrasal of the statement that on solution of \myref{evolution_equations_a_with_E} the $\mathcal{R}$ anti-linear operator becomes an involution:
\be
\mathcal{R}^2 = RR^* = E^2 = \mathrm{Id},
\ee
where the last equation holds on the space of solutions $Ea=a$. Thus, $\mathcal{R}$ is a real structure on the space of solutions, and the condition \myref{reality_condition_with_R} is a possible reality condition that can be imposed. Below we shall see that this is the physically correct condition, in particular by working out a relation to the metric description. In essence \myref{reality_condition_with_R} will then just be the statement that metric is real.

%%%%%%%%%%%%%%%%%%%%%%%%%%%%%%%%%%%%%%%%%%%%%%%%%%%%%%%%%%%
\subsection{Metric}\label{metric_subsection}
%%%%%%%%%%%%%%%%%%%%%%%%%%%%%%%%%%%%%%%%%%%%%%%%%%%%%%%%%%%
We can now re-phrase the condition \myref{reality_condition_with_R} as a statement that a certain quantity is real. Indeed, we introduce

\be\label{metric_h}
h = \f{1}{\sqrt{2} M}\bar{D}a,
\ee
where the prefactor is introduced for convenience and also in order to give $h$ the same mass dimension as $a$. Below we will show that $h$ can be viewed just as a possible new configuration variable on the phase space of the theory, with the Hamiltonian form action principle in terms of this variable taking an explicitly real form \myref{h_p_Hamiltonian}.
The evolution equation in its form \myref{evolution_equations_a_with_E} can now be re-phrased by saying that it gives the inverse relation

\be\label{evolution_equations_a_and_h}
a = \f{M}{\sqrt{2} \H^2}D h.
\ee

Taking now the Hermitian conjugate of the quantity $h$ in \myref{metric_h}, requiring it to be real
\be
h = h^\dagger
\ee
and the substituting $h=\bar{D}a/\sqrt{2}M$ into \myref{evolution_equations_a_and_h} we get precisely the reality condition \myref{reality_condition_with_R}. Thus, the essence of the condition \myref{reality_condition_with_R} imposed on the space of solutions $Ea=a$ of our theory is indeed in the statement that \myref{metric_h} is real. 

%%%%%%%%%%%%%%%%%%%%%%%%%%%%%%%%%%%%%%%%%%%%%%%%%%%%%%%%%%%
\subsection{Evolution equations for the metric}
%%%%%%%%%%%%%%%%%%%%%%%%%%%%%%%%%%%%%%%%%%%%%%%%%%%%%%%%%%%

As the last result of this section, let us use the identities derived above to obtain an evolution equation for the variable $h$. It is not hard to see that this equation is

\be\label{evolution_equations_h}
\f{1}{2\H^2}D^* D h = h.
\ee
Indeed, using \myref{identity_to_pull_2H_from_Dbar} we can rewrite this as
\be
\bar{D}\f{1}{2\H^2}Dh=h\quad or \quad \bar{D}\f{1}{2\H^2}D\bar{D}a=\bar{D}a,
\ee

where to obtain the last equation we have used the relation \myref{metric_h}. The equation obtained is just the evolution equation $Ea=a$ with the operator $\bar{D}$ applied to it. Thus, \myref{evolution_equations_h} clearly follows from \myref{evolution_equations_a}. It is also worth noting that it is a real equation, as is appropriate for a quantity that can consistently be assumed to be real.

%%%%%%%%%%%%%%%%%%%%%%%%%%%%%%%%%%%%%%%%%%%%%%%%%%%%%%%%%%%
\section{Canonical transformation to the metric variables}\label{Canonical_transformation_section}
%%%%%%%%%%%%%%%%%%%%%%%%%%%%%%%%%%%%%%%%%%%%%%%%%%%%%%%%%%%

The purpose of this section is to explicitly carry out the field redefinition \myref{metric_h} and see that it can get completed (once the momentum variable is considered) into a canonical transformation on the phase space of the theory. The content of this section was first introduced by this author together with K.Krasnov and C.Scarinci in \cite{Delfino:2012zy}.

%%%%%%%%%%%%%%%%%%%%%%%%%%%%%%%%%%%%%%%%%%%%%%%%%%%%%%%%%%%
\subsection{Canonical transformation - momentum shift}
%%%%%%%%%%%%%%%%%%%%%%%%%%%%%%%%%%%%%%%%%%%%%%%%%%%%%%%%%%%
It is very convenient to eliminate the $\pi a$ cross-term in \myref{gauge_fixed_Hamiltonia_pi_a} by shifting the momentum. Thus, we define

\be
\tl{\pi}=\pi - \im (\e \p + \im \H)a.
\ee
Because of the last, time dependent (via $\H$) term the transformation of the symplectic form gives rise to a contribution to the Hamiltonian. In other words, modulo surface terms we get

\be
\pi \dot{a}= %\tl{\pi} \dot{a}+ \im (\e \p + \im \H)a  \dot{a} =\\
\tl{\pi} \dot{a}+ \f{1}{2}\H^2 a^2. 
\ee

where we have used $\dot{\H} = \H^2.$ We now drop the tilde from the momentum variable, and write the reduced action in the Hamiltonian form as 

\be
S^{(2)} = \int dt\int d^3 x (\pi\dot{a} - H),
\ee

with the Hamiltonian given by
% given 
% \be
% H(\pi,a) = \f{1}{2}\pi^2-\im\pi(\e \p a + \im \H a) = \f{1}{2}(\pi - \im (\e \p a + \im \H a) )^2 + \f{1}{2} (\e \p a + \im \H a)^2 
% \ee
% imposing
% \be
% \pi(\tl{\pi},a)  \dot{a} - H(\pi(\tl{\pi},a) ,a) = \tl{\pi}\dot{a} -H'(\tl{\pi},a) 
% \ee
% 
% \be
% H'(\tl{\pi},a) =   \tl{\pi}\dot{a} - \pi(\tl{\pi},a)  \dot{a} + H(\pi(\tl{\pi},a) ,a) =
% \\
%  \tl{\pi}\dot{a} - (\tl{\pi} + \im (\e \p + \im \H)a) \dot{a}  +\f{1}{2}\tl{\pi}^2 + \f{1}{2} (\e \p a + \im \H a)^2 =
% \\
%    -  \im (\e \p + \im \H)a \dot{a}  +\f{1}{2}\tl{\pi}^2 + \f{1}{2} (\e \p a + \im \H a)^2 =
% \\
%      - \f{1}{2} \H^2 a^2  +\f{1}{2}\tl{\pi}^2 + \f{1}{2} (\e \p a + \im \H a)^2 
% \ee

\be
H=  \f{1}{2}\pi^2 + \f{1}{2} (\e \p a + \im \H a)^2 - \f{1}{2} \H^2 a^2.
\ee

The convenience of the new momentum variable lies in the fact that now

\be
\dot{a} = \f{\p L}{\p \pi} =\pi .
\ee

%%%%%%%%%%%%%%%%%%%%%%%%%%%%%%%%%%%%%%%%%%%%%%%%%%%%%%%%%%%
\subsection{Canonical transformation to $h$ variables}
%%%%%%%%%%%%%%%%%%%%%%%%%%%%%%%%%%%%%%%%%%%%%%%%%%%%%%%%%%%
From the previous section we know that we should be able to describe the dynamics in terms of the variable
\be
h= \f{1}{\sqrt{2}M}(\im \pi + (\e \p +\im \H)a),
\ee
and that this variable can be consistently be assumed to be real. The canonically conjugated momentum $p$ to $h$ is of course solely defined modulo $a$-dependent shifts. However, if we impose not to be any $p\cdot h$ term in the resulting Hamiltonian, then the momentum variable can be determined to be given by
\be
p = \f{M}{\sqrt{2}\H^2}(\e \p +\im\H)\pi -\im \big( (\e \p +\im \H)^2 -2\H^2\big)a.
\ee

We emphasise that this is a linear canonical transformation on the phase space of the theory.

%%%%%%%%%%%%%%%%%%%%%%%%%%%%%%%%%%%%%%%%%%%%%%%%%%%%%%%%%%%
\subsection{Metric Hamiltonian}
%%%%%%%%%%%%%%%%%%%%%%%%%%%%%%%%%%%%%%%%%%%%%%%%%%%%%%%%%%%

There are many contributions from the symplectic term $\pi\dot{a}$ to the Hamiltonian in terms of $h,p$ variables, After a rather tedious computation one finds that the action can be written as

\be
S^{(2)} = \int dt \int d^3 x (p\dot{h}-H),
\ee 
where 
\be\label{h_p_Hamiltonian}
H=\f{\H^2}{2M^2}p^2+h\f{(\e \p)^2 - 2\H^2}{2\H^2}M^2 h.
\ee
As a check, we note that this Hamiltonian goes into that for a massless field in the limit $M\rightarrow 0$. Indeed, using the explicit expression \myref{H_in_terms_of_M} for $\H$ one sees that $\H/M\rightarrow 1$ when $M\rightarrow 0$. This shows that the above Hamiltonian has the correct Minkowski limit. As for the de Sitter Hamiltonian, the above is the standard Hamiltonian for the de Sitter space spin 2 part of the metric perturbation $h_{\m\n}$ rescaled by the factor $a(t)$.

%%%%%%%%%%%%%%%%%%%%%%%%%%%%%%%%%%%%%%%%%%%%%%%%%%%%%%%%%%%
\subsection{Second-order formulation}
%%%%%%%%%%%%%%%%%%%%%%%%%%%%%%%%%%%%%%%%%%%%%%%%%%%%%%%%%%%
It is also instructive to write the above action in the second-order form, by integrating $p$ out. We get 

% 
% \be
% L= p\dot{h} - H = p\dot{h}-\f{\H^2}{2M^2}p^2-h\f{(\e \p)^2 - 2\H^2}{2\H^2}M^2 h
% \ee
% where 
% \be
% \dot{h} = \f{\H^2}{M^2}p
% \ee
% and $(\e \p)^2 = -\Delta$. Therefore
% 
% \be
% L=  \f{M^2}{2\H^2}\dot{h}^2 + h\f{\Delta + 2\H^2}{2\H^2}M^2 h.
% \ee
% we can compute $DD^*$
% \be
% DD^* = (-\im \p_t + \e\p + \im \H)(\im \p_t + \e\p - \im \H) = (\e\p)^2 + (\p_t -\H)^2 = -\Delta + \p_t^2
% \ee
% \be
% L=  - \f{M^2}{2\H^2}h \p_t \p_t h + \f{ 1 }{2\H^2}M^2 h\Delta h+h  M^2 h.
% \ee

\be\label{h_action}
\nn S^{(2)} = -M^2 \int dt \int d^3 x \f{h}{2\H^2}(D^* D - 2\H^2)h= 
\\
 -M^2 \int dt \int d^3 x \bigg( \f{1}{2\H^2} (Dh)^2 - h^2 \bigg),
\ee
where we have integrated by parts in the $(\p_t h)^2$ term to get the first expression for the action, which is explicitly real, and have used the \myref{identity_to_pull_2H_from_Dbar} together with the fact that $D$ is the adjoint of $\bar{D}$  to get the second, more symmetric expression. The first version clearly leads to \myref{evolution_equations_h} as corresponding Euler-Lagrange equations.
It is worth emphasising that the connection formalism linearised action \myref{DaDa_action} is actually simpler than the same action \myref{h_action} in the metric description. Here we are comparing only the completely symmetry reduced action, but the same holds true also about the full linearised Lagrangian \myref{PDaDa_action} is much simpler that its metric variant.  

%%%%%%%%%%%%%%%%%%%%%%%%%%%%%%%%%%%%%%%%%%%%%%%%%%%%%%%%%%%
\section{Canonical quantisation and the mode decomposition}\label{Mode_expansion_section}
%%%%%%%%%%%%%%%%%%%%%%%%%%%%%%%%%%%%%%%%%%%%%%%%%%%%%%%%%%%
We now perform all the usual steps for the canonical quantisation of the theory \myref{DaDa_action}, with the reality condition \myref{reality_condition_with_R}. Our main aim is to obtain a mode decomposition with correctly normalised creation and annihilation operators. Again, the content of this section was first introduced by this author together with K.Krasnov and C.Scarinci in \cite{Delfino:2012zy}.

%%%%%%%%%%%%%%%%%%%%%%%%%%%%%%%%%%%%%%%%%%%%%%%%%%%%%%%%%%%
\subsection{Choice of the time coordinate}
%%%%%%%%%%%%%%%%%%%%%%%%%%%%%%%%%%%%%%%%%%%%%%%%%%%%%%%%%%%
We first explicitly solve the evolution \myref{evolution_equations_a}
 for the connection, so that the linearly independent solution later will becomes the modes of the field. For this, let us first introduce a convenient parametrisation  of the $a(t)$ and $\H$ functions. We choose

\be
a(t) = \f{1}{1-Mt}
\ee

so that $a(0)=1$, i.e. we have chosen the origin of the time coordinate in such a way that $t=0$ corresponds to a conformal factor of unity. With this parametrisation we get
\be\label{H_in_terms_of_M}
\H = M a = \f{M}{1-Mt}.
\ee

%%%%%%%%%%%%%%%%%%%%%%%%%%%%%%%%%%%%%%%%%%%%%%%%%%%%%%%%%%%
\subsection{Spatial Fourier transform}
%%%%%%%%%%%%%%%%%%%%%%%%%%%%%%%%%%%%%%%%%%%%%%%%%%%%%%%%%%%

We now perform the spatial Fourier transform, and choose convenient polarization tensors. Thus, consider a mode of the form $a^{ij}_k e^{\im \vec{k}\vec{x}}$. The transverse condition $\p_i a^{ij}$ on the connection implies that the corresponding mode $a^{ij}$ is orthogonal to $k^i$. For this reason, it is very convenient to define

\be
z^i(k):=\f{k^i}{|k|},
\ee
i.e. a unit vector in the direction of the spatial momentum. We then define two (complex) vectors $m^i(k), \ \bar{m}^i(k)$ that are both orthogonal to $z^i$ and whose only non-zero scalar product is $m^i\bar{m}_i=1$.
They satisfy
\be
\im \e^{ijk} z_j m_k = m_i, \quad \e^{ijk} z_j \bar{m}_k = - \bar{m}_i, \quad \im \e^{ijk}m_j \bar{m_k} = z_i.
\ee 

Here we have omitted the momentum dependence of these vectors for brevity, but it should all time be kept in mind that they are $\vec{k}$ dependent. Thus, when we replace $\vec{k}\rightarrow -\vec{k}$ the vectors $m^i,\bar{m}^i$ get interchanged: 

\be
m^i(-k)= \bar{m}^i(k),\quad \bar{m}^i(-k)= m^i(k).
\ee
It is very important to keep these transformations in mind for the manipulations that follow.

%%%%%%%%%%%%%%%%%%%%%%%%%%%%%%%%%%%%%%%%%%%%%%%%%%%%%%%%%%%
\subsection{Polarization tensors}
%%%%%%%%%%%%%%%%%%%%%%%%%%%%%%%%%%%%%%%%%%%%%%%%%%%%%%%%%%%
The fact that $a^{ij}$ is symmetric tracefree transverse implies that every mode $e^{\im \vec{x} \vec{k}}$ comes in just two polarizations. For the corresponding polarization tensors it is convenient to choose $m^i(k)m^j(k)$ and $\bar{m}^i(k)\bar{m}^j(k)$. We shall refer to the $mm$ mode as the $negative$ helicity particle, while the $\bar{m}\bar{m}$ mode will be referred to as the positive one. We will explain a reason for this choice below.

Let us now consider the action of the operator $\e\p$ on the two polarizations. We have 

\be
(\e\p)m^i m^j a^-_k e^{\im \vec{k} \vec{x}} = \omega_k m^i m^j a^-_k e^{\im \vec{k} \vec{x}}, \quad 
(\e\p)\bar{m}^i \bar{m}^j a^+_k e^{\im \vec{k} \vec{x}} = -\omega_k \bar{m}^i \bar{m}^j a^+_k e^{\im \vec{k} \vec{x}},
\ee
where we have introduced
\be
\omega_k := |k|.
\ee
In other  words, the two modes we have introduced are the eigenvectors 
of the operator $\e\p$ with eigenvalues $\pm \omega_k$ respectively. Our choice of the name for the $mm$ mode as negative may seem unnatural at the moment (since it corresponds to the positive sign eigenvalue of $\e\p$). However, it becomes more natural if one computes the corresponding Weyl curvatures for the two modes. One finds that the negative mode has zero self-dual Weyl curvature, and is thus a purely anti-self-dual object. This is why it makes sense to refer to it as the negative helicity mode.
% 
% (computation of self-dual Weyl curvature of $mm$, works with the normal derivative..)
% \be
% (\S D a^-)^{ij} = D a^-{}^{ij} = (-\im \p_t+\e\p +\im\H) a^-{}^{ij}
% \\
% (-\im \p_t+\e\p +\im\H)m^i m^j a^-_k e^{-\im\omega_k t + \im \vec{k} \vec{x}}=
% \\
% (- \omega_k + \omega_k +\im\H)m^i m^j a^-_k e^{-\im\omega_k t + \im \vec{k} \vec{x}}=
% \ee

%%%%%%%%%%%%%%%%%%%%%%%%%%%%%%%%%%%%%%%%%%%%%%%%%%%%%%%%%%%
\subsection{Linearly independent solutions}
%%%%%%%%%%%%%%%%%%%%%%%%%%%%%%%%%%%%%%%%%%%%%%%%%%%%%%%%%%%
We now write the evolution equations \myref{evolution_equations_a} as an equation for the time evolution of the Fourier coefficients. We get, for each of the modes

\be
\p_t^2 a^-_k + (\omega_k^2 + 2\im \H \omega_k - 2\H^2)a^-_k = 0, \quad \p_t^2 a^+_k + (\omega_k^2 - 2\im \H \omega_k - 2\H^2)a^+_k = 0. 
\ee
Note that the positive helicity equation is just the complex conjugate of the negative helicity one. 
Each of the above equation is a second order ODE, and thus has a positive and negative frequency solution. It is not hard to obtain them explicitly, and they read

\be\label{a_modes}
\nn a^-_k \sim \H e^{-\im \omega_k t}, \quad a^-_k \sim \f{1}{\H} e^{i\omega_k t}\bigg( 1-\f{\im \H}{\omega_k}-\f{\H^2}{\omega_k^2} \bigg),
\\ 
a^+_k \sim \f{1}{\H} e^{-i\omega_k t}\bigg( 1+\f{\im \H}{\omega_k}-\f{\H^2}{\omega_k^2} \bigg), \quad  a^+_k \sim \H e^{\im \omega_k t},
\ee

where $\sim$ indicates that we are still able to chose a constant factor in front of the solutions. It is interesting to note that one of the modes in each case is given by a rather simple expression, with the time-dependence of the amplitude being just that of $\H$. The other mode in each case is more involved. For the negative mode it is the positive frequency solution that is simple, while for the positive mode the positive frequency solution is involved. This is a manifestation of a general pattern in our formalism, in that the negative helicity mode will always be much easier to deal with than the positive helicity one.

Another point worth emphasising is that one of the two linearly independent solutions of the connection evolution equation is actually simpler than the modes in the metric description, see \myref{h_mode_decomposition} below. This gives yet another illustration of the general statement that we would like to promote - the connection description is in may aspects simpler than the metric one.

%%%%%%%%%%%%%%%%%%%%%%%%%%%%%%%%%%%%%%%%%%%%%%%%%%%%%%%%%%%
\subsection{Action of the $\bar{D}$ operator on the modes}
%%%%%%%%%%%%%%%%%%%%%%%%%%%%%%%%%%%%%%%%%%%%%%%%%%%%%%%%%%%
It is useful to compute the action of the basic operator $\bar{D}$ on the modes \myref{a_modes}. We will need this when we impose the reality condition \myref{reality_condition_with_R}, which can be written as $a = (1/2\H^2)D(\bar{D} a)^\dagger$. We have

%remember to derivate also the \H with respect to time
\be\label{Dbar_action_on_a}
\nn \bar{D} m^i m^j \H e^{-\im \omega_k t + \im \vec{k}\vec{x}}= 2\omega_k m^i m^j\H e^{-\im \omega_k t + \im \vec{k}\vec{x}} \bigg( 1+\f{\im \H}{\omega_k} \bigg),
\\
\nn \bar{D}\bar{m}^i \bar{m}^j \f{1}{\H} e^{-\im \omega_k t + \im \vec{k}\vec{x} } \bigg( 1+\f{\im \H}{\omega_k} - \f{\H^2}{2 \omega_k^2}\bigg) = - \bar{m}^i \bar{m}^j\f{\H}{\omega_k} e^{-\im \omega_k t + \im \vec{k}\vec{x}} \bigg( 1+\f{\im \H}{\omega_k} \bigg),
\\
\nn \bar{D} \bar{m}^i \bar{m}^j \f{1}{\H} e^{\im \omega_k t - \im \vec{k}\vec{x} } \bigg( 1-\f{\im \H}{\omega_k} - \f{\H^2}{2 \omega_k^2}\bigg) =  \bar{m}^i \bar{m}^j\f{\H}{\omega_k} e^{\im \omega_k t - \im \vec{k}\vec{x}} \bigg( 1-\f{\im \H}{\omega_k} \bigg),
\\
\bar{D} m^i m^j \H e^{\im \omega_k t - \im \vec{k}\vec{x}}= -2\omega_k m^i m^j\H e^{\im \omega_k t - \im \vec{k}\vec{x}} \bigg( 1-\f{\im \H}{\omega_k} \bigg).
\ee

Now, to impose the reality condition, we take the complex conjugates of the right-hand-sides, and the apply the operator $D$ to them. We get

\be
\nn 2 \omega_k D \bar{m}^i \bar{m}^j\H e^{\im \omega_k t - \im \vec{k}\vec{x} } \bigg( 1 - \f{\im \H}{\omega_k} \bigg) = (2 \omega_k)^2 \bar{m}^i \bar{m}^j \H e^{\im \omega_k t - \im \vec{k}\vec{x}} \bigg( 1 - \f{\im \H}{\omega_k} - \f{\H^2}{2\omega_k^2} \bigg),
\\
\nn - D m^i m^j \f{\H}{\omega_k} e^{\im \omega_k t - \im \vec{k}\vec{x}} \bigg( 1-\f{\im \H}{\omega_k} \bigg) =  m^i m^j \f{\H^3}{\omega_k^2} e^{\im \omega_k t - \im \vec{k}\vec{x}},
\\
\nn  D m^i m^j \f{\H}{\omega_k} e^{-\im \omega_k t + \im \vec{k}\vec{x}} \bigg( 1+\f{\im \H}{\omega_k} \bigg) =  m^i m^j \f{\H^3}{\omega_k^2} e^{-\im \omega_k t + \im \vec{k}\vec{x}},
\\
\nn -2 \omega_k D \bar{m}^i \bar{m}^j\H e^{-\im \omega_k t + \im \vec{k}\vec{x} } \bigg( 1 + \f{\im \H}{\omega_k} \bigg) = (2 \omega_k)^2 \bar{m}^i \bar{m}^j \H e^{-\im \omega_k t + \im \vec{k}\vec{x}} \bigg( 1 + \f{\im \H}{\omega_k} - \f{\H^2}{2\omega_k^2} \bigg).
\\
\ee

%%%%%%%%%%%%%%%%%%%%%%%%%%%%%%%%%%%%%%%%%%%%%%%%%%%%%%%%%%%
\subsection{The mode expansion}
%%%%%%%%%%%%%%%%%%%%%%%%%%%%%%%%%%%%%%%%%%%%%%%%%%%%%%%%%%%
Using the above results, we can now write down a mode expansion satisfying the reality condition \myref{reality_condition_with_R}. We get

\be\label{a_mode_expansion}
\nn a^{ij}(t,\vec{x}) = \int \f{d^3k}{(2\pi)^3 2\omega_k} 
\\
\nn \bigg[  m^i m^j a^-_k \f{\H}{\sqrt{2}\omega_k} e^{-\im \omega_k t + \im \vec{k}\vec{x}} +\bar{m}^i \bar{m}^j (a^-_k)^\dagger \f{\sqrt{2}\omega_k}{\H} e^{\im \omega_k t - \im \vec{k}\vec{x}} \bigg( 1 - \f{\im \H}{\omega_k} - \f{\H^2}{2\omega_k^2} \bigg)
\\
-\bar{m}^i \bar{m}^j  \f{\sqrt{2}\omega_k}{\H} a^+_k e^{-\im \omega_k t + \im \vec{k}\vec{x}}\bigg( 1 + \f{\im \H}{\omega_k} - \f{\H^2}{2\omega_k^2} \bigg) - m^i m^j (a^+_k)^\dagger \f{\H}{\sqrt{2}\omega_k} e^{\im \omega_k t - \im \vec{k}\vec{x} }\bigg]
\ee

Here all the vectors $m^i,\bar{m}^i$ are $\vec{k}$-dependent, but this dependence is suppressed in order to have a compact expression. We could have chosen to put a plus sign in front of the positive helicity modes, but below we shall see that the above choice leads to a more symmetric expression for the metric mode expansion.

Note that the reality condition makes it unnatural to put factors of $M$ in front of the modes. Thus, as it stands, the expression \myref{a_mode_expansion} does not have a Minkowski limit $M\rightarrow 0$, because some term go to zero in this limit, and some others blow up. 
We also note that in \myref{a_mode_expansion} only the relative coefficient between the $a$,$a^\dagger$ terms in each helicity sector is fixed by the reality condition, so we could have multiplied each sector by an arbitrary constant factor. By doing so we could obtain an expression that survives in the $M\rightarrow 0$ limit. However, we are now going to show that the mode decomposition \myref{a_mode_expansion} is written in terms of canonically normalised operators. We do this by computing the commutators as implied by the canonical Poisson brackets between the connection and its conjugate momentum.

%%%%%%%%%%%%%%%%%%%%%%%%%%%%%%%%%%%%%%%%%%%%%%%%%%%%%%%%%%%
\subsection{Commutators}
%%%%%%%%%%%%%%%%%%%%%%%%%%%%%%%%%%%%%%%%%%%%%%%%%%%%%%%%%%%
We start with the relation that the equal time connection and its conjugate momentum should satisfy:
\be\label{a_da_commutator}
[a_{ij}(t,\vec{x}), \p_t a_{kl}(t,\vec{y})]= \im \d^3(x-y)P_{ijkl}.
\ee

For the conjugate momentum we have
\be
\nn \p_t a^{ij}(t,\vec{y}) = \int \f{d^3p}{(2\pi)^3 2\omega_p} 
 (-\im \omega_p)\bigg[  m^i(p) m^j(p) a^-_p \f{\H}{\sqrt{2}\omega_p} e^{-\im \omega_p t + \im \vec{p}\vec{y}}\bigg( 1+ \f{\im \H}{\omega_p}\bigg)
\\
\nn - \bar{m}^i(p) \bar{m}^j(p) (a^-_p)^\dagger \f{\sqrt{2}\omega_p}{\H} e^{\im \omega_p t - \im \vec{p}\vec{y}} \bigg( 1 - \f{\H^2}{2\omega_p^2} + \f{\im\H^3}{2\omega_p^3} \bigg)
\\
\nn -\bar{m}^i(p) \bar{m}^j(p)  \f{\sqrt{2}\omega_p}{\H} a^+_p e^{-\im \omega_p t + \im \vec{p}\vec{y}}\bigg( 1 - \f{\H^2}{2\omega_p^2} - \f{\im\H^3}{2\omega_p^3} \bigg)
\\
+ m^i(p) m^j(p) (a^+_p)^\dagger \f{\H}{\sqrt{2}\omega_p} e^{\im \omega_p t - \im \vec{p}\vec{y} }\bigg( 1 - \f{\im \H}{\omega_p}\bigg)\bigg].
\ee

Substituting this into \myref{a_da_commutator}, and using the fact that under $\vec{k}\rightarrow -\vec{k}$ the vectors $m^i,\bar{m}^i$ get interchanged, as well as the fact that for any $\vec{k}$
\be
P_{ijkl} = m_im_j\bar{m_k}\bar{m}_l + \bar{m}_i \bar{m}_j m_k m_l,
\ee
we get (in the $M\rightarrow 0$ limit)

\be
[a^\pm_k,(a^\pm_k)^\dagger ]= (2\pi)^3 2\omega_k \d^3(k-p),
\ee
which are the canonical commutation relations for the creation-annihilation operators in field theory. This gives one confirmation of the correct normalisation used in \myref{a_mode_expansion}. Another confirmation comes by computing the metric, and the associated Hamiltonian.

%%%%%%%%%%%%%%%%%%%%%%%%%%%%%%%%%%%%%%%%%%%%%%%%%%%%%%%%%%%
\subsection{Metric}
%%%%%%%%%%%%%%%%%%%%%%%%%%%%%%%%%%%%%%%%%%%%%%%%%%%%%%%%%%%
Let us now use \myref{a_mode_expansion} to obtain the mode decomposition for the metric \myref{metric_h}. The action of the operator $\bar{D}$ on all the modes has already been computed in \myref{Dbar_action_on_a}. We get

\be\label{h_mode_decomposition}
\nn h^{ij} (t,\vec{x}) = \f{\H}{M} \int \f{d^3k}{(2\pi)^3 2\omega_k}
\bigg[ (m^i m^j a^-_k+\bar{m}^i\bar{m}^j a^+_k ) e^{-\im \omega_k t + \im \vec{k}\vec{x}} \bigg( 1 + \f{\im \H}{\omega_k} \bigg)
\\
 +(\bar{m}^i\bar{m}^j (a^-_k)^\dagger + m^i m^j (a^+_k)^\dagger ) e^{\im \omega_k t - \im \vec{k}\vec{x}} \bigg( 1 - \f{\im \H}{\omega_k} \bigg) \bigg].
\ee

This expression has an obvious (correct) Minkowski limit $M\rightarrow 0 $. it is also explicitly Hermitian. It is in order to obtain the above symmetric expression that we chose to introduce the minus sign in front of the positive helicity mode in \myref{a_mode_expansion}. To compute the Hamiltonian in terms of the modes, let as also give an expression for the momentum $p=(M^2/\H^2)\p_t h$. We get

\be
\nn p^{ij} (t,\vec{x}) = \f{M}{\H} \int \f{d^3k  (-\im \omega_k)}{(2\pi)^3 2\omega_k}
\bigg[ (m^i m^j a^-_k+\bar{m}^i\bar{m}^j a^+_k ) e^{-\im \omega_k t + \im \vec{k}\vec{x}} \bigg( 1 + \f{2\im \H}{\omega_k} - \f{2 \H^2 }{\omega_k^2} \bigg)
\\
\nn -(\bar{m}^i\bar{m}^j (a^-_k)^\dagger + m^i m^j (a^+_k)^\dagger ) e^{\im \omega_k t - \im \vec{k}\vec{x}} \bigg( 1 - \f{2 \im \H}{\omega_k} - \f{2\H^2}{\omega_k^2} \bigg) \bigg].
\\
\ee 

Then the Hamiltonian \myref{h_p_Hamiltonian} reads:
\be
\nn \int H = \f{1}{2} \int \f{d^3k}{(2\pi)^3 2\omega_k}\omega_k
\\
 \bigg(a^-_k(a^-_k)^\dagger + (a^-_k)^\dagger a^-_k+a^+_k(a^+_k)^\dagger + (a^+_k)^\dagger a^+_k\bigg) \bigg( 1- \f{\H^2}{2\omega_k^2}+\f{\H^4}{\omega_k^4}\bigg).
\ee

The Hamiltonian is explicitly time dependent, as is appropriate for particles in time-dependent de Sitter Universe where the energy is not conserved. We note that it has the correct Minkowski limit $M\rightarrow 0$.

%%%%%%%%%%%%%%%%%%%%%%%%%%%%%%%%%%%%%%%%%%%%%%%%%%%%%%%%%%%
\section{Discrete symmetries}\label{Discrete_symmetries_section}
%%%%%%%%%%%%%%%%%%%%%%%%%%%%%%%%%%%%%%%%%%%%%%%%%%%%%%%%%%%
In this section we obtain the action of the discrete C, P, T symmetries on the connection field, and on the creation-annihilation operators.

%%%%%%%%%%%%%%%%%%%%%%%%%%%%%%%%%%%%%%%%%%%%%%%%%%%%%%%%%%%
\subsection{Charge conjugation}
%%%%%%%%%%%%%%%%%%%%%%%%%%%%%%%%%%%%%%%%%%%%%%%%%%%%%%%%%%%

Our fields are ``real'', in the sense that we do not have independent operators in front of the positive and negative frequency modes. The metric is explicitly real. Thus, the charge conjugation acts trivially - all operators go into themselves.

%%%%%%%%%%%%%%%%%%%%%%%%%%%%%%%%%%%%%%%%%%%%%%%%%%%%%%%%%%%
\subsection{Parity}
%%%%%%%%%%%%%%%%%%%%%%%%%%%%%%%%%%%%%%%%%%%%%%%%%%%%%%%%%%%

We could obtain the action of parity from the mode expansion for the metric, which is standard. We could also just directly define the action on the operators. Indeed, parity changes the sign of the spatial momentum, therefore interchanges the two helicities: 

\be\label{P_action_on_a}
P^\dagger a^\pm_k P= a^\mp_{-k}.
\ee

In view of \myref{h_mode_decomposition} this is equivalent to 

\be
P^\dagger h^{ij}(t,\vec{x}) P=  h^{ij}(t,-\vec{x}).
\ee

It is much more interesting to obtain the parity action on the connection field. Using \myref{P_action_on_a} and the mode decomposition \myref{a_mode_expansion} we get

\be
P^\dagger a^{ij}(t,\vec{x}) P=  - (a^{ij}(t,-\vec{x}))^\dagger.
\ee

The minus sign in this formula can be interpreted as being related to the fact that we are dealing with the spatial connection, which changes sign under parity.

%%%%%%%%%%%%%%%%%%%%%%%%%%%%%%%%%%%%%%%%%%%%%%%%%%%%%%%%%%%
\subsection{Time reversal}
%%%%%%%%%%%%%%%%%%%%%%%%%%%%%%%%%%%%%%%%%%%%%%%%%%%%%%%%%%%

Time-dependent physics in de Sitter space is not time reversal invariant. However, it can be made to be such by simultaneously reversing of the sign of the time coordinate and the sign of the parameter $M$. This sends one from one patch of de Sitter space (covered by the flat slicing) to another patch where the time flows in the opposite direction. Hence, it must be a symmetry of the theory. The action of the time reversal operator, which is anti-linear, can then be obtained by requiring

\be
T^\dagger h^{ij}(t,\vec{x}) T=  h^{ij}(-t,\vec{x})\bigg|_{M\rightarrow-M}.
\ee

This gives, at the level of the operators

\be\label{T_action_on_a}
T^\dagger a^\pm_k T= a^\pm_{-k}.
\ee

While parity flips the sign of the spatial momentum while leaving the particle spin unchanged, which results in flipping of the helicity, time reversal flips both the momentum and the spin, which does not change the helicity. At the level of the connection we get

\be
T^\dagger a^{ij}(t,\vec{x}) T=    a^{ij}(-t,\vec{x})\bigg|_{M\rightarrow-M}.
\ee

%%%%%%%%%%%%%%%%%%%%%%%%%%%%%%%%%%%%%%%%%%%%%%%%%%%%%%%%%%%
\subsection{CPT}
%%%%%%%%%%%%%%%%%%%%%%%%%%%%%%%%%%%%%%%%%%%%%%%%%%%%%%%%%%%

We now combine all of the above transformation rules into the action of the $CPT$ transformation. We see that, modulo an overall minus sign, this action is that of the spacetime inversion $(t,\vec{x})\rightarrow (-t,-\vec{x})$, as well as the Hermitian conjugation of the field. This is of course standard in field theory. Note, however, that in our case the Hermitian conjugation comes not from the charge conjugation, in spite of the fact that the field is complex. Rather, it is a part of the parity transformation. But the end result is the same: $CPT$ is Hermitian conjugation together with the spacetime inversion. This is the $CPT$ theorem for our theories - a $Hermitian$ Lagrangian will be $CPT$ invariant . At the same time, hermiticity of the Lagrangian is important for the unitarity of the theory. While we have seen this hermiticity at the linearised level (e.g. by going to the metric description), the question whether there exists an appropriate real structure on the space of solutions of the full theory that allows a real section to be taken as open.

%%%%%%%%%%%%%%%%%%%%%%%%%%%%%%%%%%%%%%%%%%%%%%%%%%%%%%%%%%%
\chapter{Graviton-graviton scattering amplitudes}\label{Amplitudes_chapter}
%%%%%%%%%%%%%%%%%%%%%%%%%%%%%%%%%%%%%%%%%%%%%%%%%%%%%%%%%%%
In a recent work by this author $et$ $al.$ \cite{Delfino:2012zy}, reviewed in chapter \myref{General_PCF_chapter}, a more systematic treatment of the kinetic sector of the theory introduced in \cite{Krasnov:2011pp} \cite{Krasnov:2011up} was given. The outcome of \cite{Delfino:2012zy} is the realisation that the reality conditions satisfied by the connection can only be properly understood for $\Lambda \neq 0$, i.e. before the Minkowski limit is taken. The previous chapter also derived the mode decomposition of the connection into creation/annihilation operators. The mode decomposition obtained demonstrates, in particular, that the gauge-theoretic description of gravitons is parity $asymmetric$, because the two helicities of the graviton are treated quite differently. One can then expect that a generic gravitational theory built using this formalism is parity-violating, which was confirmed in the paper by this author $et$ $al.$ \cite{Delfino:2012aj} by finding non-zero results for amplitudes of some parity-violating processes. This chapter will review such results.

The main objective of  this chapter is to derive the Feynman rules and to define the methods for extracting the graviton scattering amplitudes from the connection correlation functions. To illustrate how the formalism can be used for practical computation, we compute some of the simplest graviton scattering amplitudes, such as the two-to-two graviton ones.

One of the conclusions of this chapter is that the presented gauge-theoretic approach to GR works, in the sense that it can be used to reproduce the known amplitudes for the scattering of the gravitons. Most importantly, we find some new scattering amplitudes that are zero in GR, but are present in a general member of our family of theories.

However, the general message that this chapter means to get across is another.  We will show that, after one has taken care of the reality condition and understands how to take the Minkowski limit, the gauge-theoretic description of GR is a more efficient tool for practical Feynman diagram computations than the description based on the Einstein-Hilbert action. In particular, we shall see that not only the kinetic term of the Lagrangian is simpler (as we shown in chapter \myref{General_PCF_chapter}), but more importantly that also the interaction terms are much simplified with respect to EH. For instance, the cubic interaction vertex in the GR sector contains just $3$ terms, of which just a single term is responsible for the graviton-graviton scattering amplitude. The simplicity  of this approach figures even more prominently in the fourth order of the theory, where the interaction terms (in the general theory) can all be contained in a few lines instead of more than half a page of terms that appear in the full expansion (around Minkowski) of the quartic interaction in the EH picture (see appendix in \cite{Goroff:1985th}).

The organisation of this chapter is the following. We start, in section \myref{LSZ_section},  with the derivation of a prescription for how the scattering amplitudes can be obtained from the connection correlation functions. Here we also discuss the tricky issues of taking the Minkowski spacetime limit. We stress that even though the theory starts as being about gravitons in de Sitter space, the final prescription works with Minkowski space quantities, and, in particular, the usual Fourier transform is available. Then in section \myref{Gauge_fixing_propagator_section} we explain how the  gauge-fixing is done, and obtain the propagator. Section \myref{Interactions_section} computes the interaction terms of the Lagrangian, up to the quartic order. Section \myref{spinor_technology_section} reviews the necessary spinor technology that is key in this formalism. This technology is then immediately put to use in that the graviton polarization tensors are translated into the spinor language in the same section. In the following section \myref{Feynmans_rules_section} we also translate everything else in spinor terms and state the Feynman rules in their final, most useful for computational purposes, form. Section \myref{graviton_graviton_scattering_section} then computes the graviton-graviton scattering amplitudes.

%%%%%%%%%%%%%%%%%%%%%%%%%%%%%%%%%%%%%%%%%%%%%%%%%%%%%%%%%%%
\section{LSZ reduction and the Minkowski limit}\label{LSZ_section}
%%%%%%%%%%%%%%%%%%%%%%%%%%%%%%%%%%%%%%%%%%%%%%%%%%%%%%%%%%%
In this section we describe how graviton scattering amplitudes can be derived from the connection correlation functions. We will also give a detailed prescription of how the Minkowski spacetime amplitudes are extracted. We shall see that, if one is only interested in the Minkowski space graviton scattering amplitude, then all calculation can effectively be done in Minkowski space, where the usual Fourier transform (including in the time direction) is available. However, setting up the corresponding formalism requires some care, because of the blowing up of the factors $1/M$ in the interaction vertices, see below. The content of this section was first introduced by this author $et$ $al.$ in \cite{Delfino:2012aj}.

%%%%%%%%%%%%%%%%%%%%%%%%%%%%%%%%%%%%%%%%%%%%%%%%%%%%%%%%%%%
\subsection{Creation-annihilation operators}
%%%%%%%%%%%%%%%%%%%%%%%%%%%%%%%%%%%%%%%%%%%%%%%%%%%%%%%%%%%
To obtain a version of the LSZ reduction for our theory, we need expressions for the graviton creation/annihilation operators in terms of the field operator. To obtain these, let us first introduce the following definition for the polarization tensors

\be\label{polarization_tensors}
\e^-(k)_{ij}=\f{M}{\sqrt{2} \omega_k}m_i(k)m_j(k), \quad  e^+(k)_{ij}=\f{\sqrt{2} \omega_k}{M}\bar{m} _i(k) \bar{m}_j(k),
\ee

and the following modes

\be\label{u_v_modes}
u_k(x)= \f{\H}{M}e^{-\im \omega_k t + \im \vec{k}\vec{x} }, \quad v_k(x)= \f{M}{\H} e^{-\im \omega_k t + \im \vec{k}\vec{x} }\bigg( 1+ \f{\im \H}{\omega_k}-\f{\H^2}{2\omega_k^2}\bigg).
\ee

Then we can rewrite the mode expansion of the field \myref{a_mode_expansion} as 

\be\label{a_mode_expansion2}
\nn a_{ij}(t,\vec{x})= \int \f{d^3 x}{(2\pi)^3 2\omega_k} \bigg[ \e^-_{ij}(k) u_k(x) a^-_k +\e^+_{ij}(k) v_k^*(x) (a^-_k)^\dagger \quad
\\
-  \e^+_{ij}(k) v_k(x) a^+_k - \e^-_{ij}(k) u_k^*(x) (a^+_k)^\dagger \bigg].
\ee

Note that the polarization tensors $\e^\pm(k)$ and the modes $u_k(x),v_k(x)$ are $not$ complex conjugates of each other, that is because the connection is not Hermitian. 

Now we can write down the expressions for the Fourier transforms of the connection field operator. From
\ref{a_mode_expansion2} we get:

\be
\nn \int d^3 x e^{-\im\vec{k}\vec{x}}a^{ij}(t,\vec{x})= 
\f{1}{2\omega_k}
\bigg[ m^i m^j a^-_k \f{\H}{\sqrt{2}\omega_k} e^{-\im \omega_k t} + m^i m^j  (a^-_{-k})^\dagger \f{\sqrt{2}\omega_k }{\H} e^{\im \omega_k t}\bigg( 1 - \f{\im \H}{\omega_k}- \f{\H^2}{2\omega_k^2} \bigg) 
\\
\nn - \bar{m}^i \bar{m}^j a^+_k \f{\sqrt{2}\omega_k }{\H}  e^{-\im \omega_k t} \bigg( 1 + \f{\im \H}{\omega_k}- \f{\H^2}{2\omega_k^2} \bigg)  - \bar{m}^i \bar{m}^j (a^+_{-k})^\dagger  \f{\H}{\sqrt{2}\omega_k }  e^{\im \omega_k t} \bigg].
\\
\ee

We have used $m^i(-k)=\bar{m}^i(k)$ in the second and fourth terms. For compactness, the $k$ dependence of the null vector $m^i,\bar{m}^i$ is suppressed in the above formula, and it is assumed that they are all evaluated at the 3-vector $k$. We can now take the projection of the $m^i m^j$ or $\bar{m}^i \bar{m}^j$ terms, and then device an appropriate linear combination and its first time derivative to extract the creation-annihilation operators. We get 

\be
a^-_k = \im\e^+_{ij}(k) \int d^3 x \ v^*_k(x)\overleftrightarrow{\p_t} a^{ij}, \quad a^+_k = - \im\e^-_{ij}(k) \int d^3 x \ u^*_k(x)\overleftrightarrow{\p_t} a^{ij},
\\
(a^-_k)^\dagger = - \im\e^-_{ij}(k) \int d^3 x \ u_k(x)\overleftrightarrow{\p_t} a^{ij}, \quad (a^+_k)^\dagger  =  \im\e^+_{ij}(k) \int d^3 x \ v_k(x)\overleftrightarrow{\p_t} a^{ij},
\ee

where, as usual $f\overleftrightarrow{\p_t}g = f\p_t g - g\p_t f$, and $u_k(x),v_k(x)$ are the modes given by \myref{u_v_modes}. Importantly, all the creation-annihilation operators are expressed solely in terms of the field $a^{ij}$, and the complex conjugate field never appears. Thus, it is quite non-trivial to see that e.g.  $(a^-_k)^\dagger$ is the complex conjugate of $a^-_k$. This would involve using the reality condition for the field operator $a^{ij}$.
 
As usual in the proof of the LSZ reduction formulas, see e.g. \cite{Srednicki:2007qs} Chapter 5, we now take the time integrals of the time derivatives of the creation-annihilation operators. These are zero in free theory, but the corresponding expressions are used in and interacting theory to extract the scattering amplitudes.
So we have

\be
\nn a^-_k(\infty) - a^-_k(-\infty) = \int dt \ \p_t a^-_k 
% = \im \e^+_{ij}(k)\int d^4 x \p_t  \ v^*_k(x)\overleftrightarrow{\p_t} a^{ij} =  \im \e^+_{ij}(k)\int d^4 x \p_t  \ (v^*_k(x) \p_t a^{ij}-  a^{ij}\p_t v^*_k(x)) 
% \\
% = \im \e^+_{ij}(k)\int d^4 x  \ (v^*_k(x) \p_t^2 a^{ij}-  a^{ij}\p_t^2 v^*_k(x))
= \im \e^+_{ij}(k)\int d^4 x  \ v^*_k(x) \bar{D} D a^{ij},
\\
\nn a^+_k(\infty) - a^+_k(-\infty) = \int dt \ \p_t a^+_k = \im \e^-_{ij}(k)\int d^4 x  \ u^*_k(x) \bar{D} D a^{ij},
\\
\nn (a^-_k)^\dagger (\infty) - (a^-_k)^\dagger (-\infty) = \int dt \ \p_t (a^-_k)^\dagger = \im \e^-_{ij}(k)\int d^4 x  \ u_k(x) \bar{D} D a^{ij},
\\
(a^+_k)^\dagger (\infty) - (a^+_k)^\dagger (-\infty) = \int dt \ \p_t (a^+_k)^\dagger = \im \e^+_{ij}(k)\int d^4 x  \ v_k(x) \bar{D} D a^{ij},
\ee

where the definitions of the operators $D,\bar{D}$ are given in chapter \myref{General_PCF_chapter}. Note that on a connection satisfying its free theory field equation $\bar{D}Da = 0$ all these quantity are zero. We can now use these expressions to state the rules for extracting the graviton scattering amplitudes from the (interacting theory) connection correlation functions.

%%%%%%%%%%%%%%%%%%%%%%%%%%%%%%%%%%%%%%%%%%%%%%%%%%%%%%%%%%%
\subsection{LSZ reduction}
%%%%%%%%%%%%%%%%%%%%%%%%%%%%%%%%%%%%%%%%%%%%%%%%%%%%%%%%%%%

Quantum field theory in de Sitter space is an intricate subject with many subtleties. Because the background is time-dependent, one may argue that even the very in-out S-matrix is no longer defined,  see e.g. \cite{Marolf:2012kh} for a recent description of the difficulties that arise (and the possible ways to handle them). However, since in this thesis we are only interested in extracting the Minkowski limit results from our formalism, we can ignore all the subtleties and proceed in an exact analogy to what one does in Minkowski space.

We thus insert a set of graviton creation operators in the far past, and then a set of annihilation operators in the far future to form a graviton scattering amplitude.

\be
\nn \langle a^-_{k_-}(\infty)\ldots a^+_{k_+}(\infty)\ldots | (a^-_{p_-})^\dagger(-\infty)\ldots (a^+_{p_+})^\dagger(-\infty)\ldots \rangle
\\
:= \langle k_-\ldots k_+ | p_-\ldots p_+ \ldots \rangle. 
\ee
Here $k_-\ldots$ and $k_+\ldots$ are the set of $n$ negative and $m$ positive helicity outgoing graviton momenta, and $p_-\ldots p_+\ldots$ are the incoming momenta of $n'$ negative and $m'$ positive helicity gravitons. We now add the time ordering, and then express the annihilation operators in the future in terms of those in the past, and creation operators in the past in terms of those in the future via the formulas obtained in the previous subsection. This results in the following formula for the scattering amplitude  

\be\label{amplitude_from_LSZ}
\nn \langle k_-\ldots k_+ | p_-\ldots p_+  \rangle = \im^{n+n'-m-m'}\int d^4x_- \ \e^+_{ij}(k_-)v^*_{k_-}(x_-)\bar{D}D\ldots
\\
\nn 
\int d^4x_+ \ \e^-_{kl}(k_+)u^*_{k_+}(x_+)\bar{D}D\ldots
\int d^4y_- \ \e^-_{mn}(p_-)u_{p_-}(y_-)\bar{D}D\ldots 
\int d^4y_+ \ \e^+_{rs}(p_+)v_{p_+}(y_+)\bar{D}D\ldots
\\
\nn \langle T a^{ij}(x_-)\ldots a^{kl}(x_+)\ldots a^{mn}(y_-)\ldots a^{rs}(y_+) \rangle.
\\
\ee

The time-ordered correlation functions are then obtained from the functional integral via the usual perturbative expansion.

We note some unusual features of the formula \myref{amplitude_from_LSZ}. A similar formula can be written for extracting the amplitudes from the metric correlators. In this case, however, because the field equation satisfied by the metric perturbation is real, there is only one solution for each sign of the frequency. In other words, only one type of mode would appear in \myref{amplitude_from_LSZ}, together with its complex conjugate. In the case of the connection field, the field equation $\bar{D}Da=0$ is complex. This has the effect that for a given sign of the frequency, e.g. positive, there are two linearly independent solutions that we denoted by $u_k(x),v_k(x)$. This is of course just a manifestation of the parity asymmetry of our formalism, because one of the modes is used for the negative helicity and the other for the positive. 

%%%%%%%%%%%%%%%%%%%%%%%%%%%%%%%%%%%%%%%%%%%%%%%%%%%%%%%%%%%
\subsection{Minkowski limit: general discussion}
%%%%%%%%%%%%%%%%%%%%%%%%%%%%%%%%%%%%%%%%%%%%%%%%%%%%%%%%%%%
Now we would like to understand how the Minkowski spacetime graviton scattering amplitudes can be extracted from the general formula \myref{amplitude_from_LSZ}. The standard procedure for taking such limit is to keep the past and future time limits in the LSZ formula finite, take the $M\rightarrow 0 $ limit, and then take the time limits to infinity. This procedure, however, requires performing all computations in the position space, which is doable, but has the drawback of the spinor helicity formalism being not available to aid the computation. For this reason we will follow an alternative route that will eventually allow us to use the usual Fourier transform and the momentum spinors.

To set the stage, let us first discuss in more details how to take the Minkowski limit in the metric formalism. In this case one can also write a version of LSZ formula \myref{amplitude_from_LSZ}, and the amplitudes are obtained from correlation functions of the metric perturbation. For definiteness let us consider just one Feynman diagram contributing to some scattering amplitude. In this diagram, every graviton (on internal or external line) is characterised by its energy $\omega_k$, and we would like to concentrate on processes for which all gravitons $\omega_k \gg M$. A systematic way to do this is to expand all building blocks of this Feynman diagram (i.e. external wave function, propagators and vertices) in powers of $M/\omega_k$. The leading order term in this expansion is the desired Minkowski limit.

Then, instead of first computing the result in the de Sitter and the taking a limit, one can reduce the problem to a computation in Minkowski space. To see this, we need the above mentioned rule that the time intervals are taken to be finite, the  $M\rightarrow 0 $ limit is computed and then the time interval is set to infinity. Indeed, consider the time dependence of the external wave-functions and the propagators. For $\omega_k \gg M$ this can be separated into the ``fast'' time dependence coming from the exponent $e^{\im \omega_k t}$ and ``slow'' coming from the factors of $\H$. The covariant derivatives present in the vertices act on the external wave-functions and the propagators, and the result is sensitive to both the fast and slow dependence. Let us assume that all the time derivatives have been evaluated. Then, bearing in mind that we take the limit $M\rightarrow 0 $ while keeping all time intervals finite, we can switch off the ``slow'' time dependence by setting $M t\rightarrow 0 $ everywhere. In physical terms, this corresponds to assuming that the duration of any process is much shorter that the de Sitter time scale. With the slow time dependence switched off, the space and time dependence of all the quantities is as in flat space, and the Fourier transform becomes available. Effectively, the above discussion implies that we can just do the computation in Minkowski space, using the Minkowski limits of the vertices, propagators and the external wave-functions.

The situation in our formalism is not so simple. The source of complication, absent in the metric formulation is that our interaction vertices have powers of $1/M$ in front of them, see below. There are similar factors of $M$ and $1/M$ in the polarisation tensors also. Thus, in computing Feynman diagrams and applying \myref{amplitude_from_LSZ} we face the problem that we can only take the limit  $M\rightarrow 0 $ if the whole scattering amplitude, but not of the pieces that it is built from, as some of these pieces diverge in the limit, while some tend to zero. This appears to prevent us from doing the computation in Minkowski space.

We can, however, circumvent this problem by taking the limit in two steps. First, we can compute all time derivatives (coming from the vertices) present in a given Feynman diagram, and then switch off the slow time dependence. At this stage all the quantities are still $M$-dependent, only the time dependence became the fast time dependence of the exponential $e^{\im \omega_k t}$. We can then expand all the quantities, i.e. the external wave-functions and propagators (possibly with derivatives applied to them) in powers of $M/\omega_k$. At this stage, while $M$-dependence is still there, we can already do computations in Minkowski space. In the second step, after the Feynman diagram is evaluated, one keeps only the leading, zeroth order in $M$ term in the result.

An apparent difficulty with this prescription is that it seems hard to decide how many orders of the expansion in powers of $M/\omega_k$ we need to keep for each quantity. Indeed, in principle there could be powers of $1/M$ coming from the (positive) polarisation tensors and the vertices cancelling the powers of $M$ coming from $(M/\omega_k)^n$, and producing a finite result. This difficulty is resolved by the following consideration. Let us start by considering the leading order terms in the expansion of all the quantities. Let us call the result of the computation of the Feynman diagram where only the leading orders are kept its $leading$ $order$ $part$. The first assumption we need to make is that for physically relevant diagrams (i.e. those arising in computations of scattering amplitudes) it is never the case that the leading order parts blows up in the $M\rightarrow 0 $ limit. We will see that this is true in our case, however see some further discussion about this point below.

Let us now consider some amplitude with the leading part surviving in the limit $M\rightarrow 0 $. For such amplitude it is clear that the corrections obtained by keeping the subleading terms in the expansion of all its building blocks will be vanishing in the $M\rightarrow 0 $ limit. This establishes that it is sufficient to consider only the leading parts in the expansion of the (derivatives of the) wave-functions and the propagators.

Let us now come back to our assumption that the leading order part is never singular in the $M\rightarrow 0 $ limit. What we will see below when we do the computations is that the potentially singular leading order of the amplitudes, i.e. containing a negative power of $M$ in front of them are always zero for reasons of spinor contractions. In other words, these amplitudes are such that the spinors coming from the helicity tensor inserted on the external likes contract to zero. It could, in principle happen that there are $subleading$ terms in $M/\omega_k$ for which the spinor contractions do not give a null result, and this conspires to give a non-zero amplitude in the $M\rightarrow 0 $ limit. If this was the case, it would not be sufficient to just work with the leading parts of all the building blocks of the diagrams, which would make the analysis significantly more complicated. While we cannot give a general proof that this does not happen, the reasons for why the spinor contractions give a null result for potentially ``dangerous'' amplitudes are always very general, and appear to hold independently of any expansion in powers of $M/\omega_k$. We thus assume that the ``dangerous'' diagrams that have zero leading parts for reasons of spinor contractions are zero precisely, i.e. even before  any expansion in $M/\omega_k$ is performed. This allows us to work with only the leading parts of all the objects. As we have said, to us this assumption appears to be well motivated by the details of the computations that are made below. However, it would be comforting to either perform explicit checks (by doing computations in de Sitter space), or find another direct argument proving this. At the moment, all our analysis is based on this assumption. Our results, and in particular the fact that the GR amplitudes are correctly reproduced, seems to justify the assumption, but  more direct arguments would be very welcome. 

%%%%%%%%%%%%%%%%%%%%%%%%%%%%%%%%%%%%%%%%%%%%%%%%%%%%%%%%%%%
\subsection{Minkowski limit: prescription for the leading order parts}
%%%%%%%%%%%%%%%%%%%%%%%%%%%%%%%%%%%%%%%%%%%%%%%%%%%%%%%%%%%% 
Let us see how this works in practice. From the above discussion we know that we need to keep the leading order of the external wave-functions and propagators, as well as their derivatives (as coming from the vertices), when everything is expanded in powers of $M/\omega_k$. For propagators this is easy, as the leading order of the propagator itself is just its Minkowski limit, and the leading order of a covariant derivative of the propagator is just the usual partial derivative of the Minkowski limit. In physical terms, this can be rephrased by saying that in Minkowski all the internal lines, as well as the derivatives acting on them, are unaware of the fact that they are actually objects in de Sitter space. Thus, we only need to worry about the leading order parts of the external wave-functions and their covariant derivatives.

The covariant derivative that is applied to an external graviton wave-function appears in our interaction vertices only in the combination $D_{[\m}a^i_{\n]}$. This Lie algebra-valued two-form can be further decomposed into its self-dual and anti-self-dual parts as in \myref{Da_decomposition}. From the Hamiltonian analysis in \myref{General_PCF_chapter}, see formula \myref{aPSDa}, we know that the self-dual part of $D_{[\m}a_{\n]}^i$ is essentially given by the action of the operator $D$ on the spin 2 component of $a^{ij}$ of the spatial connection. This is modulo the term involving the derivative of the temporal component of the connection $a_0^i$ (shifted by $c^i$, see \myref{aPSDa}), which is set to zero in the Hamiltonian treatment. Similarly, it is clear that the anti-self-dual part of $D_{[\m}a_{\n]}^i$ is obtained by applying the derivative operator $\bar{D}$ to $a^{ij}$,  see formula \myref{DDbar_operators} for a definition of both operators. Thus, we have to consider the action of $D,\bar{D}$ on all the wave-functions that appear in \myref{amplitude_from_LSZ}, i.e. on $\e^-(k)u_k(x)$ and $\e^-(k)u_k^*(x)$, as well as on $\e^+(k)v_k(x)$ and $\e^+(k)v_k^*(x)$. Moreover, as we discussed above, we are only interested in the leading-order behaviour of these derivatives in the limit $M\rightarrow 0$. We get 

\be
\nn D\e^-(k)u_k = D \e^-(k)u_k^* = 0,
\\
\nn D\e^+(k)v_k \rightarrow -2\omega_k \e^+(k)e^{-\im \omega_k t + \im \vec{k}\vec{x}}, 
\quad 
D\e^+(k)v_k^* \rightarrow 2\omega_k \e^+(k)e^{\im \omega_k t - \im \vec{k}\vec{x}}.  
\ee

The action of the $\bar{D}$ operator has been worked out in \myref{Dbar_action_on_a}. Keeping the leading order, in the limit $M\rightarrow 0$, we can write

\be
\nn \bar{D}\e^-(k)u_k \rightarrow 2\omega_k \e^-(k)e^{-\im \omega_k t + \im \vec{k}\vec{x}},
\quad
\bar{D}\e^-(k)u_k^* \rightarrow -2\omega_k\e^-(k)e^{\im \omega_k t - \im \vec{k}\vec{x}},
\\
\nn \bar{D}\e^+(k)v_k \rightarrow -\f{M^2}{\omega_k}\e^+(k)e^{-\im \omega_k t + \im \vec{k}\vec{x}}, 
\quad 
\bar{D}\e^+(k)v_k^* \rightarrow \f{M^2}{\omega_k}\e^+(k)e^{\im \omega_k t - \im \vec{k}\vec{x}}.  
\ee

The idea now is to devise some Minkowski spacetime wave-functions that give exactly the same leading order results when the operators $lim_{M\rightarrow 0}D=-\im\p_t+\e\p, lim_{M\rightarrow 0}\bar{D}=\im\p_t+\e\p$ are applied. We also note that the limiting case operators are just the complex conjugates of each other.
 For the modes involving $u_k$ and its complex conjugate the answer is obvious - one should just take $lim_{M\rightarrow 0}u_k$ as the corresponding wave-function. Thus, we set 

\be\label{Minkowski_wave_function_u}
u_k^M(x):= e^{-\im \omega_k t + \im \vec{k}\vec{x}},
\ee

and use this wave-function (and its complex conjugate) instead of $u_k(x)$ and $u_k^*(x)$ every time it appears in the LSZ formula \myref{amplitude_from_LSZ}. the operators that act on $u_k^M$ (and its complex conjugate) are the Minkowski ones $ -\im\p_t+\e\p$ and $\im\p_t+\e\p$.

For the $v_k$ mode the situation is more non-trivial. If we choose it to be just the ordinary plane wave we will correctly reproduce the leading order of the action of the $D$ operator. However, the $ lim_{M\rightarrow 0}\bar{D}=\im\p_t+\e\p$ operator will give zero. The reason why one gets a non-zero answer when acting on the full wave-function $v_k(x)$ is that this has a non-trivial time-dependent factor multiplying $e^{-\im \omega_k t + \im \vec{k}\vec{x}}$. The idea is then to change the frequency $\omega_k$ in the plane wave to model this non-trivial time-dependent factor. This is achieved by the following plane waves

\be\label{Minkowski_wave_function_v}
v_k^M(x):= e^{-\im \tilde{\omega}_k t + \im \vec{k}\vec{x}},
\ee
where

\be\label{massive_omega}
\tl{w}_k := \sqrt{\omega_k^2-2M^2}=\omega_k \bigg( 1-\f{M^2}{\omega_k^2}+O(\f{M^4}{\omega_k^4}) \bigg).
\ee
Indeed, modulo the higher order corrections this choice of the wave-function gives precisely the required 

\be
\nn (\im\p_t+\e\p)\e^+(k)v^M_k(x)=-\f{M^2}{\omega_k}\e^+(k)v^M_k(x),
\\
 (\im\p_t+\e\p)\e^+(k)v^{*M}_k(x)=\f{M^2}{\omega_k}\e^+(k)v^{*M}_k(x).
\ee

At the same time, the leading order term in the result of the action of $-\im\p_t+\e\p$ on this mode is unchanged by this shift of the frequency.

Thus, the choice of \myref{Minkowski_wave_function_u}, \myref{Minkowski_wave_function_v} satisfies the requirement that when acted upon by the operators $-\im\p_t+\e\p$ and $\im\p_t+\e\p$
one reproduces the leading behaviour of the derivatives of the wave-functions with the full time-dependence.

Therefore, if we are interested in the Minkowski limit up to the first derivatives of the wave-functions, it is sufficient to replace the full wave-functions by \myref{Minkowski_wave_function_u}, \myref{Minkowski_wave_function_v}, and the operators $D,\bar{D}$ by the corresponding $M\rightarrow 0$ limit operators.

%%%%%%%%%%%%%%%%%%%%%%%%%%%%%%%%%%%%%%%%%%%%%%%%%%%%%%%%%%%
\subsection{The prescription for the Minkowski limit amplitudes}
%%%%%%%%%%%%%%%%%%%%%%%%%%%%%%%%%%%%%%%%%%%%%%%%%%%%%%%%%%%

All in all, we see that the computation of the Minkowski graviton amplitudes can be reduced to computations with the Minkowski $1/k^2$ propagator \myref{propagator}, the vertices obtained by replacing the covariant derivatives with partial ones everywhere, with the polarization tensors \myref{polarization_tensors}, and with the wave functions \myref{Minkowski_wave_function_u}, \myref{Minkowski_wave_function_v}. The rule for the wave-functions is that the on-shell condition for the negative helicity graviton is the usual one $\omega = \pm \omega_k$, while the positive helicity graviton should be taken to have a small mass $\omega = \pm \tl{\omega}_k$, with $\tl{\omega}_k$ given by \myref{massive_omega}. In doing the computation one keeps all the $M$-dependent prefactors, and at the end takes the limit $M\rightarrow 0$ (if this exists).

Let us write the above prescription in terms of a formula. Given that our Minkowski space wave-functions \myref{Minkowski_wave_function_u}, \myref{Minkowski_wave_function_v} are just the standard plane waves, we can immediately write down the momentum space LSZ prescription. Indeed, we note that the position space integrals give us the Fourier transforms of the time-ordered correlation function. As usual we will be assigning an arrow to each line in the Feynman diagram, with the arrows on incoming lines pointing towards the diagram, and the arrows on outgoing lines pointing away from it.  Then as usual the 4-momentum of each particle is to be understood as in the direction of the arrow. This prescription takes care of the factors of $u^M_k, v^M_k$ and their complex conjugates. The factors of $\bar{D}D$ will then amputate the propagators on the external lines. This will also absorb the overall factors of $\im$ in the formula \myref{amplitude_from_LSZ}. However, with our conventions for the mode decomposition there are some signs that are left over. Indeed, we have $\bar{D}D=\p_t^2-\Delta=-\p^\m\p_\m=-\Box$. At the same time we have $-\Box e^{\im k x}=k^2e^{\im k x}$, and so the operator $\bar{D}D$ will cancel $k^2$ from each external line. However, in our conventions the propagator is $1/\im k^2$, and so we will have a prefactor of $(-1)^{m+m'}$ left, where $m,m'$ are the numbers of incoming and outgoing $positive$ helicity gravitons. This nicely follows the pattern that positive helicity gravitons are a source of headache in the formalism. However these minus signs in the amplitudes are convention dependent and are not of any physical significance. We finally have

\be\label{amplitude_from_LSZ_prescription}
 \langle k_-\ldots k_+ | p_-\ldots p_+  \rangle = (-1)^{m+m'} (2\pi)^4\d^4\bigg( \sum k - \sum p\bigg)\quad
\\
\nn \e^+_{ij}(k_-)\e^-_{kl}(k_+)\e^-_{mn}(p_-)\e^+_{rs}(p_+)\langle T a^{ij}(k_-)\ldots a^{kl}(k_+)\ldots a^{mn}(p_-)\ldots a^{rs}(p_+) \rangle_{amput},
\ee
where the momentum space amplitude is amputated from its external line propagators. We should add to this formula the prescription that the positive helicity incoming particles 4-momenta, and the 4-momenta of the negative helicity outgoing particles, are slightly massive, see \myref{massive_omega}, while the 4-momenta of the incoming negative and outgoing positive helicity are massless $\omega = \omega_k = |k|$.

The above prescription is guaranteed to reproduce the correct leading order $M$-dependence. Thus, it is only consistent if the answers one gets turn out to have a non-zero limit as $M\rightarrow 0$, i.e. if all factors of $M$ cancel out from the end result. Below we will see that this is the case for the graviton-graviton amplitudes.

%%%%%%%%%%%%%%%%%%%%%%%%%%%%%%%%%%%%%%%%%%%%%%%%%%%%%%%%%%%
\subsection{Crossing symmetry}
%%%%%%%%%%%%%%%%%%%%%%%%%%%%%%%%%%%%%%%%%%%%%%%%%%%%%%%%%%%

We can now ask about an analogue of the field theory crossing symmetry relation for our scattering amplitudes. We shall discuss the crossing symmetry in the Minkowski space limit only. We recall that the usual QFT crossing symmetry arises if one takes an incoming particle of momentum $\vec{k}$ and energy $\omega_k$, and analytically continues the amplitude to energy $-\omega_k$. The amplitude can then be interpreted as that of an outgoing anti-particle of momentum $-k$.

Let us see what happens in our case. Since the field is electrically neutral, our particles are their own anti-particles, so we should only expect the helicity to change if we flip one graviton from the initial to the final state. That is indeed the case that is seen from our formula \myref{amplitude_from_LSZ_prescription}. Indeed, to make an outgoing particle incoming one should just flip the direction of the arrow on Feynman diagram external line corresponding to that particle. This will correctly continue $\omega_k\rightarrow-\omega_k$ and $\tl{\omega}_k\rightarrow-\tl{\omega}_k$, as well as change the sign of the corresponding 3-momentum. Note that we do not touch the helicity states. It is clear that if we apply the crossing symmetry to a negative helicity outgoing graviton, we produce a positive helicity incoming one, and similarly if we make a positive helicity outgoing graviton to be incoming, we will get a negative helicity one. This is clear from the fact that we are projecting on the negative/positive polarization tensors for the incoming negative/positive helicity gravitons, but do the reverse projection for the outgoing ones. Since by flipping one graviton from outgoing to incoming state we always change the total number of positive helicity gravitons in the amplitude (negative becomes positive, positive becomes negative, so there is always a change in the number of total positive helicity particles), then any crossing symmetry flip always introduces a minus sign coming from the $(-1)^{m+m'}$ prefactor. %Such minus sign is appropriate for fermions, and here we see it occurring for purely bosonic particles, which again signifies the analogy between our treatment and that of fermions.

Thus, we see that the crossing symmetry operates within our formalism, and we can from now on restrict our attention to all gravitons being e.g. incoming. Realistic scattering amplitudes can then be obtained from the by applying the crossing symmetry relations.

%%%%%%%%%%%%%%%%%%%%%%%%%%%%%%%%%%%%%%%%%%%%%%%%%%%%%%%%%%%
\section{Gauge-fixing and the propagator}\label{Gauge_fixing_propagator_section}
%%%%%%%%%%%%%%%%%%%%%%%%%%%%%%%%%%%%%%%%%%%%%%%%%%%%%%%%%%%

We will now derive the Feynman rules, starting with the propagator (that was already referred to in the previous section), and then finishing with the interaction vertices. The propagator in the pure connection formalism was obtained in \cite{Krasnov:2011up}. Here we repeat the analysis, by a slightly simpler method, and add some details such as the gauge-fixed linearised action before the Minkowski limit is taken. We first work in de Sitter space and then take the Minkowski limit, as is explained in the previous section.

%%%%%%%%%%%%%%%%%%%%%%%%%%%%%%%%%%%%%%%%%%%%%%%%%%%%%%%%%%%
\subsection{Diffeomorphisms}
%%%%%%%%%%%%%%%%%%%%%%%%%%%%%%%%%%%%%%%%%%%%%%%%%%%%%%%%%%%
As is discussed in more details in \myref{symmetries_section}, at the linearised level, diffeomorphisms act as $\d_\xi a^i_\m= \xi^\a \S^i_{\a\m}$. It is not hard to see what this action is by decomposing the field $a^i_\m$ into its irreducible components with respect to the action of the Lorentz group. Thus, let us introduce the following operators

\be\label{Lorentz_representations_projectors}
P^{(3,1)}_{\mu i| \nu j} := \frac{2}{3} \left( \delta_{ij} g_{\mu\nu} - \frac{1}{2} \epsilon_{ijk} \Sigma^k_{\mu\nu} \right), \qquad P^{(1,1)}_{\mu i| \nu j} := \frac{1}{3} \left( \delta_{ij} g_{\mu\nu}+ \epsilon_{ijk} \Sigma^k_{\mu\nu} \right).
\ee

Both act on pairs $\m i$ of a spacetime index and internal one. The projector $P^{(3,1)}$ is on the irreducible component $S^3_+\otimes S_-$, and $P^{(1,1)}$ is on $S_+\otimes S_-$ in the spinor representation of $S^3_+\otimes S_+\otimes S_-$ that the pair $\m i$ lives in. Here $S_+, S_-$ are the two 2-dimensional fundamental representations of the Lorentz group (i.e. the representations realised by unprimed and by primed 2-component spinors). The diffeomorphisms are then simply shifts of the (1,1) component. They can be completely gauge-fixed by requiring

\be
a_\mu^i=\epsilon^{ijk} \Sigma^j_{\mu}{}^\nu a_\nu^k,
\ee

or 

\be\label{gauge_fix_condition_diffeos}
P^{(1,1)}_{\m,i|\n,j} a^{\n j} = 0.
\ee

It is important to stress that the gauge-fixing condition for the diffeomorphisms does not contain derivatives (as is appropriate for a transformation that is merely a shift of the field in some direction in the field space).

We could now add the square of this gauge-fixing condition to the action (with some gauge-fixing parameter) and allow to propagate only the corresponding components of the connection. However, this would have the effect that some components of the connection have the $1/k^2$ propagator, while pure diffeomorphism gauge modes have a mode-independent, algebraic propagator. This would require dealing with the two components separately, which would make the formalism very cumbersome. To avoid this, we fix the gauge \myref{gauge_fix_condition_diffeos} sharply, i.e. work in the corresponding Landau gauge. We shall later see that the gauge fixing condition \myref{gauge_fix_condition_diffeos} is particularly transparent when expressed in spinor terms. It will simply state that everything but the completely symmetric (in spinor indices) component of the connection is zero. This condition will be then easy to impose and it will simplify the computation significantly.
On the other hand, the remaining gauge freedom, namely the usual $SO(3)$ transformations, will be gauge fixed (in the next subsection) as in Yang-Mills theory, i.e. by adding a $D^\m a_\m^i$ term squared to the Lagrangian. It is interesting to note that the way the gauge is fixed in our pure connection approach is opposite to that used in e.g. the first order formulation of general relativity that possesses both the internal as well as diffeomorphism gauge symmetry, as in our case. In the first-order formulation the gauge symmetry corresponding to $SO(1,3)$ local gauge transformations is fixed by requiring the tetrad perturbation $h^{IJ},(I,J=0,1,2,3)$ to be symmetric, i.e. by projecting away some irreducible components of the field with respect to the action of the Lorentz group. The diffeomorphisms are then fixed the usual derivative way, using the de Donder gauge. What happens in our formalism is precisely the opposite. The diffeomorphisms are fixed in a non-derivative way by projecting away some irreducible component of the field. The gauge rotations are then fixed by adding to the action the square of a term containing the derivative of the field.
%TO CHECK

%%%%%%%%%%%%%%%%%%%%%%%%%%%%%%%%%%%%%%%%%%%%%%%%%%%%%%%%%%%
\subsection{Gauge-fixing the rotations}
%%%%%%%%%%%%%%%%%%%%%%%%%%%%%%%%%%%%%%%%%%%%%%%%%%%%%%%%%%%
We now add to the Lagrangian the gauge-fixing term

\be
\L_{gr}= -\f{\a}{2}\left( D^\m P_{\m i|\n j }^{(3,1)}a^l{}^\n\right)^2,
\ee

%TO CHECK
where the projector is inserted to make the gauge-fixing condition diffeomorphism-invariant. As in \cite{Krasnov:2011up}, we could have then done the transformation in full generality, without imposing the gauge-fixing \myref{gauge_fix_condition_diffeos}. We would find that (for a choice of $\a$ to be given below) the gauge-fixed linearised Lagrangian (modulo the background curvature term) is just a multiple of $aD^\m D_\m P^{(3,1)}a$. However, we can simplify the computation significantly by imposing the gauge condition \myref{gauge_fix_condition_diffeos} from the very beginning.

The simpler computation is as follows. First, we note that the gauge-fixed connections satisfy $\S^{\m\n i}a^i_\n=0$.% indeed the gauge fixing condition \myref{gauge_fix_condition_diffeos} implies
% 
% \be
% a^{\m i} = \e_{ijk}\S^k_{\m\n}  a^{\n j},
% \ee
% 
% therefore
% 
% \be
% \nn \S^{\m\n i}a^i_\n= \S^{\m\n i}\S^k_{\n\r} \e_{ijk} a^{\r j} =
% \\
% \left( -\d^{ij}\eta^\m_\r+\e^{ijk}\S^{k \ \m}_\r \right)\e_{ijk} a^{\r j} = 0
% \ee

%where we used the algebra of the self-dual forms \myref{sigma_algebra}.
We can then ignore the $\d_{ij}\d_{kl}$ term in the projector $P_{ijkl}$, and write the Lagrangian \myref{PDaDa_action} as

\be\label{PDaDa_action_reduced}
\L^{(2)}=-\f{1}{2}\d_{ik} \d_{jl}\ (\S^{(i\m\n}D_\m a^{j)}_\n) (\S^{(ki\r\s}D_\r a^{l)}_\s).
\ee

We now use the gauge-fixing condition \myref{gauge_fix_condition_diffeos} to obtain the following identity

\be\label{anti_symmetric_part_of_SDa}
\Sigma^{i\mu\nu} D_\mu a_\nu^j - \Sigma^{j\mu\nu} D_\mu a_\nu^i=  \epsilon^{ijk} D^\mu a_{\mu}^k.
\ee

In other words %split symmetric plus antisymmetric parts

\be\label{SDa_symmetric_part}
\Sigma^{(i\mu\nu}D_\mu a^{j)}_\nu = \Sigma^{i\mu\nu}D_\mu a^{j}_\nu - \frac{1}{2} \epsilon^{ijk} D^\mu a_{\mu}^k.
\ee
In the derivation of \myref{anti_symmetric_part_of_SDa} we have secretly extended the covariant derivative with respect to the background connection to a derivative that also acts on the spacetime indices, so that $\S^{\m\n i}$ can be taken through the covariant derivative. Thus, from now on there is also the usual Christoffel symbol inside $D_\m$.

We now substitute \myref{SDa_symmetric_part} into the Lagrangian \myref{PDaDa_action_reduced}, and use the gauge-fixing condition once more to convert the result into a sum of just two terms:

\be
\L^{(2)} = -2 P^{+\m\n\r\s} \d_{ij}D_\m a_\n^i D_\r a_\s^j+\f{1}{4}\d_{ij}(D^\m a_\m^i)(D^\n a_\n^j),
\ee

where $P^{+\m\n\r\s}$ is a projector onto the self-dual part, defined as following

\be
\S^{\m\n i} \S^{\r\s i}=4P^{+\m\n\r\s} = g^{\m\r}g^{\n\s}-g^{\m\s}g^{\n\r}-\im \e^{\m\n\r\s},
\ee
which is just a multiple of the self-dual projector.
 We now represent the self-dual projector as

%note the indices in P^-
\be
4P^{+\m\n\r\s}= g^{\m\r}g^{\n\s}-g^{\m\n}g^{\r\s}+4P^{-\m\r\n\s},
\ee

where $P^-$ is the anti-self-dual projector. Now we integrate by parts in the term proportional to $P^-$, then, in terms that are anti-symmetric in the covariant derivatives, we express the commutator of two covariant derivatives via the curvature tensors. We have

\be
D_{[\m}D_{\r]}a^i_\s=-\f{1}{2}R_{\m\r\s}{}^\a a^i_\a + \f{1}{2}\e^{ijk} F_{\m\r}^j a^k_\s,
\ee

where $R_{\m\n\s}{}^\a$ is the Riemann curvature. We can now use the fact that the last term here is a purely self-dual quantity, and thus drop this part as it will be multiplied by $P^{-\m\r\n\s}$. We get:

%using the first Bianchi identity and the symmetry properties of the Riemann tensor
\be
4P^{-\m\r\n\s}D_{[\m}D_{\r]}a^i_\s=-\f{1}{2} (g^{\m\n}g^{\r\s}-g^{\m\s}g^{\n\r}+\im \e^{\m\r\n\s})  R_{\m\r\s}{}^\a a^i_\a = R_\n{}^\a a^i_\a,
\ee

where $R_\n{}^\a := g^{\s\r}R_{\s\n\r}{}^\a$ is the Ricci tensor. We finally get 

\be
\L^{(2)} = -\f{1}{2} \d_{ij} g^{\s\r} D_\m a_\r^i D_\m a_\s^j+\f{3}{4}\d_{ij}(D^\m a_\m^i)(D^\n a_\n^j)+\f{1}{2}\d_{ij}R^{\m\n} a^i_\m a^j_\n.
\ee

It is now clear that the choice $\a=3/2$ gives 

\be\label{second_order_lagrangian_with_R}
\L^{(2)} +\L_{gf} = -\f{1}{2} \d_{ij} g^{\s\r} D_\m a_\r^i D_\m a_\s^j+\f{1}{2}\d_{ij}R^{\m\n} a^i_\m a^j_\n.
\ee

which is just the scalar field Lagrangian for every component of $a^i_\m$ (projected onto the (3,1) representation by the gauge-fixing condition \myref{gauge_fix_condition_diffeos}), plus a curvature term.

We can further rewrite \myref{second_order_lagrangian_with_R} by using the fact that the background metric is Einstein:

%this is the Einstein equation $R_{\m\n}-\Lambda g_{\m\n} = 0$
\be
R_{\m\n} = 3 M^2 g_{\m\n}.
\ee

This gives for the gauge-fixed Lagrangian (after integrating by parts)

\be\label{second_order_lagrangian_with_M}
\L^{(2)} +\L_{gf} = \f{1}{2} a^{\m i} (D^2+3M^2)a^i_\m,
\ee

where $D^2=D^\m D_\m$. Note the ``wrong'' sign in front of the ``mass'' term here. It is not a source of any inconsistencies,  as the squared covariant derivative contains additional terms of the order $M^2$, and it overall gives rise to the mode behaviour in  chapter \myref{General_PCF_chapter}. This is of course also just the appropriate mass term for a massless spin two field in de Sitter space.

What is significant about the gauge-fixed Lagrangian \myref{second_order_lagrangian_with_M} is that all modes appear in it with the same sign in front of their kinetic term, unlike in the metric-based description that exhibits the conformal factor problem. This is also the reason why the Lagrangian \myref{second_order_lagrangian_with_M} gives rise to simpler Feynman rules than in the metric case. Indeed, in our case all the fields are treated uniformly, while in the metric case one often meets (e.g. in the vertices) the tracefree part of $h_{\m\n}$ and its trace separately, which makes computations more involved.

%%%%%%%%%%%%%%%%%%%%%%%%%%%%%%%%%%%%%%%%%%%%%%%%%%%%%%%%%%%
\subsection{Minkowski space propagator}
%%%%%%%%%%%%%%%%%%%%%%%%%%%%%%%%%%%%%%%%%%%%%%%%%%%%%%%%%%%
The quadratic form in the gauge-fixed Lagrangian \myref{second_order_lagrangian_with_R} can be inverted in full generality, using the relevant de Sitter space modes to construct the associated Green's function. However, as we have already stated above, effectively we are doing all our graviton scattering calculations in Minkowski space. Then the Lagrangian \myref{second_order_lagrangian_with_R} admits an obvious Minkowski spacetime limit:

\be
\L^{(2)} +\L_{gf}  = -\f{1}{2}(\p_\m a^i_\n)^2.
\ee 
The propagator is now easily obtained by going to the momentum space, and is obviously a multiple of the projector $P^{(3,1)}$ times $1/k^2$. To get all the factors right we introduce into the action a source term, and integrate out the connection

\be
S^{(2)}[a,J]=\int \f{d^4 k}{(2\pi)^4}\left[ -\f{1}{2}a^{\m i}(-k)k^2 a^i_\m(k) + J^{\m i}(-k) a^i_\m(k) \right].
\ee

Integrating out the connection we get

\be
S^{(2)}[J]= \int \f{d^4 k}{(2\pi)^4} \f{1}{2}J^{\m i} (-k)\f{ P^{ (3,1)}_{\m i|\n j}  }{k^2}J^{\n j}(k),
\ee

where the usual $\im\e$ prescription is implied. The propagator is then

\be\label{propagator}
\langle a_{\m i}(-k) a_{\n j}(k) \rangle = \f{1}{\im}\f{\d}{\d J^{\m i}(-k)} \f{1}{\im}\f{\d}{\d J^{\n j}(k)} e^{\im S[J]}\bigg|_{J=0} = \f{ P^{ (3,1) }_{\m i|\n j} }{\im k^2}.
\ee

%%%%%%%%%%%%%%%%%%%%%%%%%%%%%%%%%%%%%%%%%%%%%%%%%%%%%%%%%%%
\section{Interactions}\label{Interactions_section}
%%%%%%%%%%%%%%%%%%%%%%%%%%%%%%%%%%%%%%%%%%%%%%%%%%%%%%%%%%%
Having derived the propagator we only need the Feynman rule vertex factors, as well as the polarisations to be used to project the external legs of the diagrams onto physical graviton scattering amplitudes. We have already gave the expression for the latter \myref{polarization_tensors} when spelling out the mode decomposition. However, we will also need the covariant versions, and these will require introducing spinors. At the same time we can get sufficiently far in the analysis of the interaction without using spinors. Thus, we first work out the interactions. We first compute the full de Sitter interactions, and will specialise to the interaction vertices as relevant in the Minkowski limit in the section that computes their spinor versions.

%%%%%%%%%%%%%%%%%%%%%%%%%%%%%%%%%%%%%%%%%%%%%%%%%%%%%%%%%%%
\subsection{Decomposition of $Da$}
%%%%%%%%%%%%%%%%%%%%%%%%%%%%%%%%%%%%%%%%%%%%%%%%%%%%%%%%%%%
Before we do the algebra that exhibits the structure of the interaction vertices, let us introduce a convenient representation for the Lie-algebra valued two-form $D_{[\m}a^i_{\n]}$. We can write

\be\label{Da_decomposition}
D_{[\m}a^i_{\n]} = \f{1}{4}(Da)^{ij}\S^i_{\m\n} + \f{1}{8}\e^{ijk}(D^\r a_\r^j)\S^k_{\m\n}+(\widetilde{Da^i})_{\m\n},
\ee

where 
\be\label{Da_traceless_definition}
(Da)^{ij}:= \S^{\m\n (i}D_\m a^{j)}_\n, \quad (Da)^{ij}= (Da)^{(ij)}, \quad \Tr{Da}=0 
\ee

is the symmetric tracefree matrix that encodes the self-dual components of  $D_{[\m}a^i_{\n]}$, and $(\widetilde{Da^i})_{\m\n}$ stands for the anti-self-dual part

\be
(\widetilde{Da^i})_{\m\n}:= (D_{[\m}a^i_{\n]})^{asd}.
\ee

Our gauge-fixing condition \myref{gauge_fix_condition_diffeos}, together with the fact that the propagator contains the $P^{(3,1)}$ projector, implies that in all vertices and on all the lines, internal and external, the connection $a^i_\m$ can be taken to belong to just its $S^3_+\otimes S_-$ irreducible component. Indeed, on the internal lines this projection is carried out by the propagator. On the external lines it will be performed by the polarization tensors, see below. We can thus use the gauge-fixing condition for $a^i_\m$. Then the matrix $\S^{i \m\n}D_\m a_\n^j$ is traceless, and we have denoted its symmetric part by $(Da)^{ij}$ and wrote the anti-symmetric part as a separate (second) term in \myref{Da_decomposition}. As we shall see below, the above components of $D_{[\m}a^i_{\n]}$ encode different information, and this is why it is convenient to separate the self and anti-self-dual parts of $D_{[\m}a^i_{\n]}$ in the vertices.

Let us use the above expansion of $Da$ to rewrite some terms that frequently appear in the interaction vertices. We have

\be\label{eDaDa}
\nn \f{1}{\im}\e^{\m\n\r\s} D_\m a_\n^i D_\r a^j_\s = 
\f{1}{2}(Da)^{ik} (Da)^{kj}-\f{1}{2}(Da)^{k(i}e^{j)kl}(D^\m a^l_\m)
\\
+\f{1}{8}(\d^{ij}\d^{kl}-\d^{ik}\d^{jl})(D^\m a^k_\m)(D^\n a^l_\n)-2(\widetilde{Da^i})^{\m\n}(\widetilde{Da^j})_{\m\n}
\ee

and

\be\label{eDaaa}
\nn \f{1}{\im}\e^{\m\n\r\s} D_\m a_\n^i \e^{ikl} a^k_\r a^l_\s= \f{1}{2}(Da)^{ik}(\S\e aa)^{kj} - \f{1}{4} \e^{ikl}(\S\e aa)^{kj}(D^\m a^l_\m)- 2 (\widetilde{Da^i})^{\m\n} \e^{ikl}a^k_\m a^l_\n,
\\
\ee

where we have introduced 

\be\label{Seaa}
(\S\e aa)^{ij} := \S^{\m\n i} \e^{jkl} a^k_\m a^l_\n.
\ee
We note that $(\S\e aa)^{ij}$ is automatically symmetric as a consequence of \myref{gauge_fix_condition_diffeos}.

%%%%%%%%%%%%%%%%%%%%%%%%%%%%%%%%%%%%%%%%%%%%%%%%%%%%%%%%%%%
\subsection{Cubic interaction}
%%%%%%%%%%%%%%%%%%%%%%%%%%%%%%%%%%%%%%%%%%%%%%%%%%%%%%%%%%%

The cubic interaction vertex is obtained from the third order terms in the expansion of the action \myref{variations_gen_action}. Dividing the third variation by $3!$, and rescaling the variation of the connection $\d A^i_\m=(\im/\sqrt{g^{(2)}})a^i_\m$ we get the following third order Lagrangian

\be
\nn 3\im M^2 (g^{(2)})^{3/2}\L^{(3)}= g^{(3)}(Da)^{ij}(Da)^{jk}(Da)^{ki}
\\
\nn -\f{3 g^{(2)} }{2} \bigg( \f{1}{\im} \e^{\m\n\r\s} D_\m a^i_\n D_\r a^j_\s - M^2 (\S\e aa)^{ij} \bigg) (Da)^{ij} - \f{1}{\im} f(\d)M^2 \e^{\m\n\r\s} D_\m a^i_\n \e^{ijk} a^j_\r a^k_\s.
\\
\ee

We now use \myref{eDaDa} and \myref{eDaaa} to rewrite the above as

\be\label{3_vertex}
\nn 3\im M^2 (g^{(2)})^{3/2}\L^{(3)}= \left( g^{(3)} -\f{3 g^{(2)} }{4} \right) (Da)^{ij}(Da)^{jk}(Da)^{ki} + \f{3 g^{(2)} }{16} (Da)^{ij} (D^\m a^i_\m) (D^\n a^j_\n)
\\
\nn 3 g^{(2)} (Da)^{ij}(\widetilde{Da^i})^{\m\n}(\widetilde{Da^j})_{\m\n}+ \f{M^2}{2}(3 g^{(2)}-f(\d)) (Da)^{ij} (\S\e aa)^{ij} + 2M^2 f(\d)(\widetilde{Da^i})^{\m\n}\e^{ijk}  a^j_\m a^k_\n.
\\
\ee

We note that in the case of GR, see \myref{GR_constants}, the first term in the line 1 and the second term in the line 2 above are absent and we get simply

\be\label{GR_3_vertex}
\nn \im M_p M \L^{(3)}_{GR} = (Da)^{ij}(\widetilde{Da^i})^{\m\n}(\widetilde{Da^j})_{\m\n}+\f{1}{16}(Da)^{ij} (D^\m a^i_\m) (D^\n a^j_\n)+2M^2 (\widetilde{Da^i})^{\m\n}\e^{ijk}  a^j_\m a^k_\n.
\\
\ee

Below we shall see that only the first of these 3 terms in the cubic GR Lagrangian is important for the scattering of two gravitons of opposite helicities. Note that the terms which are cubic in the derivatives of the connection blow up in the limit $M\rightarrow 0$, both in the general theory case as in GR. Thus, care will have to be taken when going to this limit. Note also that the cubic interaction starts with $(\p a)^3$ terms, and thus seems to very different from the $(\p h)^2 h$ cubic vertex in the metric formulation. Still, we will see that in the case of \myref{GR_3_vertex} one is working with just a different description of the same GR interactions of gravitons. 

We would like to emphasise how much simpler the cubic vertex \myref{GR_3_vertex} is as compared to the 13 terms one finds in the expansion of the Einstein-Hilbert Lagrangian around the Minkowski background metric, see \cite{Goroff:1985th} formula (A.5) of the Appendix. The cubic vertex \myref{GR_3_vertex} is still more complicated than the one in the case of Yang-Mills theory, but we shall see that in many cases (e.g. for purposes of computing GR amplitudes) one effectively needs only the first term, which is of the same degree of complexity as in the Yang-Mills case. The analogy with Yang-Mills will become more striking when we write down the spinor expression for this cubic vertex below.

%%%%%%%%%%%%%%%%%%%%%%%%%%%%%%%%%%%%%%%%%%%%%%%%%%%%%%%%%%%
\subsection{Quartic interaction}
%%%%%%%%%%%%%%%%%%%%%%%%%%%%%%%%%%%%%%%%%%%%%%%%%%%%%%%%%%%
We now work out the quartic term. Dividing the fourth variation of the action from \myref{variations_gen_action} by $4!$ we get

\be\nn
-12 M^4 (g^{(2)})^2 {\cal L}^{(4)} = - g^{(4)}  (Da)^{ij} (Da)^{ij} (Da)^{kl} (Da)^{kl}  
\\ \nn 
+ 6 g^{(3)} P_{ijkl} (Da)^{im} (Da)^{mj} \left( \frac{1}{\im}\epsilon^{\mu\nu\rho\sigma} D_\mu a_\nu^k D_\rho a_\sigma^l - M^2 (\Sigma\epsilon  aa)^{kl} \right) 
\\ \nn
+\frac{6}{\im}   g^{(2)} M^2 (Da)^{ij}  \epsilon^{\mu\nu\rho\sigma} D_\mu a_\nu^i \epsilon^{jkl} a_\rho^k a_\sigma^l + g^{(2)} (Da)^{ij}(Da)^{ij} \left( \frac{1}{\im}\epsilon^{\mu\nu\rho\sigma} D_\mu a_\nu^k D_\rho a_\sigma^k - M^2 (\Sigma\epsilon  aa)^{kk} \right) 
\\ \nn
- \frac{3g^{(2)}}{2} P_{ijkl} \left( \frac{1}{\im}\epsilon^{\mu\nu\rho\sigma} D_\mu a_\nu^i D_\rho a_\sigma^j - M^2 (\Sigma\epsilon  aa)^{ij} \right)\left( \frac{1}{\im}\epsilon^{\mu\nu\rho\sigma} D_\mu a_\nu^k D_\rho a_\sigma^l - M^2 (\Sigma\epsilon  aa)^{kl} \right).
\ee

For purposes of this chapter we will only need the 4-vertex when evaluated completely on-shell. Thus, let us use the Lorentz gauge condition $D^\mu a_\mu^i=0$. This simplifies both \myref{eDaDa} and \myref{eDaaa}. Using these expansions, and collecting the terms we get

\be\label{4_vertex}
\nonumber
-12 M^4 (g^{(2)})^2 {\cal L}^{(4)} = \left( - g^{(4)} + \f{1}{2}g^{(3)} + \frac{7g^{(2)}}{16} \right) (Da)^{ij} (Da)^{ij} (Da)^{kl} (Da)^{kl}
\\ %\label{L4}
\nn -3(4g^{(3)} - g^{(2)}) (Da)^{ik} (Da)^{kj} (\widetilde{Da^i})^{\mu\nu} (\widetilde{Da^j})_{\mu\nu} + (4g^{(3)} - 3g^{(2)}) (Da)^{ij} (Da)^{ij} (\widetilde{Da^k})^{\mu\nu} (\widetilde{Da^k})_{\mu\nu} \\ \nonumber
- \frac{3M^2}{2}( 4g^{(3)} - 3g^{(2)}) \left( (Da)^{ik}(Da)^{kj} -\frac{1}{3} {\rm Tr}((Da)^2) \delta^{ij}\right) (\Sigma \epsilon aa)^{ij} \\ \nonumber
-\frac{3g^{(2)}}{2} P_{ijkl} \left( 2(\widetilde{Da^i})^{\mu\nu} (\widetilde{Da^j})_{\mu\nu} + M^2  (\Sigma \epsilon aa)^{ij}\right)\left( 2(\widetilde{Da^k})^{\mu\nu} (\widetilde{Da^l})_{\mu\nu}+ M^2  (\Sigma \epsilon aa)^{kl}\right)  
\\ \nn
- 12 M^2 g^{(2)} (Da)^{ij} (\widetilde{Da^i})^{\mu\nu} \epsilon^{jkl} a_\mu^k a_\nu^l.
\\
\ee

In the case of GR many of these terms become zero and we get a much simpler (on-shell $D^\mu a_\mu^i=0$) 4-vertex for GR:

\be\label{L4-GR}
\nn 2M_p^2 M^2 {\cal L}^{(4)}_{\rm GR} = (Da)^{ik} (Da)^{kj} (\widetilde{Da^i})^{\mu\nu} (\widetilde{Da^j})_{\mu\nu} + 2 M^2  (Da)^{ij} (\widetilde{Da^i})^{\mu\nu} \epsilon^{jkl} a_\mu^k a_\nu^l \\  
+ P_{ijkl} \left( (\widetilde{Da^i})^{\mu\nu} (\widetilde{Da^j})_{\mu\nu} + \frac{M^2}{2}  (\Sigma \epsilon aa)^{ij}\right)\left( (\widetilde{Da^k})^{\mu\nu} (\widetilde{Da^l})_{\mu\nu}+ \frac{M^2}{2}  (\Sigma \epsilon aa)^{kl}\right).
\ee

This should be compared with the much more formidable expression in the case of the metric-based GR, see \cite{Goroff:1985th}, formula (A.6). Even with the graviton field on-shell and the background metric taken to be flat, this occupies about half a page, as compared to just to lines in \myref{L4-GR}. We also note that both the GR 4-vertex as well as the general vertex \myref{4_vertex} start with terms $(\p a)^4$, to be compared with just two derivatives present in the metric-based vertex $(\p h)^2 hh$. this is part of a general pattern, and in our gauge-theoretic description the order $n$ vertex starts from $(\p a)^n$ terms.

%%%%%%%%%%%%%%%%%%%%%%%%%%%%%%%%%%%%%%%%%%%%%%%%%%%%%%%%%%%
\section{Spinor technology and the helicity spinors}\label{spinor_technology_section}
%%%%%%%%%%%%%%%%%%%%%%%%%%%%%%%%%%%%%%%%%%%%%%%%%%%%%%%%%%%

As is common to any modern derivation of the scattering amplitudes, the formalism of helicity states turns out to be extremely convenient. Indeed, scattering amplitudes are most efficiently described using spinors, or, as some literature calls them, twistors. The recent wave of interest into the spinor helicity methods originates in \cite{Witten:2003nn}. The method itself is, however, at least twenty years older, see e.g. \cite{Gunion:1985vca}, \cite{Chalmers:1997ui}. We start by listing some formulas involving spinors, mainly to establish the conventions.

%%%%%%%%%%%%%%%%%%%%%%%%%%%%%%%%%%%%%%%%%%%%%%%%%%%%%%%%%%%
\subsection{Soldering form}
%%%%%%%%%%%%%%%%%%%%%%%%%%%%%%%%%%%%%%%%%%%%%%%%%%%%%%%%%%%

The soldering form provides a map from the space of vectors to the space of rank two spinors (with two indices of opposite types). We use the conventions with a Hermitian soldering form:

\be
(\t^{AA'}_\m)^*=\t^{AA'}_\m.
\ee

The metric is obtained as a square of the soldering form:

\be\label{eta_with_thetas}
\eta_{\m\n}= -\t_{\m A}{}^{A'}\t_{\n B}{}^{B'} \e^{AB}{}_{A'B'},
\ee
where the minus sign is dictated by our desire to work with an Hermitian soldering form, while at the same time have signature $(-,+,+,+)$. We can also rewrite this formula as

\be
\eta_{\m\n}=  \t^{A}{}_{\m A'}\t_{\n A}{}^{A'},
\ee

so that the minus sign disappears. The contraction that appears in this formula, i.e. unprimed indices contracting bottom left to top right, and the primed indices contracting oppositely, will be referred to as $natural$ contraction. We will sometimes use index free notation and then the natural contraction will be implied.

%%%%%%%%%%%%%%%%%%%%%%%%%%%%%%%%%%%%%%%%%%%%%%%%%%%%%%%%%%%
\subsection{The spinor basis}
%%%%%%%%%%%%%%%%%%%%%%%%%%%%%%%%%%%%%%%%%%%%%%%%%%%%%%%%%%%

It is very convenient to introduce in each spinor space $S_+,S_-$ a certain spinor basis. Since each space is (complex) 2-dimensional we need two basis vectors for each space. Let us denote these by

\be
o_A,\i_A \in S_+, \quad o^{A'},\i^{A'}\ \in S_-.
\ee

Note that we shall assume that the basis in the space of primed spinors is the complex conjugate of the basis in the space $S_+$:

\be
\i^{A'} = (\i^A)^*, \quad o^{A'} = (o^A)^*.
\ee

The basis vectors are pronounced as ``omicron'' and ``iota''. Since the norm of every spinor is zero, we cannot demand that each of the basis vectors is normalised. However, we can demand that the product between the two basis vectors in each space is unity. Thus, the basis vectors satisfy the following normalisation:

\be
\i^A o_A = 1, \quad \i^{A'}o_{A'} = 1.
\ee

Of course, a spinor basis in each space $S_+,S_-$ is only defined up to an $SL(2,\mathbb C)$ rotation. Any $SL(2,\mathbb C)$ rotated basis gives an equally good basis, and it can be seen that any two bases can be related by a (unique) $SL(2,\mathbb C)$ rotation.

Once a spinor basis is introduced, we have the following expansion of the spinor metric, that is the $\e_{AB}$ symbol

\be
\e_{AB}=o_A \i_B- \i_A o_B.
\ee
A similar formula is also valid for $\e_{A'B'}$. To lower and raise indices we use follow the convention

$$V^M \e_{MN}=V_N, \quad V^{M'} \e_{M' N'}=V_{N'}, \quad \e^{NM}V_M=V^N, \quad  \e^{N' M'} V_{M'}=V^{N'}.$$

%%%%%%%%%%%%%%%%%%%%%%%%%%%%%%%%%%%%%%%%%%%%%%%%%%%%%%%%%%%
\subsection{The soldering form in the spinor basis}
%%%%%%%%%%%%%%%%%%%%%%%%%%%%%%%%%%%%%%%%%%%%%%%%%%%%%%%%%%%

The following explicit expression for the soldering form $\t_{\m}{}^{AA'}$ in terms of the basis one-forms $t_\m$ and $x_\m, y_\m, z_\m$, as well as the spinor basis vector $o^A,o^{A'},\i^A,\i^{A'}$ can be obtained:

\be
\nn \t^{AA'}_\m = \f{t_\m}{\sqrt{2}}(o^A o^{A'}+\i^A\i^{A'})+ \f{z_\m}{\sqrt{2}}(o^A o^{A'}-\i^A\i^{A'})+ \f{x_\m}{\sqrt{2}}(o^A \i^{A'}+\i^A o^{A'})+ \f{\im y_\m}{\sqrt{2}}(o^A \i^{A'}- \i^A o^{A'}).
\\
\ee
Note that the above expression is explicitly Hermitian.

Collecting the components in front of equal spinor combinations in the above formula for the soldering form we can rewrite it as:

\be\label{theta_with_null_tetrad}
\t^{AA'}_\m = l_\m o^A o^{A'}+n_\m \i^A \i^{A'} + m_\m o^A \i^{A'} + \bar{m}_\m \i^A o^{A'},
\ee

where

\be
l_\m = \f{t_\m+z_\m}{\sqrt{2}},\quad n_\m = \f{t_\m-z_\m}{\sqrt{2}},\quad m_\m = \f{x_\m+\im y_\m}{\sqrt{2}},\quad \bar{m}_\m = \f{x_\m-\im y_\m}{\sqrt{2}}.
\ee

Note that $l,n$ are real one-forms, while $\bar{m}_\m= m_\m^*$. The above collection of one-forms is known as \textit{doubly null tetrad}. Indeed, it is easy to see that all 4 one-forms introduced above are null\footnote{Recall also that we are using the metric signature $(-,+,+,+)$.}, e.g. $l^\m l_\m=0$. The only non-zero products are

\be
l^\m n_\n=-1, \quad m^\m \bar{m}_\m= 1.
\ee

Thus, the Minkowski metric can be written in terms of a doubly null tetrad as

\be
\eta_{\m\n} = - l_\m n_\m - n_\m l_\n + m_\m\bar{m}_\n+\bar{m}_\m m_\n,
\ee

which can also be verified directly by substituting \myref{theta_with_null_tetrad} into the formula \myref{eta_with_thetas} for the metric.

We can also verify the contraction of two soldering forms

\be
\t^\m_{AA'}\t^\m_{BB'}=-\e_{AB}\e_{A'B'}.
\ee

%%%%%%%%%%%%%%%%%%%%%%%%%%%%%%%%%%%%%%%%%%%%%%%%%%%%%%%%%%%
\subsection{Self-dual two forms}
%%%%%%%%%%%%%%%%%%%%%%%%%%%%%%%%%%%%%%%%%%%%%%%%%%%%%%%%%%%

The self-dual two-forms that play the central role in this chapter can be written down more naturally (i.e. without any reference to the time plus space decomposition of the tetrad internal index) in terms of spinors. We use the following definition:

\be\label{S_AB_with_thetas}
\S^{AB}:= \f{1}{2}\t^A{}_{A'}\wedge \t^{BA'},
\ee

or, without the form notation

\be
\S^{AB}_{\m\n}= \t_{\m A'}{}^{(A} \t_{\n}{}^{B)A'},
\ee

where we used the fact that symmetrisation on the unprimed spinor indices has the same effect as the anti-symmetrisation on the spacetime indices. 

Explicitly, in terms of the null tetrad and the spinor basis we get

\be
\S^{AB}= l\wedge m o^A o^B+ \bar{m}\wedge n \i^A \i^B + (l\wedge n - m\wedge \bar{m}) \i^{(A} o^{B)}.
\ee
Using $\epsilon_{\mu\nu\rho\sigma}=24\im \, l_{[\mu} n_{\nu} m_{\rho} \bar{m}_{\sigma]}$ it can be checked that these forms are indeed self-dual $$\epsilon_{\mu\nu}{}^{\rho\sigma} \Sigma_{\rho\sigma}^{AB} = 2\im \Sigma_{\mu\nu}^{AB}.$$

Let us also give the following useful formula for the decomposition of a contraction of two soldering forms (via a primed index) in terms of the metric and the self-dual two-forms:

\be
\t_{\m A'}{}^{A} \t_{\n}{}^{BA'}= -\f{1}{2}\e^{AB}g_{\m\n}+\S^{AB}_{\m\n}.
\ee

This is easily checked by either contracting with $\e_{AB}$, which produces minus the metric on both sides, or by symmetrising with respect to $AB$, which reproduces \myref{S_AB_with_thetas}.

%%%%%%%%%%%%%%%%%%%%%%%%%%%%%%%%%%%%%%%%%%%%%%%%%%%%%%%%%%%
\subsection{$SU(2)$ spinors}
%%%%%%%%%%%%%%%%%%%%%%%%%%%%%%%%%%%%%%%%%%%%%%%%%%%%%%%%%%%

We need to introduce the notion of $SU(2)$ spinors when we consider the Hamiltonian formulation of any fermionic theory. In our case, we need this notion to establish a relation between our polarization tensors \myref{polarization_tensors} and some spacetime covariant expression that we shall write down below.

To define $SU(2)$ spinors we need a Hermitian positive-definite form on spinors. This is a rank 2 mixed spinor $G_{A'A}:G^*_{A'A}=G_{A'A}$, such that for any spinor $\l^A$ we have  $(\l^*)^{A'}\l^A G_{A'A}>0$. Here $(\l^*)^{A'}$ is the complex conjugate of $\l^A$. We can define the $SU(2)$ transformations to be those $SL(2,\mathbb C)$ ones that preserve the form $G_{A'A}$. Then $G_{A'A}$ defines an anti-linear operation $\star$ on spinors via:

\be
(\l^\star)_A := G_{AA'}(\l^*)^{A'}.
\ee

We require that the anti-symmetric rank 2 spinor $\e_{AB}$ is preserved by the $\star$-operation:

\be
(\e^\star)_{AB} = \e_{AB},
\ee

which implies the following normalisation condition

\be\label{G_normalisation_condition}
G_{AA'} G^{A'}{}_B= \e_{AB}.
\ee

Using the normalisation condition we find that $(\l^{\star \star})^A = - \l^A$, or

\be
\star^2 = -1.
\ee

Thus, the $\star$-operation so defined is similar to a ``complex structure'', except for the fact that is anti-linear:

\be
(\a \l^A +\b\eta^A)^\star= \bar{\a}(\l^\star)^A+\bar{\b}(\eta^\star)^A.
\ee

Now for the purpose of comparing to results of the 3+1 decomposition, we need to introduce a special Hermitian from that arises once a time vector field is chosen. We can consider the zeroth component of the soldering form

\be
\t^{AA'}_0 = \t^{AA'}_\m \left( \f{\p}{\p t} \right)^\m = \f{1}{\sqrt{2}}\left(  o^A o^{A'}+  \i^A \i^{A'}\right) .
\ee

It is Hermitian, and so we can use a multiple of $\t^{AA'}_0$ as $G^{AA'}$. It remains to satisfy the normalisation condition \myref{G_normalisation_condition}. This is achieved by

\be
G^{AA'}= \sqrt{2}\t^{AA'}_0.
\ee

We then define the spatial soldering form via

\be
\s^{i \ AB} := G^A{}_{A'}\t^{i \ BA'},
\ee

which is automatically symmetric $\s^{i \ AB} = \s^{i \ (AB)} $ because its anti-symmetric part is proportional to the product of the time vector with a spatial vector, which is zero. Explicitly, in terms of the spinor basis introduced above we have

\be\label{spatial_soldering_form_in_spinors}
\s^{i AB}= m^i o^A o^B-\bar{m}^i \i^A \i^B - \f{z^i}{\sqrt{2}}(\i^A o^B+o^A \i^B). 
\ee

The action of the $\star$-operation on the basis spinor is as follows:

\be\label{star}
(o^\star)^A= -\iota^A, \quad (\i^\star)^A=o^A.
\ee

It is then easy to see from \myref{spatial_soldering_form_in_spinors} that the spatial soldering form so defined is anti-Hermitian with respect to the $\star$ operation:

\be
(\s^{i\star})^{AB}=-\s^{i AB}.
\ee
It is not hard to deduce the following property of the product of two spatial soldering forms:

\be
\s^i_A{}^B\s^j_B{}^C= \f{1}{2}\d^{ij}\e_A{}^C + \f{\im}{\sqrt{2}}\e^{ijk}\s^{k C}_A.
\ee

%%%%%%%%%%%%%%%%%%%%%%%%%%%%%%%%%%%%%%%%%%%%%%%%%%%%%%%%%%%
\subsection{Converting to the spinor form}
%%%%%%%%%%%%%%%%%%%%%%%%%%%%%%%%%%%%%%%%%%%%%%%%%%%%%%%%%%%

We are now ready to use the spinor objects introduced above. First, let us discuss how the expressions written in $SO(3)$ notation used so far can be converted into spinor notation. Indeed, we have so far worked with the connection perturbation being $a^i_\m$. It is now convenient to pass to the spinor description, in which all indices of $a^i_\m$ are converted into spinor ones. This is done with the soldering form from the spacetime index, and with Pauli matrices for the internal one.

To fix the form of the multiple of the Pauli matrices that is relevant here, we will require that under this map the identity matrix $\d^{ij}$ becomes the matrix $\e^{(A|C|}\e^{B)D}$ in spinor notation. Indeed, both matrices have trace 3. Thus, we denote the map from objects with $SO(3)$ indices to those with pairs of unprimed spinor indices by $T^{i AB}$ and require it to have the property:

\be
\d^{ij}T^{i AB}T^{j CD}= \e^{(A|C|}\e^{B)D}.%checked
\ee

This fixes $T^{i AB}$ up to a sign.

Now, to determine what multiple of this object appears in the relation between $\S^i_{\m\n}$ and $\S^{AB}_{\m\n}$, both of which have been defined before, we need to look into the algebra satisfied by them. We have the algebra used many times in the preceding text:

\be\label{Sigma_algebra_converting_to_spinor_form_section}
\S^i_\m{}^\r \S^j_{\r \n}= -\d^{ij}\eta_{\m\n}-\e^{ijk}\S^k_{\m\n}.
\ee 

At the same time, a simple computation of the same contraction for $\S^{AB}_{\m\n}$ gives

\be\label{Sigma_algebra_in_spinors}
\S^{AB}_\m{}^\r \S^{CD}_{\r \n}= - \f{1}{2}\e^{(A|C}\e^{B)D}\eta_{\m\n}+\f{1}{2}\e^{A(C}\S^{D)B}_{\m\n} +\f{1}{2}\e^{B(C}\S^{D)A}_{\m\n}.
\ee 

The coefficient in front the first term here is half of that in \myref{Sigma_algebra_converting_to_spinor_form_section}. We thus learn that there is a factor of $\sqrt{2}$ in the conversion of an $SO(3)$ index into a pair of symmetric spinor ones:

\be\label{spin-con-1}
\f{1}{\sqrt{2}}\S^i_{\m\n}T^{i AB}=\S^{AB}_{\m\n}.
\ee

To fix the sign of the quantity $T^{i\, AB}$ we just need to compare the terms containing $\Sigma$ in \myref{Sigma_algebra_converting_to_spinor_form_section} and \myref{Sigma_algebra_in_spinors}. Substituting \myref{spin-con-1} into \myref{Sigma_algebra_in_spinors} we get
\be\label{s-a-3}
\sqrt{2} \epsilon^{ij}{}_{k} T^{i\, AB} T^{j\, CD} = - \epsilon^{A(C} T_k^{D)B} - \epsilon^{B(C} T_k^{D)A}.
\ee
We thus see that the matrices $T^{i AB}$ satisfy the following algebra:
\be
T^{i AB} T^{j}_B{}^C = -\frac{1}{2} \delta^{ij} \epsilon^{AC} + \frac{1}{\sqrt{2}}\epsilon^{ijk} T^{k AC},
\ee
which fixes them uniquely. We see that these quantities are just 
$$T^{i AB}=-\im \sigma^{i AB},$$
where $\sigma^{i AB}$ are the spatial soldering forms introduced above. Explicitly, in terms of the spinor basis, as well as a basis $m^i,\bar{m}^i,z^i$ in ${\mathbb R}^3$ we have:
\be\label{T}
T^{i\,AB}= -\im m^i o_A o_B + \im \bar{m}^i \iota_A \iota_B+\frac{\im}{\sqrt{2}} z^i (\iota_A o_B+o_A\iota_B ).
\ee
Note that $T^{i AB}$ is $\star$-Hermitian, i.e. $(T^{i\star})^{AB} = T^{i AB}$. 

We can now write down the conversion rule of the $\epsilon^{ijk}$ tensor. Thus, introducing
$$ \epsilon^{(AB) \ (CD) \ (EF)}: = \epsilon^{ijk} T^{i \, AB} T^{j\, CD} T^{k \, EF}$$ 
we get from (\ref{s-a-3})
\be\label{eps-conv}
\sqrt{2}\ \epsilon^{(AB) \ (CD)}{}_{(EF)}\:X^{(EF)}=-\epsilon^{A(C} X^{D)B}-\epsilon^{B(C} X^{D)A}.
\ee
This can be rewritten more conveniently as a rule for the commutator
\be
\sqrt{2} \epsilon^{(AB)}{}_{(CD)(EF)} X^{CD} Y^{EF}= X^{AE} Y_{E}{}^B+ X^{BE}Y_E{}^A,
\ee
where the spinor contraction is in a natural order.

%%%%%%%%%%%%%%%%%%%%%%%%%%%%%%%%%%%%%%%%%%%%%%%%%%%%%%%%%%%
\subsection{Further on the spinor conversion}
%%%%%%%%%%%%%%%%%%%%%%%%%%%%%%%%%%%%%%%%%%%%%%%%%%%%%%%%%%%

Let us now discuss the rules for dealing with the spacetime indices. Each such index has to be converted into a mixed type pair of spinor indices using the soldering form $\t^{\m AA'}$. We shall refer to the operator of the partial derivative with its spacetime index converted into a pair of spinor indices as the Dirac operator:

\be
\p_\m := \t_{\m AA'}\p^{AA'}. 
\ee

Note that, because of our signature choice, and thus the minus sign in \myref{eta_with_thetas}, we have $\p^{AA'}= - \t^{\m AA'}\p_\m$. One has to be careful about these minus signs. %checked

We now come to objects that have both types of indices, spacetime and internal. The conversion of these is that we write them as the corresponding soldering forms times objects with only spinor indices. Thus, e.g. for the connection we write

\be
a^i_\m = T^{i AB}\t^{MM'}_\m a_{AB\ MM'},
\ee

which defines what we mean by the connection with all its indices translated into the spinor ones. This choice of the normalisation factor in the above formula is convenient, because as we already discussed before the Kronecker delta $\d^{ij}$ goes under this map into the object $\e^{(A}{}_C \e^{B)}{}_D$, which is the identity map on the space of symmetric rank 2 spinors. The only unusual translation rule is

\be\label{S_spinor_translation}
\S^i_{\m\n} = \sqrt{2}T^{i AB} \t^{MM'}_{\m} \t^{NN'}_\n \S^{AB}_{MM' \ NN'},
\ee

where

\be
\S^{AB}_{MM' \ NN'}= \e^{(A}{}_M \e^{B)}{}_N \e_{M'N'}.
\ee

The reason for putting the factor of $\sqrt{2}$ was explained in the previous section.

The final useful formula for the purpose of conversion is 

\be\label{theta_theta_into_self_antiself_dual_parts}
\t^{MM'}_{[\m} \t^{NN'}_{\n]} = \f{1}{2} \e^{NM}\S^{M'N'}_{\m\n} + \f{1}{2}\e^{M'N'}\S^{MN}_{\m\n},
\ee
where

\be
\S^{MN}_{\m\n} = \t^{(M}_{\m M'} \t^{N)M'}_\n, \quad \S^{M'N'}_{\m\n}  = \t^{M(M'}_\m \t^{N')}_{\n M}.
\ee

Note that the natural contractions appear in these definitions, and, as a result, the anti-self-dual two-forms $\Sigma^{M'N'}_{\mu\nu}$ are {\it minus} the complex conjugates of the self-dual ones $\Sigma^{MN}_{\mu\nu}$. The formula \myref{theta_theta_into_self_antiself_dual_parts} then implies

\be
V_{[\m}U_{\n]} = \f{1}{2}V_{MM'}U_N{}^{N'}\S^{MN}_{\m\n}+\f{1}{2} V^M{}_{M'}U_{NN'}\S^{M'N'}_{\m\n}, 
\ee

where again the natural contractions appear. The first term here is the self-dual part, and the second is the anti-self-dual part of the two form $V_{[\m}U_{\n]}$.

\subsection{Momentum spinors}

Consider a massless particle of a particular 3-momentum vector $\vec{k}$. The 4-vector $k^{\mu}=(|k|,\vec{k})$ is then null. As such, it can be written as a product of two spinors $k^A k^{A'}=\theta^{\mu}_{AA'}k_\mu $. In the case of Lorentzian signature the spinors $k^A, k^{A'}$ must be complex conjugates of each other (so that the resulting null 4-vector is real). It is then clear that $k^A$ is only defined modulo a phase. Moreover, as the vector $\vec{k}$ varies, i.e. as $\vec{n}=\vec{k}/|\vec{k}|$ varies over the sphere $S^2$, there is no continuous choice of the spinor $k^A$. We make the following choice:
\be\label{kA}
k^A \equiv k^A(\vec{k}) := 2^{1/4} \sqrt{\omega_k} \left( \sin(\theta/2) e^{-\im\phi/2} \iota^A + \cos(\theta/2) e^{\im\phi/2} o^A\right),
\ee
where $o^A,\iota^A$ is a basis in the space of unprimed spinors, and $\omega_k=|k|$. Here $\theta,\phi$ are the usual coordinates on $S^2$ so that the momentum vector in the direction of the positive z-axes corresponds to $\theta=\phi=0$. We see that the corresponding spinor is $2^{1/4} \sqrt{\omega_k}o^A$. The formula \myref{kA} can be checked using the expression \myref{theta_with_null_tetrad} for the soldering form.

We can now see effects of the change of the momentum vector direction. Consider, for example, what happens when the momentum direction gets reversed. This corresponds to $\theta\to \pi-\theta$ and $\phi\to \phi+\pi$. We get
\be
k^A(-\vec{k}) = \im \, 2^{1/4} \sqrt{\omega_k} \left( - \cos(\theta/2) e^{-\im\phi/2} \iota^A + \sin(\theta/2) e^{\im\phi/2} o^A\right).
\ee
We now note that 
\be
k^A(-k) = \im (k^\star)^A(k),
\ee
where the action of the $\star$-operation on the basis spinors is given in \myref{star}. Now, using the fact that $\star^2=-1$ it is easy to see that flipping the sign of the momentum twice we get minus the original momentum spinor. In other words, $k^A$ takes values in a non-trivial spinor bundle over $S^2$.

\subsection{Helicity spinors}

The aim of this subsection is to use the rules for the $i\to (AB)$ conversion deduced above, as well as the definition \myref{kA} of the momentum spinors $k^A$ to write down convenient expressions for the polarization tensors \myref{polarization_tensors} in the spinor language. 

Our polarization tensors are built from the vectors $m^i(k),\bar{m}^i(k)$, where the direction of the $z^i$ axes is chosen to be that of the momentum 3-vector $\vec{k}$. Thus, let us start by assuming that $\vec{k}$ points along the positive $z$-direction. Then from \myref{T} we have $T^{i AB} m^i =  \im\, \iota^A \iota^B, T^{i AB} \bar{m}^i =-\im \, o^A o^B$, and therefore
\be\label{polars-1}
m^i m^j \to - \iota^A \iota^B \iota^C \iota^D, \qquad \bar{m}^i \bar{m}^j \to - o^A o^B o^C o^D
\ee
when converted to spinor notations. We can, however, use the available freedom of gauge ${\rm SO}(3)$ rotations and consider polarization tensors (spinors) more general than those above. Indeed, we can always shift our (spatial projection of the) connection by a gauge transformation $a_{ij}\to a_{ij} + (\partial_{(i} \phi_{j)})^{tf}$, where also the tracefree part needs to be taken in order to preserve the tracelessness of the $a_{ij}$. Such a shift being pure gauge, it does not have any effect on the scattering amplitudes. So, we can freely add to both polarization tensors an object of the form $z_{(i} \phi_{j)}$, where again a tracefree part is assumed. Moreover, the vector $\phi_i$ can be different for the positive and negative helicity polarizations. Using the spinor conversion rules written above, it is not hard to see that this means that one will obtain correct scattering amplitudes when using instead of \myref{polars-1} the following expressions
$$
\iota^A \iota^B \iota^C \iota^D \to (\iota+ \alpha o)^{(A} (\iota+\beta o)^B (\iota+\gamma o)^C \iota^{D)}, \quad
o^A o^B o^C o^D \to (o+\alpha'\iota)^{(A} (o+\beta'\iota)^B (o+\gamma'\iota)^C o^{D)},
$$
for arbitrary coefficients $\alpha,\beta,\gamma,\alpha',\beta',\gamma'$. For reasons to become clear below, the most convenient choice is
\be\label{polars-2}
\iota^A \iota^B \iota^C \iota^D \to \frac{q^{(A} q^B q^C \iota^{D)}}{(q^E o_E)^3}, \qquad o^A o^B o^C o^D \to \frac{o^{(A} o^B o^C (p^\star)^{D)}}{(\iota^E (p^\star)_E)},
\ee
where $q^A, p^A$ are arbitrary spinors, and $\star$ is the operation on ${\rm SU}(2)$ spinors introduced above. Note that while in the choice of the first polarization spinor we have replaced as many as 3 copies of $\iota^A$ by an arbitrary reference spinor $q^A$, in the second polarization we only changed a single copy of $o^A$ to $(p^\star)^A$. The reason for this will become clear below. 

Let us now rewrite the spinor expressions for the full polarization tensors \myref{polarization_tensors} using the spinors $k^A= 2^{1/4}\sqrt{\omega_k} o^A$ and $(k^\star)^A = - 2^{1/4}\sqrt{\omega_k} \iota^A$. We get
\be\label{hel-1}
\e^{- ABCD}(k) = M \frac{ q^{(A} q^B q^C (k^\star)^{D)}}{(q^E k_E)^3}, \qquad \e^{+ ABCD}(k) = \frac{1}{M} \frac{k^{(A} k^B k^C (p^\star)^{D)}}{(k^*)^E (p^\star)_E}.
\ee
Note that all the annoying factors of $\sqrt{2}$ in the original formulas \myref{polarization_tensors}, as well as some minus signs present in the intermediate expressions, have now cancelled. Note also that while the previous spinor expressions were only valid in a frame where the 3-momentum was pointing in the $z$-direction, the expressions \myref{hel-1} are valid in an arbitrary frame. 

It remains to observe that one will obtain \myref{hel-1}  as the spin 2 parts of the spatial projections of the following mixed spinors:
\be\label{helicity-spin}
\e^{- ABCA'}(k) = M \frac{ q^A q^B q^C k^{A'}}{\ket{q}{k}^3}, \qquad \e^{+ ABCA'}(k) = \frac{1}{M} \frac{k^A k^B k^C p^{A'}}{\bra{p}{k}},
\ee
where we have introduced the usual notations for the spinor contractions
\be
\ket{\lambda}{\eta} := \lambda^A \eta_A, \qquad \bra{\lambda}{\eta} = \lambda_{A'} \eta^{A'}.
\ee
We note that the helicity spinors are normalised so that
\be
\e^{-\, ABC A'} \epsilon^{+}_{ABC A'}=1.
\ee

The expressions \myref{helicity-spin} are the main outcome of this heavy in conventions section. We note that these expressions could have been guessed as the only ones with the correct dimensions, as well as with the right homogeneity degree zero dependence on the reference spinors $q^A, p^{A'}$, and the right degree of homogeneity under the rescaling of the momentum spinors $k^A\to t k^A, k^{A'}\to t^{-1} k^{A'}$. Indeed, it is clear that under these rescalings (keeping the 4-momentum $k^Ak^{A'}$ unchanged) we get
\be
\e^{- ABCA'}(k)  \to t^{-4} \e^{- ABCA'}(k),\qquad \e^{+ ABCA'}(k)\to t^4 \e^{+ ABCA'}(k).
\ee
However, under any such a guess possibly important numerical factors could have been missed, and it is gratifying to see that after establishing all the conversion formulas, the helicity spinors turned out to be just the simplest expressions possible, without any complicating numerical prefactors. We note that the final spacetime covariant expressions \myref{helicity-spin} explain our choice \myref{polars-2} at the level of the spatially projected expressions. 

The only complication that remains to be discussed is the fact that the positive helicity gravitons have to be taken to be slightly massive, as we have seen in the section on the Minkowski space limit. Because of this, the meaning of the spinor $k^A$ that is used in the positive helicity spinor in \myref{helicity-spin} is not yet defined. To settle this, we shall represent the massive 4-vector $k^2=2M^2$ of the positive helicity gravitons as follows
\be\label{mass-shell}
k^{AA'}= k^A k^{A'} + M^2 \frac{p^A p^{A'}}{\ket{p}{k}\bra{p}{k}}.
\ee
This gives precisely the required $k^2=-k^{AA'} k_{AA'}=2M^2$. Here $p^A p^{A'}$ are a reference spinor and its complex conjugate. At this point it can be arbitrary, but it is convenient to take it to be the same as the one that appears in the positive helicity spinor in \myref{helicity-spin}. It is now the spinor $k^A$ that appears in the decomposition \myref{mass-shell} is what one has to use in the positive helicity spinor in \myref{helicity-spin}. We emphasise that only the positive helicity momentum 4-vectors should be taken to be massive, while the negative helicity does not need these complications, and the corresponding momenta 4-vectors come directly as a product of two spinors. 

\subsection{A relation to the metric helicity states}

It is instructive to see how the metric description helicity spinors can be obtained from the expressions \myref{helicity-spin}. For these we need to recall the passage to the metric perturbation variable that was explained in great detail in \myref{metric_subsection}. In that section we have seen that the metric perturbation is obtained by applying to the connection the operator $\bar{D}$. One should also rescale by $1/M$ to keep the mass dimension correct. At the spacetime covariant level the operator $\bar{D}$ corresponds to the operation of taking the anti-self-dual part of the two-form $da^i$. In the spinor notations, this boils down to the following expression for the metric perturbation
\be\label{h-a}
h_{AB\, A'B'} \sim \frac{1}{M} \partial^E_{A'} a_{B' E AB},
\ee
where $\partial_{AA'}= - \theta^\mu_{AA'} \partial_\mu$ is the Dirac operator, and the $\sim$ sign means that we are only interested in this relation modulo numerical factors. Applying this to the helicity spinors \myref{helicity-spin}, and ignoring the arising numerical factors, one immediately sees that the usual metric spinor helicity states get reproduced:
\be
h^-_{AA'BB'}(k) \sim \frac{q_A k_{A'} q_B k_{B'}}{\ket{q}{k}^2}, \qquad h^+_{AA'BB'}(k) \sim \frac{k_A p_{A'} k_B p_{B'}}{\bra{p}{k}^2}.
\ee
Note that for the negative helicity the metric formulation helicity spinor arises by a single $q$ spinor in the numerator of \myref{helicity-spin} contracting with the momentum $k$ spinor, removing one of the factors of $\ket{q}{k}$ from the denominator. There is also the cancellation of the factor of $M$ in the connection helicity spinor with $1/M$ in the passage to the metric perturbation. For the positive helicity the mechanism of obtaining the usual metric formulation helicity spinor is more subtle. Indeed, if the positive helicity graviton 4-momentum was null, then we would be contracting two momentum $k$ spinors, which would give a zero result. Instead, it is the second, mass term in \myref{mass-shell} that gives a non-zero contribution. The factor of $M^2$ in this second term then is nicely cancelled by the $1/M$ in the helicity spinor and the additional factor of $1/M$ in \myref{h-a}. We therefore see that it is essential that the positive helicity graviton is kept massive till the Minkowski limit can be taken. 

Once again, the fact that the usual metric helicity states get reproduced could be taken as the sufficient reason to work with \myref{helicity-spin}. However, we find the given above derivation of \myref{helicity-spin} that does not involve any reference to the metric more self-contained.

%%%%%%%%%%%%%%%%%%%%%%%%%%%%%%%%%%%%%%%%%%%%%%%%%%%%%%%%%%%
\section{Feynman rules in the spinor form}\label{Feynmans_rules_section}
%%%%%%%%%%%%%%%%%%%%%%%%%%%%%%%%%%%%%%%%%%%%%%%%%%%%%%%%%%%

Now that we understand how expressions can be converted into the spinor language, we can write down the derived above Feynman rules in the spinor form. Here we will also pass to the form relevant in the Minkowski limit, i.e. replace all the derivatives by the partial ones. As we shall see, there are many advantages in working with spinors, as some operations that are not easy to deal with in the ${\rm SO}(3)$ notation become elementary once one expresses them using spinors. The prime example is the projection on the $S_+^3\otimes S_-$ representation of the Lorentz group that appeared in our derivation of the propagator. In the spinor language this simply corresponds to the symmetrisation on the unprimed spinor indices. We shall also see that the interaction vertices take a particularly simple form once the spinor notations are applied.

%%%%%%%%%%%%%%%%%%%%%%%%%%%%%%%%%%%%%%%%%%%%%%%%%%%%%%%%%%%
\subsection{Propagator}
%%%%%%%%%%%%%%%%%%%%%%%%%%%%%%%%%%%%%%%%%%%%%%%%%%%%%%%%%%%

We have previously found the propagator to be given by $1/\im k^2$ times the projector $P^{(3,1)}_{\m i|\n j}$, given in \myref{Lorentz_representations_projectors}, on the $S^3_+\otimes S_-$ irreducible components of objects of type $a^i_\m$ with one spacetime and one internal index. To find what this projector becomes once converted into the spinor form one can multiply the spacetime indices with the soldering forms, and the internal indices with $T^{i AB}$. However, one does not need to do this computation as it is clear that this projector is simply the product of the identity operator acting on the primed index, times the operator of symmetrisation of the 3 unprimed spinor indices. Thus, we can write

\be
\langle a_{EFGE'}(-k)a^{ABCA'}(k) \rangle = \f{1}{\im k^2} \e_E{}^{(A} \e_F{}^B \e_G{}^{C)} \e_{E'}{}^{A'}.
\ee

%%%%%%%%%%%%%%%%%%%%%%%%%%%%%%%%%%%%%%%%%%%%%%%%%%%%%%%%%%%
\subsection{Pieces of the interaction vertices}
%%%%%%%%%%%%%%%%%%%%%%%%%%%%%%%%%%%%%%%%%%%%%%%%%%%%%%%%%%%

Here we develop a dictionary translating the various blocks that appear in the interaction vertices into the spinor form. As we recall from \myref{Da_decomposition}, one of the main building blocks of the vertices is the two-form $D_{[\m}a^i_{\n]}$, and various quantities constructed from it. We recall that from now on we replace the covariant derivative by the partial one.

The first block to be translated is the $(Da)^{ij}$ symmetric tracefree matrix, whose definition is given in \myref{Da_traceless_definition}. Applying the rules given above we get:

\be
(Da)^{ij}\rightarrow \sqrt{2}\p^{(A}{}_{M'} a^{BCD)M'},%checked
\ee
where the result is easy to understand, and the factor of $\sqrt{2}$ comes from the same factor in the translation \myref{S_spinor_translation} of $\S^i_{\m\n}$. Further, we have

\be
D^\m a^i_\m \rightarrow \p^M_{M'} a^{AB}{}_M{}^{M'}.
\ee

Finally, we have

\be\label{DaTilde_spinor_conversion}
 (\widetilde{Da^i})^{\mu\nu} \rightarrow \f{1}{2} \S^{M'N'}_{\m\n}\partial^M{}_{M'} a^{AB}{}_{MN'}.
\ee

We also need the spinor representation of the two-form $\e^{ijk}a^j_\m a^k_\n$. Its self-dual part encoded in the matrix \myref{Seaa} is given by

\be
(\S\e a a)^{ij}\rightarrow  a^{CE(A}{}_{M'}a^{B)}{}_E{}^{DM'} + a^{DE(A}{}_{M'}a^{B)}{}_E{}^{CM'}.
\ee

It is $(AB)\rightarrow (CD)$ symmetric, but it has trace given by the full contraction of the two connections. The anti-self-dual part of $\e^{ijk}a^j_\m a^k_\n$ is given by

\be
 (\e^{ijk} a^j_\m a^k_\n )_{asd} \rightarrow \f{1}{ \sqrt{2} } \S^{M'N'}_{\m\n} a^{EF(A}{}_{M'} a^{B)}{}_{EFN'}.
\ee

We now introduce some notations to simplify the above spinorial expressions. The idea behind this notation is that we omit pairs of naturally contracted indices. Thus, we define

\be
\nn (\partial a)^{ABCD}:=\partial^{(A}{}_{M'}a^{BCD)M'},\quad (\partial a)^{M'N'AB}:=\partial^{C(M'}a_C{}^{ABN')},\quad (\partial a)^{AB}:=\partial^{M}_{M'} a^{AB}{}_{M}{}^{M'},
\\ \nonumber
(aa)^{ABCD}:=a^{ABM}{}_{M'}a^{CD}{}_M{}^{M'},\quad (aa)^{M'N'CD}:=a^{CD(AM'}a_{CD}{}^{B)N'}.
\\
\ee
The blocks appearing in the interaction vertices can then be written very compactly in terms of these quantities.

%%%%%%%%%%%%%%%%%%%%%%%%%%%%%%%%%%%%%%%%%%%%%%%%%%%%%%%%%%%
\subsection{Cubic interaction}
%%%%%%%%%%%%%%%%%%%%%%%%%%%%%%%%%%%%%%%%%%%%%%%%%%%%%%%%%%%

As we discussed in the section on the Minkowski limit, we keep only the leading order terms of all the quantities in the limit $M\to 0$. Converting all the terms in \myref{3_vertex} into spinor form we get
\be \label{L3-M}
\nn 3\im M^2 (g^{(2)})^{3/2} {\cal L}^{(3)} =  2^{3/2} \left(g^{(3)} - \frac{3g^{(2)}}{4}\right) (\partial a)_{AB}{}^{CD} (\partial a)_{CD}{}^{EF} (\partial a)_{EF}{}^{AB} \\ \nonumber
+ 2^{-1/2} 3 g^{(2)} (\partial a)^{ABCD} (\partial a)_{M'N'AB} (\partial a)^{M'N'}{}_{CD}  + 2^{1/2} \frac{3g^{(2)}}{16}  (\partial a)^{ABCD} (\partial a)_{AB} (\partial a)_{CD} \\ \nonumber
- 2^{1/2} M^2 (3g^{(2)}-f(\delta)) (\partial a)^{ABCD} (aa)_{ABCD} + 2^{1/2} M^2 f(\delta) (\partial a)^{M'N' AB} (aa)_{M'N' AB},
\\
\ee

where an additional factor of $1/2$ in the first term in the second line came from the factors of $1/2$ in \myref{DaTilde_spinor_conversion}, with one of them being cancelled by the factor of $2$ that appears in the contraction of two $\Sigma$'s. 

The last term in the second line in (\ref{L3-M}) is only (possibly) relevant for loop computations, for in any tree diagram at least one of the two factors of $(\partial a)^{AB}$ gets hit by an external state, which gives zero. Thus, we shall ignore this term in this thesis. Let us now consider the second term in (\ref{L3-M}) in more details, which can be seen to be the only relevant term in the $M\to 0$ limit of the GR cubic interaction.

%%%%%%%%%%%%%%%%%%%%%%%%%%%%%%%%%%%%%%%%%%%%%%%%%%%%%%%%%%%
\subsection{The parity-preserving cubic vertex}\label{parity_preserving_cubic_vertex_section}
%%%%%%%%%%%%%%%%%%%%%%%%%%%%%%%%%%%%%%%%%%%%%%%%%%%%%%%%%%%

In the case of GR the coefficients in front of the term in the first line, and the first term in the third line in (\ref{L3-M}) become zero, and we are left with the simple
\be\label{cubic-GR}
\im \mathcal{L}^{(3)}_{\rm GR}=\frac{1}{2} \frac{\kappa}{M}(\partial a)^{ABCD} (\partial a)_{M'N'AB}(\partial a)^{M'N'}{}_{CD} + \kappa M (\partial a)^{M'N' AB} (aa)_{M'N' AB},
\ee
where we have used (\ref{GR_constants}). Here we introduced the usual notation
\be
\kappa^2:=32\pi G, \qquad \kappa= \frac{\sqrt{2}}{M_p}.
\ee

Below we shall see that it is the first term in (\ref{cubic-GR}) that gives most of the interesting physics, and, in particular, is the one relevant for computations of the GR amplitudes. The second term seems to be suppressed by a factor of $M^2/E^2$, where $E$ is energy,  as compared to the first. However, due to the subtleties of taking the Minkowski limit we cannot just throw it away, and it turns out to give a contribution to some scattering amplitudes. For now, let us analyse the much more interesting first term and come back to the second piece of the GR cubic vertex later. 

First, we would like to make a pause and emphasise the similarity of the first term in (\ref{cubic-GR}) to the vertex of Yang-Mills theory, when rewritten in the spinor notations. Thus, one starts with the Yang-Mills Lagrangian in the form ${\cal L}_{\rm YM}\sim (F_{sd})^2$, where $F_{sd}$ is the self-dual part of the curvature. Applying the described above spinor conversion rules, at the linearised level this gives 
\be
{\cal L}^{(2)}_{\rm YM} \sim (\partial^{(A}_{A'} A^{B) A'})^2,
\ee
where $A_{AA'}$ is the connection, and we omitted the Lie-algebra indices. Our linearised Lagrangian (\ref{PDaDa_action}), when converted into the spinor form, is precisely analogous, except that the connection field in the case of gravity has two more unprimed spinor indices. Let us now look at the Yang-Mills cubic interaction. Again translating into the spinor form we get
\be\label{cubic-YM}
{\cal L}^{(3)}_{\rm YM} \sim (\partial^{(A}_{A'} A^{B) A'}) A_{AB'} A_B{}^{B'}.
\ee
Again, the analogy to the first term in (\ref{cubic-GR}) is striking. Basically, the gravitational interaction described by the first term in (\ref{cubic-GR}) is the only possible one that can be constructed following the Yang-Mills pattern, but now with more unprimed indices on the connection. Indeed, generalising the first block $ (\partial^{(A}_{A'} A^{B) A'})$ to the case with more indices one gets $(\partial^{(A}_{A'} a^{BCD) A'})\equiv(\partial a)^{ABCD}$. We would now like to have a symmetric block involving two connections with free indices being 4 unprimed spinor indices $ABCD$. Thus, two indices must come from one connection, and the other two from the other. For reason to become clear below, we also want to have some quantities constructed out of the connections that contract only in the primed indices. This means that we have to convert one of the unprimed indices in each connection to a primed using the Dirac operator. This results precisely in (\ref{cubic-GR}). The other possible choice of having connections contracting directly, with no Dirac operators involved, i.e. $a_{AB}{}^E{}_{A'} a_{CDE}{}^{A'}$, can be easily seen not to give the desired on-shell amplitudes, see below. We thus learn that (\ref{cubic-GR}) is the only possible generalisation of the Yang-Mills cubic vertex that gives the correct on-shell amplitudes. The second term in (\ref{cubic-GR}) is superficially more analogous to (\ref{cubic-YM}) than the first, as it involves just one derivative. However, we will see that it does not give the standard answers for the on-shell amplitudes, and so is not the ``right'' generalisation.  

Let us now check that the vertex given by (\ref{cubic-GR}), when evaluated on the graviton helicity spinors gives just the required $--+$ and $++-$ amplitudes, i.e. the squares of these amplitudes for spin one particles (it is easy to check that the $---$ and $+++$ amplitudes resulting from it are zero). 

Let us first compute the $--+$ amplitude. The required helicity states are given in (\ref{helicity-spin}). We first note that the second term in (\ref{cubic-GR}) cannot contribute to this amplitude. Indeed, no matter which pair of legs of this term the two negative helicity states are inserted, there is always a contraction of the reference spinors $q^A$ and the result is zero. Thus, we only have to consider the first term in (\ref{cubic-GR}) for this helicity combination. Let us now recall that the combination $(\partial a)^{ABCD}$ only gives a non-zero result when applied to a positive helicity state. Thus, we must insert the positive helicity wave-function in this leg. The other two legs are symmetric, and we insert into them the remaining two negative helicity states. Let us denote the negative helicity momenta by $k_1, k_2$, and the positive momentum by $k_3$. We will assume that all the momenta are incoming. After applying all the derivatives present in the vertex to their corresponding states we obtain the following contraction
\be\label{3--+}
\im \frac{\kappa}{M} \frac{1}{M} 3^A 3^B 3^C 3^D M \frac{q_A q_B 1_{M'} 1_{N'}}{\ket{1}{q}^2} M \frac{q_C q_D 2^{M'} 2^{N'}}{\ket{2}{q}^2} = \im \kappa \bra{1}{2}^2 \frac{\ket{3}{q}^4}{\ket{1}{q}^2\ket{2}{q}^2},
\ee
where we have used the usual notation $k_1^A \equiv 1^A$, etc. We can now use the momentum conservation equation, which we contract with the reference spinor $q_A$ to get $\ket{1}{q} 1^{A'} + \ket{2}{q} 2^{A'} + \ket{3}{q} 3^{A'} = 0$. This immediately gives $\ket{3}{q}^2/\ket{1}{q}^2 = \bra{1}{2}^2/\bra{3}{2}^2, \ket{3}{q}^2/\ket{2}{q}^2 = \bra{1}{2}^2/\bra{3}{1}^2$, which allows us to rewrite (\ref{3--+}) as
\be\label{--+}
{\cal M}^{--+} = \im \kappa \frac{\bra{1}{2}^6}{\bra{1}{3}^2 \bra{2}{3}^2}.
\ee 
This is just the expected square of the spin one result. 

The opposite $++-$ helicity configuration is computed similarly, but now we cannot ignore the second term in (\ref{cubic-GR}). So, the computation is somewhat more involved. Let us consider the first term in (\ref{cubic-GR}) first. We note that if two positive helicity gravitons hit the two symmetric legs of this vertex, the reference spinors $p^{A'}$ will contract, and this gives zero. Thus, one of the positive helicity gravitons must necessarily be inserted into the leg $(\partial a)^{ABCD}$. This of course also comes from the fact that $(\partial a)^{ABCD}$ gives zero when applied to a negative helicity state. Thus, the negative helicity graviton must hit one of the two symmetric legs. If we choose the state inserted into $(\partial a)^{ABCD}$ to be the graviton number two, we get the following contraction
\be
- \im\frac{\kappa}{M} \frac{1}{M} 2^A 2^B 2^C 2^D M \frac{1_A 1_B p_{M'} p_{N'}}{\bra{1}{p}^2} M \frac{q_C q_D 3^{M'} 3^{N'}}{\ket{3}{q}^2} = -\im \kappa \ket{1}{2}^2 \frac{ \bra{3}{p}^2 \ket{2}{q}^2}{\ket{3}{q}^2 \bra{1}{p}^2}.
\ee
To this we must add the contribution from $1$ and $2$ gravitons exchanged, which gives the overall contribution from the first term in (\ref{cubic-GR}) as
\be\label{3-contr-1}
-\im \kappa  \ket{1}{2}^2\frac{\bra{3}{p}^2}{\ket{3}{q}^2}\left(\frac{\ket{2}{q}^2}{\bra{1}{p}^2} + \frac{\ket{1}{q}^2}{\bra{2}{p}^2} \right).
\ee
To this, we must add the contribution from the second term in (\ref{cubic-GR}). In this case, the positive helicity gravitons can only be inserted into the two symmetric legs of this vertex. This gives the following contraction
\be\label{3-contr-2}
\nn 2\im\kappa M 3^{M'} 3^E \, M \frac{q_E q^A q^B 3^{N'}}{\ket{q}{3}^3} \frac{1}{M} \frac{1^E 1^F 1_A p_{M'}}{\bra{p}{1}} \frac{1}{M} \frac{2_E 2_F 2_B p_{N'}}{\bra{p}{1}} = -2\im \kappa \ket{1}{2}^2 \frac{\bra{3}{p}^2}{\ket{3}{q}^2} \frac{\ket{1}{q}\ket{2}{q}}{\ket{3}{q}^2}.
\\
\ee
Adding (\ref{3-contr-1}) and (\ref{3-contr-2}) we get
\be
{\cal M}^{++-}=-\im \kappa  \ket{1}{2}^2\frac{\bra{3}{p}^2}{\ket{3}{q}^2}\left(\frac{\ket{2}{q}}{\bra{1}{p}} + \frac{\ket{1}{q}}{\bra{2}{p}} \right)^2.
\ee
We would like to rewrite the answer in terms of the angle brackets only. To this aim we take the $\bra{3}{p}$ into the brackets, then use the momentum conservation and then the Schouten identity

$$\ket{i}{j}\ket{k}{l}+\ket{i}{k}\ket{l}{j}+\ket{i}{l}\ket{j}{k}=0.$$
 This gives the desired
\be\label{++-usual}
{\cal M}^{++-} = -\im \kappa \frac{\ket{1}{2}^6}{\ket{1}{3}^2 \ket{2}{3}^2}.
\ee 
We thus see that the GR cubic vertex is parity-invariant, as it should be of course, in the sense that the amplitude for an opposite configuration of helicities is given by the complex conjugate of the original amplitude. Note, however, that to obtain this result both terms in (\ref{cubic-GR}) were used in an essential way. This can be viewed as a justification for the presence of the second term in (\ref{cubic-GR}), while it is the first term only that is important for many calculations, such as e.g. that of the amplitude with two positive incoming gravitons and two positive outgoing.

Let us now discuss the corresponding Feynman rules. When we write the vertex factor corresponding to the first term in (\ref{cubic-GR}) we obtain 3 terms, since the first instance of $(\partial a)$ corresponding to the self-dual part of the two-form $da$ can be applied to any of the 3 legs of the vertex. We will not write the vertex factor as we will never need it in this thesis, and because its explicit form containing many $\epsilon$ symbols and symmetrisations is not very illuminating. Instead, we shall draw a picture of the contractions involved in the vertex for just one of the terms, when momenta $k_{1,2}$ that we assume are incoming are on the lines corresponding to $(\partial a)_{asd}$, and the momentum $k_{12}=k_1+k_2$ is on the line corresponding to $(\partial a)_{sd}$. In this picture the dashed lines correspond to primed, and black to unprimed spinor indices. The symbols of momenta in circles stand for the factors of $k^{AA'}$. This part of the full vertex is given by
\be\label{V-GR}
\frac{\im \kappa}{M} \quad \lower0.6in\hbox{\includegraphics[width=1.2in]{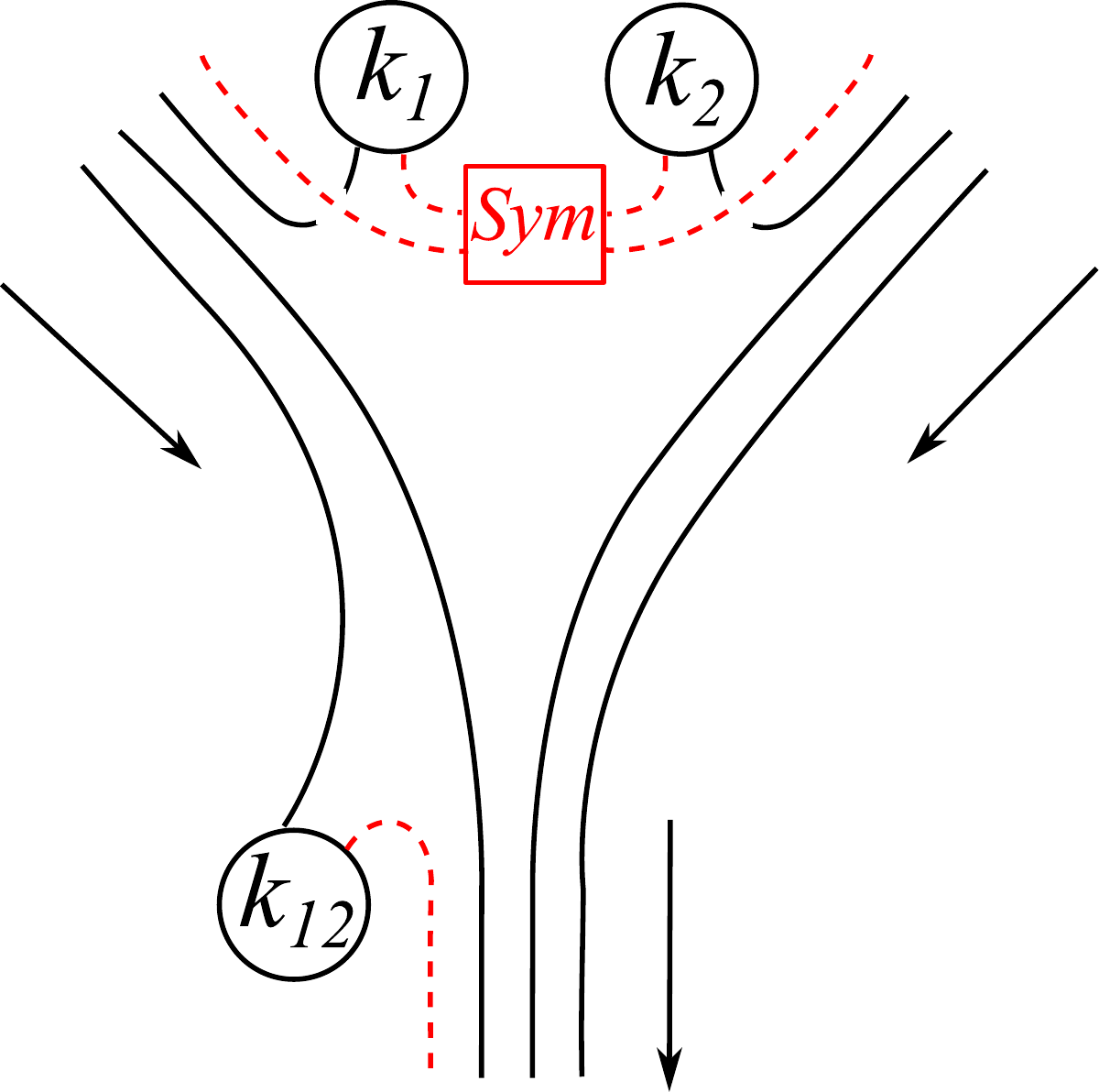}}
\ee
where one also has to symmetrise over the two external legs that can get contracted to $k_{12}$. As we shall see later, for the purpose of computing the most interesting (graviton-graviton) amplitudes this will be the only surviving contribution to the full vertex. 

Let us now briefly discuss the case of a general theory. In this case the vertex of interest in the tree level computations has several pieces. The most ``interesting'' part is still essentially the most interesting part of the GR vertex, i.e. the first term in (\ref{cubic-GR}), but with the different prefactor 
\be\label{V-1}
\frac{\im \sqrt{2}}{M^2 \sqrt{g^{(2)}}}  \quad  \lower0.6in\hbox{\includegraphics[width=1.2in]{3vertexASD.pdf}}
\ee
We can therefore expect that the Newton constant measuring the strength of interactions of gravitons in the general theory is
\be\label{Mp}
\frac{1}{16\pi G} = M^2 g^{(2)},
\ee
which is essentially the coupling constant $g^{(2)}$, expressed in the units of $M$, the only dimensionful parameter present in the theory. This expectation will be confirmed below when we compute the parity-preserving graviton scattering amplitude. 

Let us also give a pictorial representation for the second term in (\ref{cubic-GR}). We have
\be\label{V-strange}
- 2 \im \kappa M \quad \lower0.6in\hbox{\includegraphics[width=1.3in]{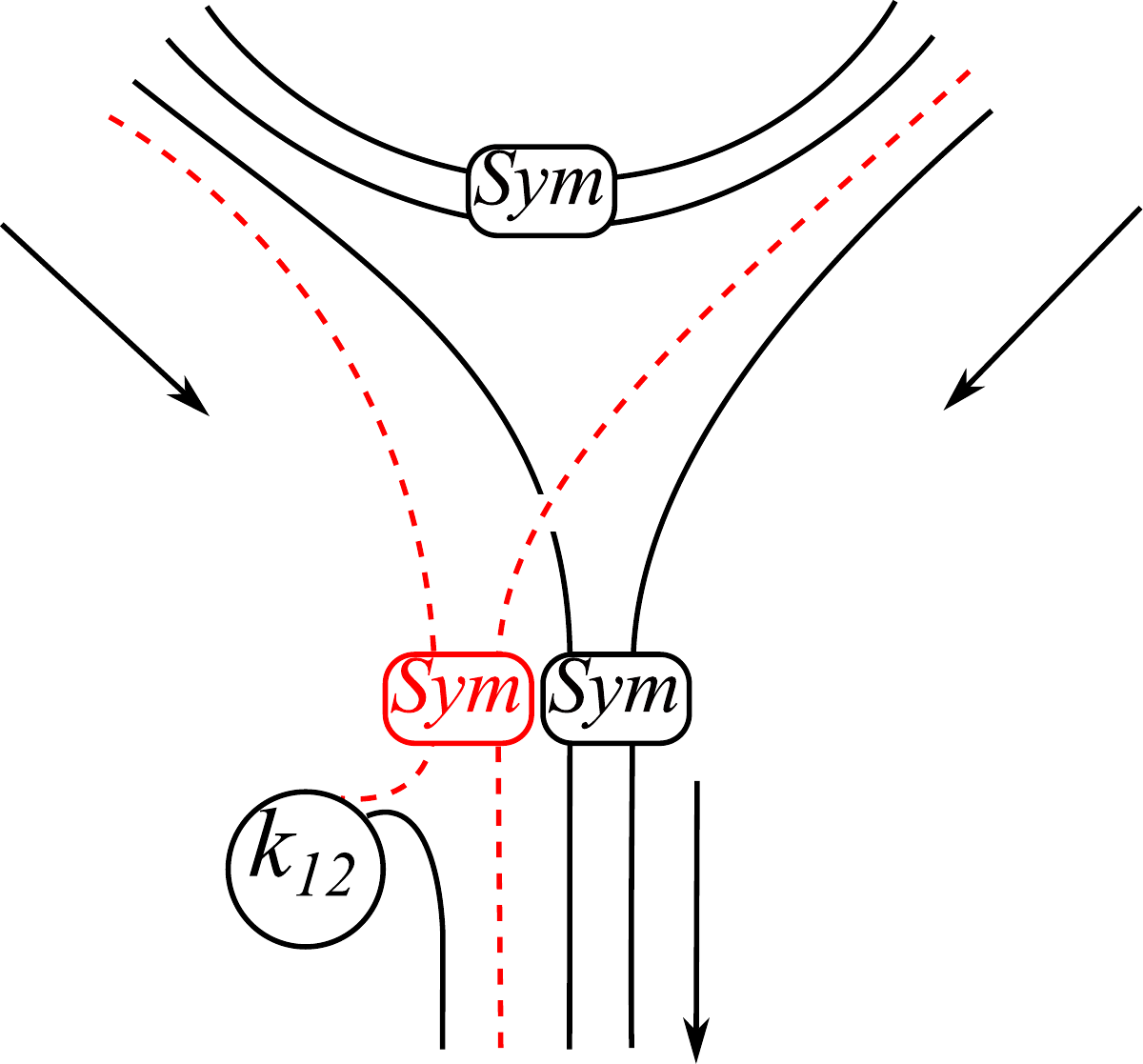}} 
\ee
where the factor of two in front comes from the two ways that the symmetric legs can be applied. The prefactor here is as is relevant for the case of GR. Let us also give a general expression for this vertex. From (\ref{L3-M}) we get the following graphical representation 
\be\label{V-3}
- \frac{\im}{3} \kappa^3 M^3 f(\delta) \quad \lower0.6in\hbox{\includegraphics[width=1.3in]{3vertex1derASD.pdf}} 
\ee
Here we have already summed over the two ways that the symmetric legs can be applied.

\subsection{The parity-violating cubic vertex}

For a general theory there are two more vertices. One comes from the first term in (\ref{L3-M}). Assuming again the convention that the two momenta are incoming and one is outgoing, and taking into account the symmetry factor of $3!$, the vertex can be graphically represented as
\be\label{V-2}
\frac{\im \sqrt{2} (4g^{(3)}-3g^{(2)})}{M^2 (g^{(2)})^{3/2}} \quad  \lower0.6in\hbox{\includegraphics[width=1.2in]{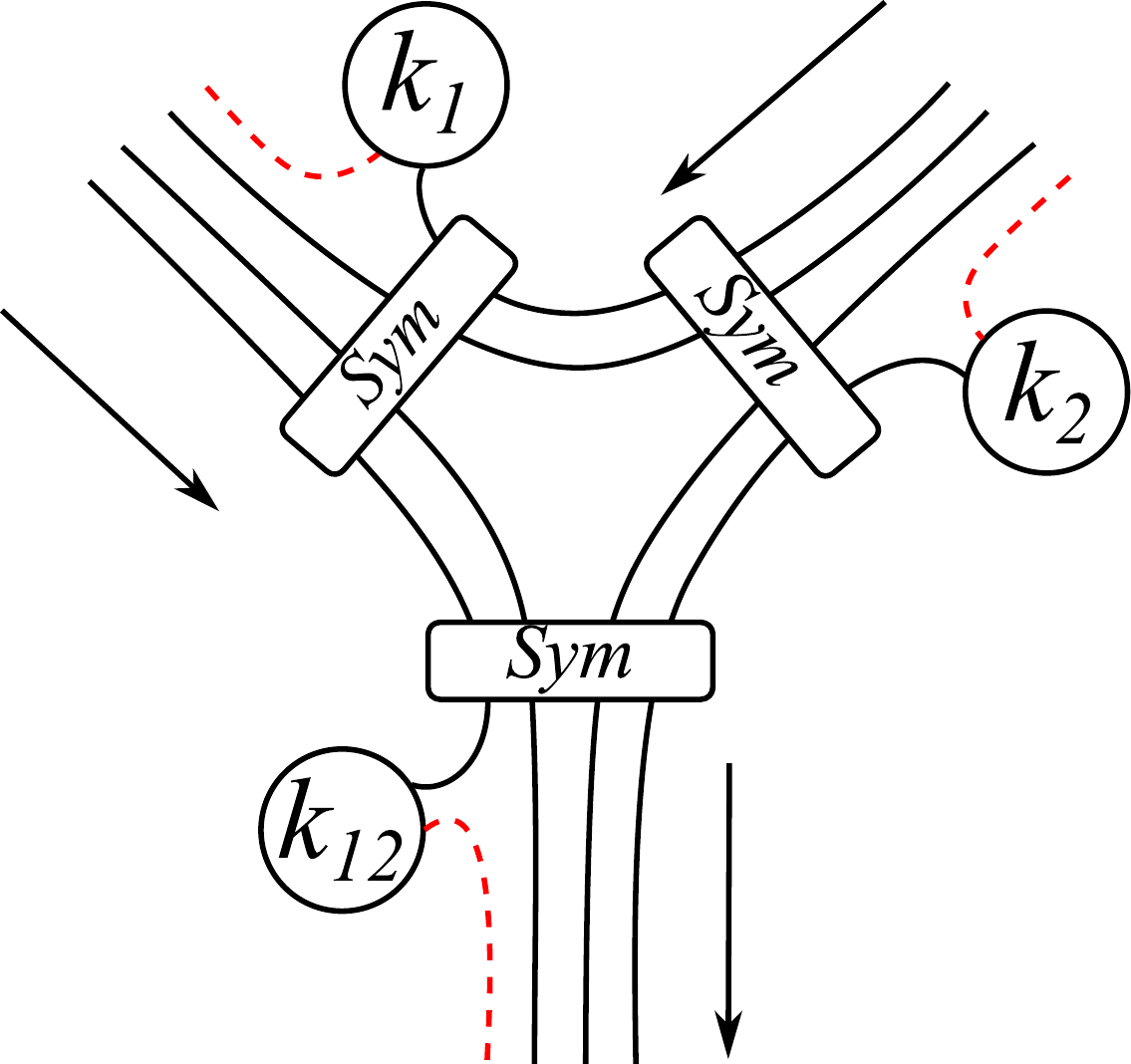}}
\ee

To see what kind of interaction this generates, let us evaluate this vertex on the graviton polarization spinors. It is immediately clear that it only produces a non-zero result when all the helicities are positive, because the negative helicity inserted into the combination $(\partial a)^{ABCD}$ gives a zero result. After applying all the derivatives to the external states, we get for the amplitude coming from this vertex
\be
{\cal M}^{+++} = \f{1}{\im}\frac{\sqrt{2} (4g^{(3)}-3g^{(2)})}{M^5 (g^{(2)})^{3/2}}  \ket{1}{2}^2 \ket{2}{3}^2 \ket{3}{1}^2.
\ee
Using the already known to us fact (\ref{Mp}) that the Planck mass $M_p^2=1/16\pi G$ equals $M^2 g^{(2)}$ we can express the quantity $g^{(2)}$ here in terms of the Planck mass. Also, to understand the behaviour of this amplitude in the $M\to 0$ limit we specialise to a family of theories considered in the Appendix, which are guaranteed to have only Planckian modifications of GR. For this family the difference $4g^{(3)}-3g^{(2)}= -(27/4)\beta^2 M^2/M_p^2$, and the amplitude becomes, modulo phase factors
\be\label{3-par-viol}
{\cal M}^{+++} =\im \frac{27\kappa^5\beta^2}{16} \ket{1}{2}^2 \ket{2}{3}^2 \ket{3}{1}^2.
\ee
Here $\beta$ is a parameter controlling deviations from GR, see (\ref{mod-action}). 

We note that the amplitude of the type (\ref{3-par-viol}) can arise in a gravity theory with the $(Riemann)^3$ term in the Lagrangian, see e.g. \cite{ArkaniHamed:2008gz}, discussion following formula (57). However, a theory with this term in the Lagrangian would be parity-preserving, and thus would also have a non-zero ${\cal M}^{---}$ amplitude. We stress that in our theories this is not the case, with only one chiral half of this amplitude being present, and the amplitude ${\cal M}^{---}$ continuing to vanish as in GR. We thus learn that the higher order in curvature correction that is present in a general member of our family of theories is not $(Riemann)^3$ but rather $(Weyl_{sd})^3$, in other words the cube of the self-dual part of Weyl curvature. No anti-self-dual part cubed is present, and this is why the 3 negative helicity amplitude continues to be zero. Below we shall discuss implications of this fact in more details. 

\subsection{Another first derivative vertex}

We finally consider the last remaining piece of a general cubic vertex. It comes from the first term in the last line of (\ref{L3-M}). Graphically, it can be represented as
\be\label{V-4}
\im \sqrt{2} \, \frac{3g^{(2)}-f(\delta)}{3(g^{(2)})^{3/2}} \quad  \lower0.6in\hbox{\includegraphics[width=1.2in]{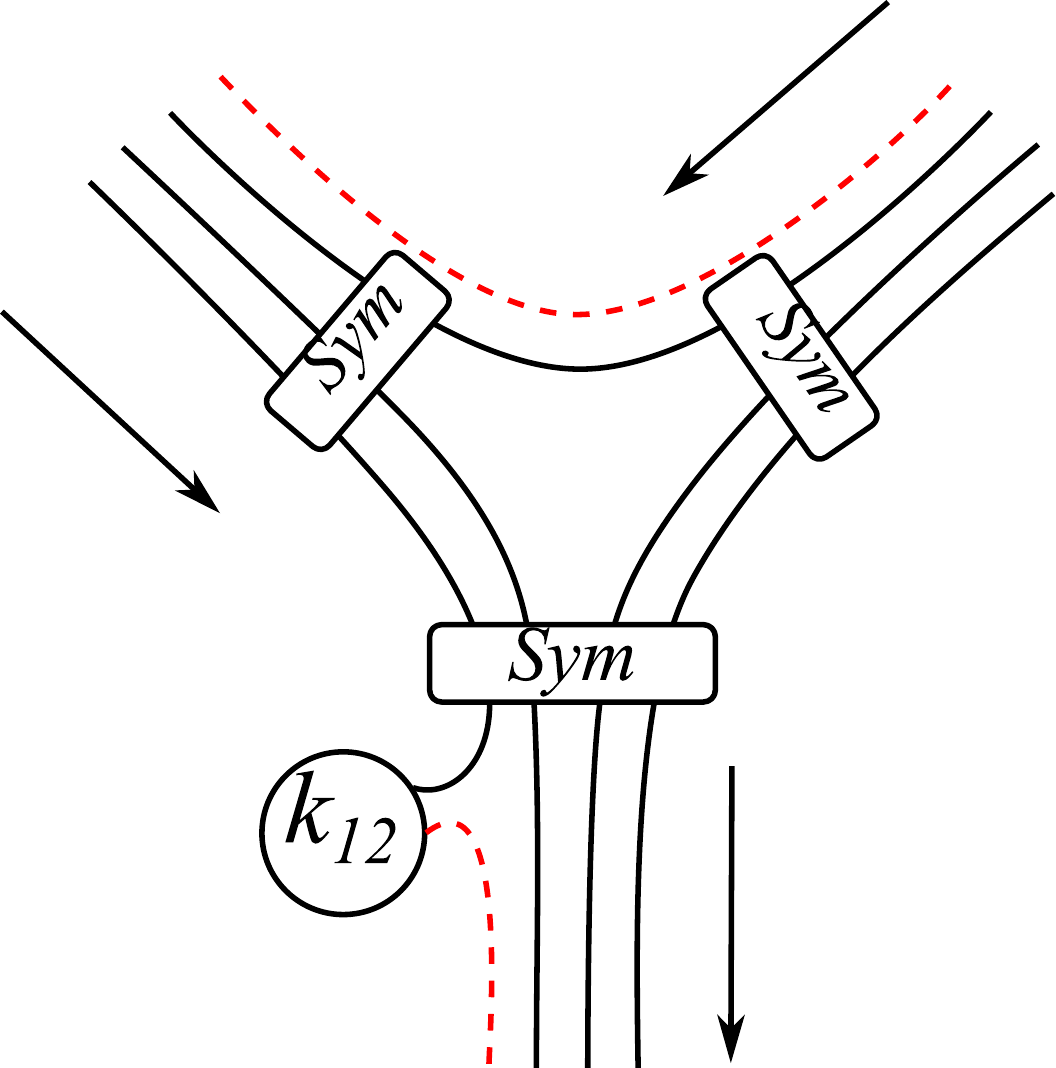}}
\ee

As before, let us understand what kind of interaction this vertex represents by evaluating it on-shell. It is easy to see that there can be at most a single negative helicity graviton. Indeed, any pair of legs in this vertex is connected by a black line. Thus, if there are two negative helicity states, there will be a contraction of the reference $q^A$ spinors, which results in a null amplitude. Thus, the only possible amplitude is $++-$. We also see that one of the plus helicities must be inserted into the bottom leg. This is because the derivative present in the bottom leg would give zero when applied to a negative helicity, and also because if two positive helicity states are inserted from the top, there would be a contraction of the reference spinors $p^{A'}$ along the dashed lines. For this configuration of helicities, however, there is no need to compute the result, as it can be seen to go to zero as $M\to 0$. Indeed, we know that $g^{(2)} = M_p^2/M^2$. From the consideration of the Appendix we know, see (\ref{g2-f}), that the difference $3g^{(2)}-f(\delta)=(27/4)\alpha\beta^2 $ goes to zero as $M^2/M_p^2$. This means that the prefactor in this vertex goes as $M^5/M_p^5$, which is too fast to give any surviving contribution when contracted with the $1/M$ factor coming from the helicity states. So, there is no surviving in the $M\to 0$ limit amplitude that this vertex produces. It turns out that this vertex does not contribute to the graviton-graviton amplitudes either, so we can safely ignore it for the rest of this chapter.

\subsection{Quartic interaction}

As for the cubic vertex, we now take the full expression (\ref{4_vertex}), replace all the derivatives by the partial ones, and convert everything into spinors. Our life is considerably simplified by an observation that in order to contribute to the graviton-graviton amplitudes the 4-vertex needs to have at least 4 derivatives. This will be demonstrated below. Taking this observation into account, we can ignore the lower derivative parts of the vertex (\ref{4_vertex}). The 4-derivative part that we keep, converted into the spinor form, reads
\be\label{L4-M}
\nonumber
-4! M^4 (g^{(2)})^2 {\cal L}^{(4)} =  \left(- 8 g^{(4)} +4 g^{(3)} + (7/2) g^{(2)}\right) (\partial a)_{AB}{}^{CD} (\partial a)_{CD}{}^{AB}  (\partial a)_{EF}{}^{MN} (\partial a)_{MN}{}^{EF} \\ 
\nn -24(4g^{(3)} - g^{(2)}) (\partial a)^{ABEF} (\partial a)_{EF}{}^{CD} (\partial a)_{M'N' AB} (\partial a)^{M'N'}{}_{CD} 
\\ \nonumber
+8(4g^{(3)} - 3g^{(2)}) (\partial a)^{ABCD} (\partial a)_{ABCD} (\partial a)_{M'N'}{}^{EF} (\partial a)^{M'N'}{}_{EF} 
\\ \nonumber
-12 g^{(2)} (\partial a)_{M'N'}{}^{(AB} (\partial a)^{CD)M'N'} (\partial a)_{E'F'AB} (\partial a)^{E'F'}{}_{CD}  \, .
\\
 \ee
We emphasise that this is an on-shell vertex, with all legs satisfying $\partial^\mu a_\mu^i=0$ condition, and that only the 4-derivative part was kept. Thus, some terms potentially relevant for loop computations have not been written down. Only two of these terms survive in the GR case, and we get
\be\label{L4-GR-M}
 {\cal L}^{(4)}_{\rm GR} =\frac{\kappa^2}{M^2} (\partial a)^{ABEF} (\partial a)_{EF}{}^{CD} (\partial a)_{M'N' AB} (\partial a)^{M'N'}{}_{CD} 
\\ \nonumber
+\frac{\kappa^2}{4M^2} (\partial a)_{M'N'}{}^{(AB} (\partial a)^{CD)M'N'} (\partial a)_{E'F'AB} (\partial a)^{E'F'}{}_{CD}  \, .
 \ee

We will only write down the vertex factors corresponding to the first term in (\ref{L4-M}), the reason being that the other terms cannot contribute to the graviton-graviton scattering (for a particular convenient choice of the reference spinors). This will become clear in the next section. Thus, in particular the GR 4-vertex present in (\ref{L4-GR-M}) is not as relevant as far as the graviton-graviton scattering is concerned, and we have written it only for reference.  

For the vertex factor we will use the rule that all 4 momenta are incoming. Taking into account the symmetry factors, the associated vertex is 
\be\label{4vert-sd-2}
\frac{\im ( 8g^{(4)} - 4g^{(3)} -(7/2) g^{(2)} )}{3 M^4 (g^{(2)})^2}   \quad  \lower0.6in\hbox{\includegraphics[width=1.2in]{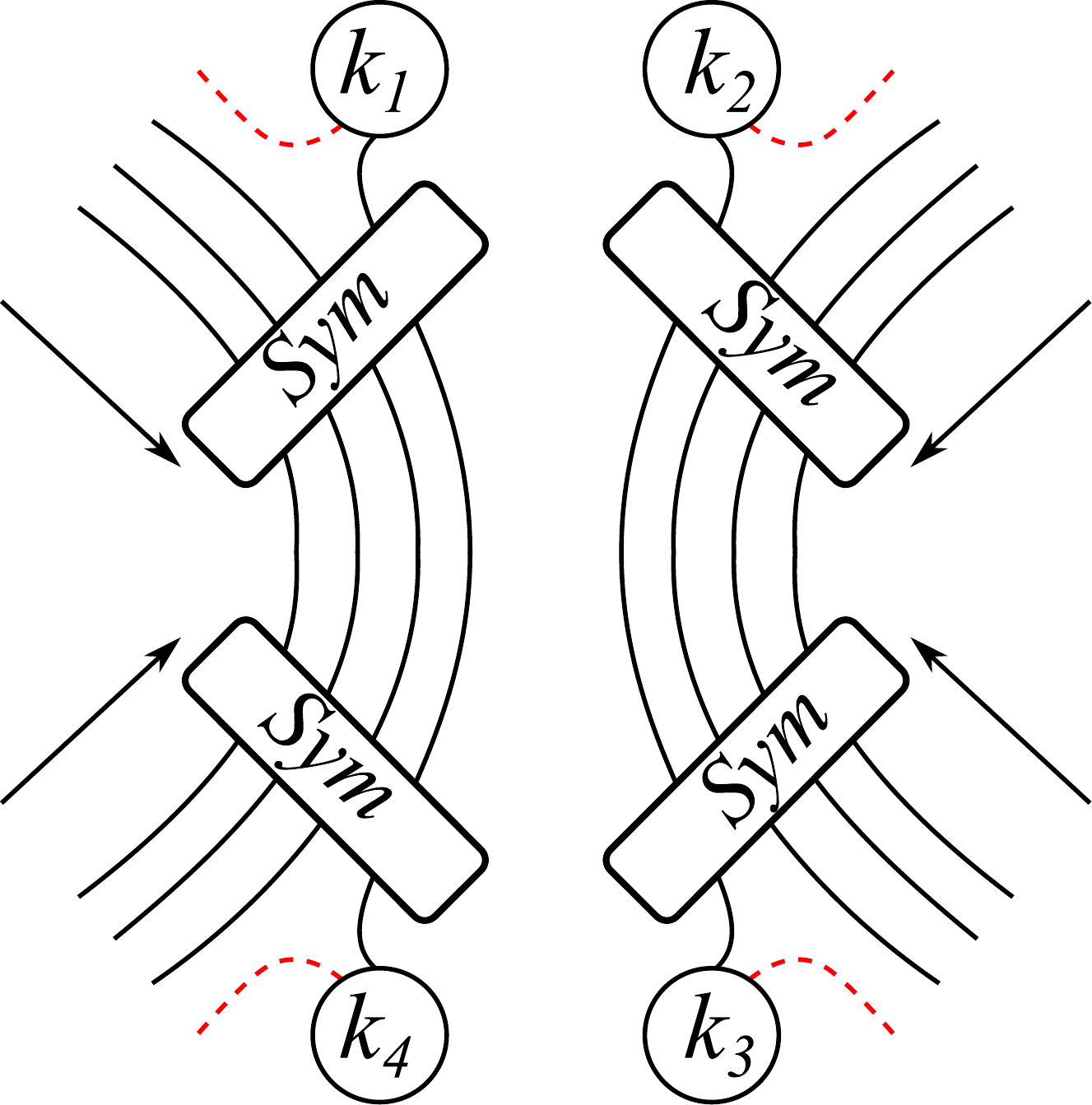}} \, ,
\ee
where one needs to the sum over possible permutations. There are 3 terms that are numbered e.g. by what $k_1$ gets connected to. 

Just as a reference, we will also give pictures of the index contractions present in the other 3 terms appearing in (\ref{L4-M}). As we have said, these other terms will not contribute to any computations in this thesis, but the pictures below will be instrumental in seeing this. We give them in the order that they appear in (\ref{L4-M}), without any associated prefactors.
\be\label{4vert-asd}
\quad  \lower0.6in\hbox{\includegraphics[width=1.2in]{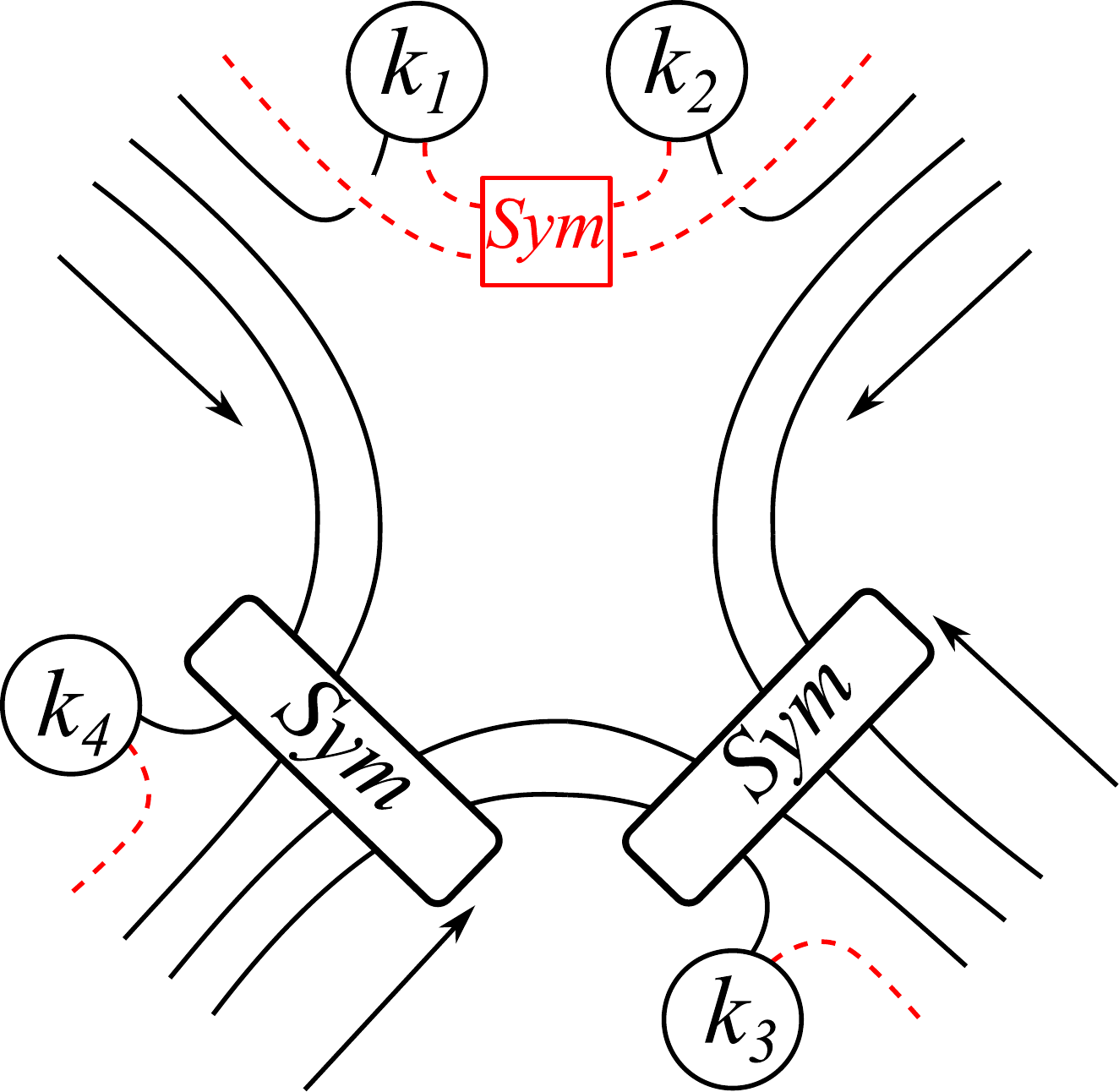}}\quad  \lower0.6in\hbox{\includegraphics[width=1.2in]{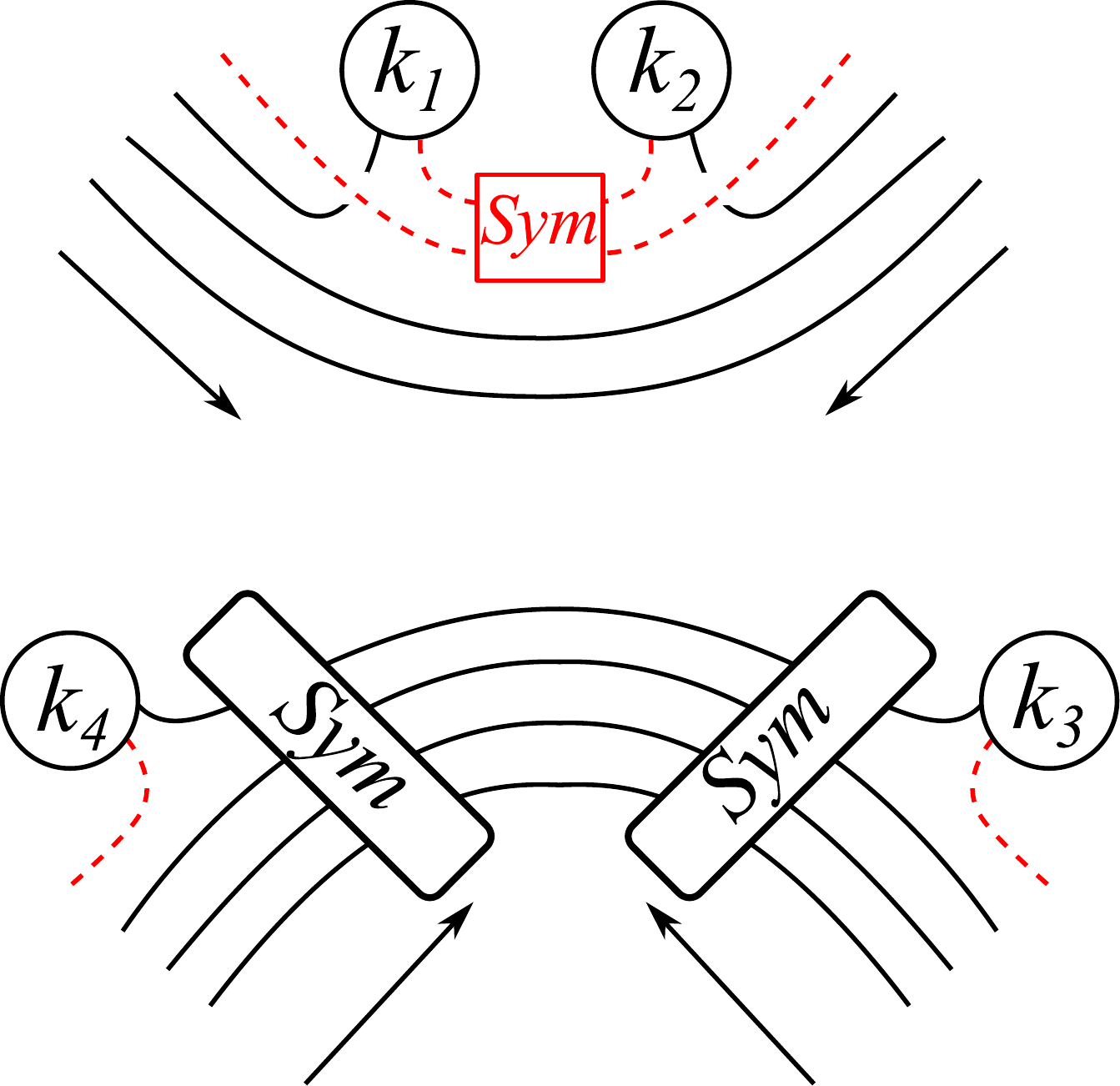}}\quad  \lower0.6in\hbox{\includegraphics[width=1.2in]{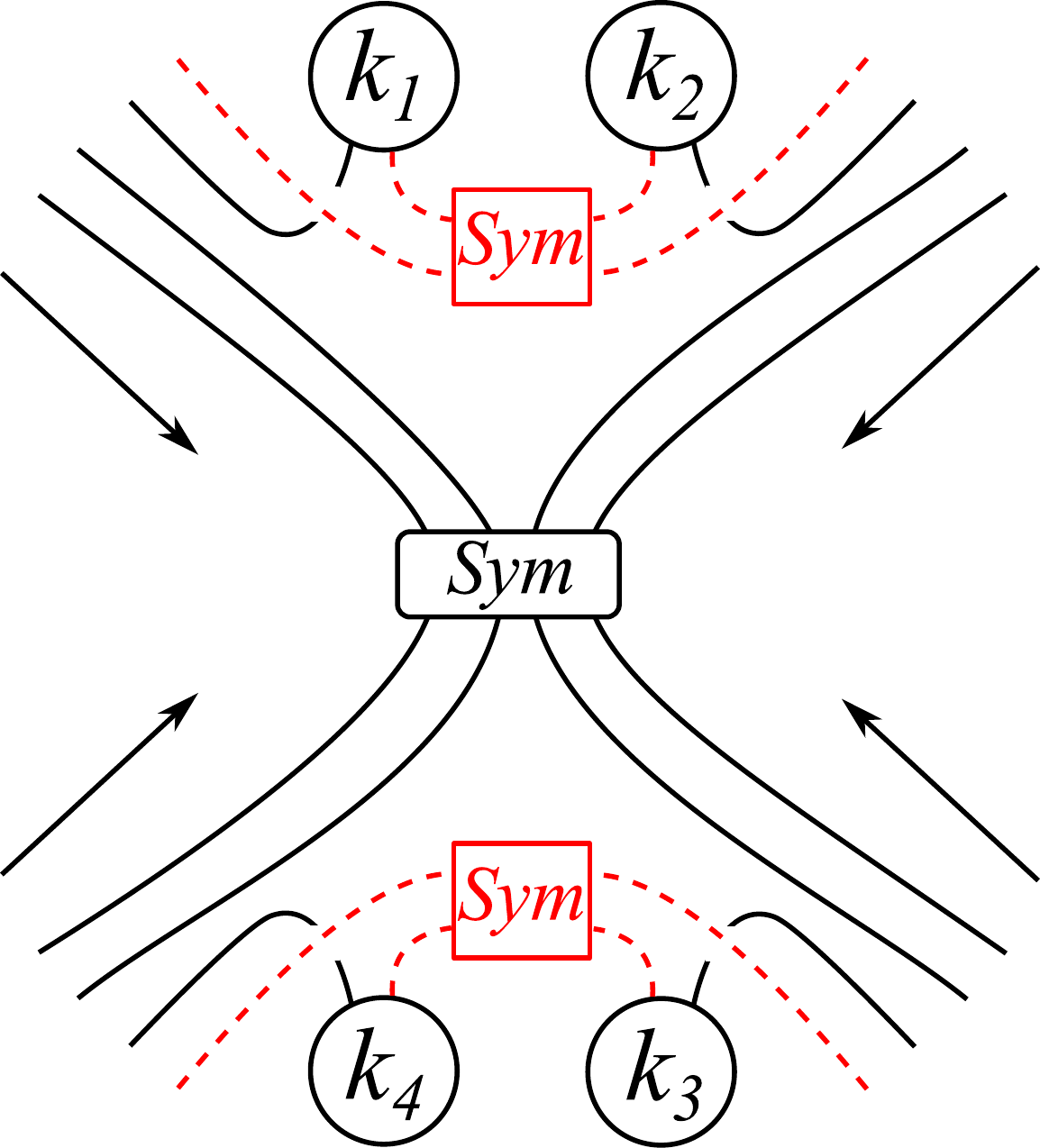}}
\ee
The first and the last of these are the ones that appear in (\ref{L4-GR-M}).

%%%%%%%%%%%%%%%%%%%%%%%%%%%%%%%%%%%%%%%%%%%%%%%%%%%%%%%%%%%
\section{Graviton-Graviton Scattering}\label{graviton_graviton_scattering_section}
%%%%%%%%%%%%%%%%%%%%%%%%%%%%%%%%%%%%%%%%%%%%%%%%%%%%%%%%%%%

This is the last section of the chapter, where we put to use all the technology that we developed. %As is the case with the metric-based GR, even the simplest $--++$ amplitude is somewhat easier to compute as just the first non-trivial case of a more general MHV amplitude with all but two positive helicity gravitons. For such amplitudes one can write the so-called BCFW recursion relation \cite{Britto:2005fq}, and then the $--++$ amplitude is given by a sum of two terms, both involving just the known $--+$ amplitudes (\ref{--+}). However, here we avoid developing the technology of recursion relations, postponing it to the next paper. Instead, 
 We compute all amplitudes of interest by directly evaluating the relevant Feynman diagrams, of course using the spinor helicity technology that we already have developed. We will see that the $--++$ amplitude receives contributions from only tri-valent graphs with vertices (\ref{V-GR}), and there are two diagrams to compute. The degree of complexity of this computation is very similar to the analogous textbook computation in Yang-Mills theory, see e.g. \cite{Srednicki:2007qs}. We first do the computation of the $--++$ amplitude, and then consider somewhat simpler cases of the $-+++$ and $++++$ amplitudes (which are only non-zero in our more general parity-violating theories).\footnote{In this chapter we consider all the gravitons having $incoming$ momentum. That means that, for instance, the $\mathcal{M}^{++--}$ below is effectively the amplitude of the process $++\to++$ where the last two particle are intended as outgoing.}

\subsection{The all-negative and all-negative-one-positive helicity amplitudes}

The amplitudes with at most one positive helicity graviton can be shown to vanish in full generality, just by a simple count of the number of derivatives present in any given diagram. 

Let us first consider the all minus case, with $n$ external legs. In this case the helicity states (\ref{helicity-spin}) each carry 3 copies of the negative helicity reference spinor $q^A$. This spinor can then be chosen to be the same for all the gravitons. Thus, we have $3n$ copies of $q^A$ in the diagram, and we have to contract them all with some momenta that appear as a result of evaluating in the derivatives present in the vertices. It is then easy to see that there are not enough derivatives to avoid contracting the $q^A$'s between themselves. Indeed, the largest possible number of derivatives is in a diagram with 3-valent vertices only. There are $n-2$ such vertices, and thus at most $3(n-2)$ derivatives present (as the largest power of the derivative in each vertex is 3). Since each derivative can only eat one copy of $q^A$, there are not enough derivatives for these amplitudes to be non-zero. 

The argument with all minus one plus proceeds similarly, but in this case one chooses all the reference spinors of the negative helicity gravitons to be the momentum spinor $k^A$ of the positive helicity graviton. Then there are again $3n$ instances of $k^A$ and only at most $3(n-2)$ derivatives, which is not enough to avoid the $k^A$'s contracting. 

\subsection{The $--++$ amplitude}

This is the only non-vanishing amplitude in the case of GR, as we shall also explicitly see using our methods. 

Let us first verify that the 4-vertex containing diagrams cannot contribute to this amplitude. This is an argument of the same type that we already gave to show that the all minus and all minus one plus amplitudes are zero. Indeed, we now have 6 copies of $q^A$ coming from the two negative helicity states, and we can choose them to be equal to the momentum spinor $k_3^A$ of one of the positive helicity graviton. We thus get 9 copies of $k_3^A$ that need to contract with something else than themselves. Let us count other object available. We have 3 copies of the other positive helicity momentum spinor $k_4^A$ coming from its helicity spinor (\ref{helicity-spin}). We thus need at least 6 derivatives to absorb the remaining 6 copies of $k_3^A$. These can only come from tri-valent diagrams, as the other diagrams contain less derivatives. Incidentally, the same argument shows that only the 3-valent graphs contribute to any amplitude with at most two plus helicity gravitons. 

It remains to see that only the vertices (\ref{V-1}) can contribute to this amplitude. As we just saw, we need to have at least 6 derivatives, so the only other possible vertex is (\ref{V-2}). We cannot use this vertex twice, because it can only take the positive helicity states. Thus, let us assume that just one of the vertices used is (\ref{V-2}). Since this vertex can only take the positive helicity gravitons, both available positive helicity states must go in it. The other vertex is then (\ref{V-1}), with two negative helicity states inserted in it. It is then a simple verification to see that in the internal edge of the diagram we will have $q^A q^B q^C$ coming from the negative helicities contracting with $k_3^{(A} k_3^{B} k_4^{C)}$ or with $k_3^{(A} k_4^{B} k_4^{C)}$ coming from the positive helicities. Since we have chosen $q^A$ to be the momentum spinor of one of the positive helicity gravitons, this contraction is zero. Again, precisely the same argument works for a general amplitude with at most two plus helicity gravitons, and establishes that only the vertex (\ref{V-1}) is relevant in any of these amplitudes. This means that after the identification (\ref{Mp}) is made, all such amplitudes for a general member of our family of theories are the same as in GR. In particular, the $--++$ amplitude defining the Newton's constant is the same, and so in the following considerations we can assume the form (\ref{V-GR}) of the relevant vertex. 

Thus, we only need to consider the 3-valent graphs with vertices (\ref{V-1}). There are 3 such graphs ($s,t,u$ channels), and each vertex factor has 3 terms. Thus, there are in principle 27 terms to consider. However, most of them are zero. 

A very convenient way to organise the computation is to consider the possible ways of putting two negative helicities into the vertex (\ref{V-GR}). It is clear that they must go into the two legs in which the vertex is symmetric. Indeed, any pair of external legs containing the $(\partial a)^{ABCD}$ leg (this is the bottom leg in the pictorial representation) is connected by a black line. Thus, the insertion of a pair of negative helicity states in any other way but from the top gives a zero result. Recalling that the top legs came from the anti-self-dual part of the two-form $da^i$, we shall refer to them as the ASD legs. Similarly, the bottom leg will be referred to as SD. 

We can now recall our choice $q^A=k_3^A$. It means that for many purposes the positive helicity graviton of momentum $k_3$ behaves like a negative helicity one. In particular, if we are to put this positive helicity graviton into the vertex (\ref{V-GR}) together with some negative helicity graviton, it is easy to see that both necessarily must go into the ASD legs. Indeed, if the positive helicity goes into the bottom leg in the picture (\ref{V-GR}) it is easy to see that at least one of the $q^A$'s from the negative helicity graviton will contract with $k_3^A$ along the black lines. Thus, in case of the graviton of momentum $k_3$ and some negative helicity graviton, both must go into the upper, ASD legs in the figure (\ref{V-GR}). 

Let us now compute the result of such insertion of two negative or one negative one positive of momentum $k_3$ helicity states into the vertex (\ref{V-GR}). For two negative states that we label by $1$ and $2$, we get the following quantity
\be\label{--a}
\im M\kappa \, q^A q^B q^C  (k_1+k_2)^{DD'} q_D \frac{\bra{1}{2}^2}{\ket{1}{q}^2\ket{2}{q}^2} .
\ee
Similarly, when the states $1$ and $3$ are put together into this vertex we get
\be\label{-+a}
- \im M\kappa \, q^A q^B q^C  (k_1+k_3)^{DD'} q_D \frac{\bra{1}{p}^2}{\ket{1}{q}^2\bra{3}{p}^2} ,
\ee
where an extra minus is from $\partial^E_{M'} a_{EABN'}$ where we get $k_1^E q_E=\ket{1}{q}$, but in the denominator of these helicity states we have $\ket{q}{1}^3$, and this different order of the contraction produces an extra minus sign. Two of such minus signs have cancelled each other in (\ref{--a}). We remind the reader that in (\ref{-+a}) we have used $k_3^A=q^A$. 

The important point about the results (\ref{--a}) and (\ref{-+a}) is that they can now be put together with some other state via a 3-vertex (\ref{V-GR}) only in a single way, again from the upper two legs in the picture. The reason for this is that both of them, as the original helicity states, contain 3 copies of $q^A$, thus, the argument we gave above about the only possible way to couple such states applies. This means that the vertex (\ref{V-GR}) works only ``in one direction'' coupling the negative helicities or the positive helicity graviton of momentum $k_3$ by taking them in its ASD legs. As an aside remark, we note that this immediately implies that even in a general N-amplitude with only two plus helicity gravitons, the vertex (\ref{V-2}) cannot be used. Indeed, it could only be used to couple the two positive helicity gravitons to all negative gravitons already coupled in some way with (\ref{V-1}). However, the one leg off-shell current with any number of negative helicity gravitons is necessarily proportional to $q^A q^B q^C$, and this will vanish when contracted with what comes from the vertex (\ref{V-2}), as we have already discussed above. 

The above picture of the vertex (\ref{V-GR}) working as a coupler of states just in one direction gives 3 possibilities to consider. One can either couple first the two negative helicity gravitons $1$ and $2$, and then couple the result to $3$, or first couple $1$ and $3$, and then couple to $2$, or first couple $2$ to $3$ and then to $1$. 
\be\nonumber
\includegraphics[height=1in]{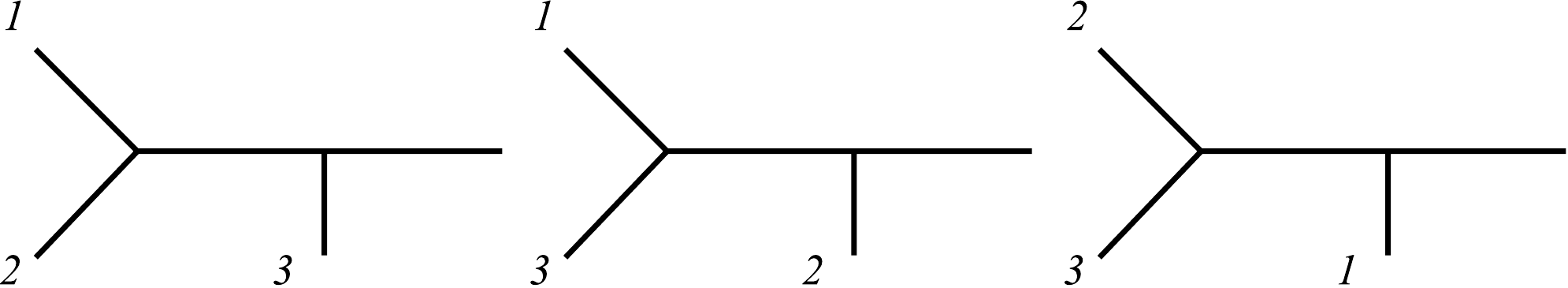}
\ee
In all these cases one couples to the positive helicity graviton $4$ at the very end. These are of course just the 3 different $s,t,u$ channels, but we now have just 3 terms to consider instead of 27. 

The final simplification comes from the availability to choose the positive helicity reference spinor $p^{A'}$ conveniently. We recall that we have not yet made any choice of this in the analysis so far, so we are now free to make the most convenient choice. We choose $p^{A'}=2^{A'}$, which eliminates the possibility (\ref{-+a}) to couple $3$ directly to $2$. This leaves just the first two terms in the above picture to consider. 

Let us compute the diagram when $1$ and $2$ get coupled first, and then couple to $3$. We have computed the result of coupling $1$ and $2$ in (\ref{--a}). We then multiply this result by the propagator $1/\im k^2$, where $k^2=k^A{}_{A'} k_A{}^{A'}$ computes, using the fact that both $k_1$ and $k_2$ are null, to just $k_{12}^2=2\ket{1}{2}\bra{1}{2}$. We now couple together (\ref{--a}) with the propagator added at the end with the positive helicity state $3$. After applying the derivatives present in the vertex we get the following contraction
\be\nonumber
(\im)^2 \kappa^2 \, q_A q_B (k_1+k_2)^{E}{}_{M'}  q_E  (k_1+k_2)^{F}{}_{N'} q_F (k_1+k_2+k_3)^D{}_{D'} \frac{\bra{1}{2}^2}{\ket{1}{q}^2\ket{2}{q}^2} \frac{1}{2\im \ket{1}{2}\bra{1}{2}} M \frac{q_{C} q_{D} p^{M'} p^{N'}}{\bra{3}{p}^2}
\\ \nonumber
= 2\im M \left(\frac{\kappa}{2}\right)^2 q_A q_B q_C (k_1+k_2+k_3)^D{}_{D'} q_D \frac{\bra{1}{2}}{\ket{1}{2}} \frac{\langle q| k_1+k_2| p]^2}{\ket{1}{q}^2\ket{2}{q}^2\bra{3}{p}^2}.
\ee
We finally insert into this result the last remaining positive helicity $(1/M) 4^A 4^B 4^C p^{D'}/\bra{p}{4}$. We can also use the momentum conservation to replace $k_1+k_2+k_3$ by $-k_4$ (all momenta are incoming). Substituting also $q^A=3^A$ and $p^{A'}=2^{A'}$ we get, overall
\be\label{amp-1}
2\im \left(\frac{\kappa}{2}\right)^2  \frac{\bra{1}{2}^3}{\ket{1}{2}} \frac{\ket{3}{4}^4}{\ket{2}{3}^2\bra{2}{3}^2}.
\ee

We now compute the other non-vanishing diagram, where $1$ gets first connected to $3$, and then the result connects to $2$. Adding to (\ref{-+a}) the propagator, taking the helicity state for $2$ and applying all the derivatives gives the following contraction
\be\nonumber
(\im)^2 \kappa^2 \, q_A q_B (k_1+k_3)^{E}{}_{M'}  q_E  (k_1+k_3)^{F}{}_{N'} q_F (k_1+k_2+k_3)^D{}_{D'} \frac{\bra{1}{p}^2}{\ket{1}{q}^2\bra{3}{p}^2} \frac{1}{2\im \ket{1}{3}\bra{1}{3}} M \frac{q_{C} q_{D} p^{M'} p^{N'}}{\ket{2}{q}^2}
\\ \nonumber
= 2\im M \left(\frac{\kappa}{2}\right)^2 q_A q_B q_C (k_1+k_2+k_3)^D{}_{D'} q_D \frac{\bra{1}{p}^2}{\ket{1}{3}\bra{1}{3}} \frac{\langle q| k_1+k_3| p]^2}{\ket{1}{q}^2\ket{2}{q}^2\bra{3}{p}^2}.
\ee
We now put in this the last remaining state $4$, and use the values of $q^A, p^{A'}$ to get
\be\label{amp-2}
2\im \left(\frac{\kappa}{2}\right)^2  \frac{\bra{1}{2}^4}{\ket{1}{3}\bra{1}{3}} \frac{\ket{3}{4}^4}{\ket{2}{3}^2\bra{2}{3}^2}.
\ee
Adding (\ref{amp-1}) and (\ref{amp-2}) and using the momentum conservation we get
\be
- 2\im \left(\frac{\kappa}{2}\right)^2 \frac{\ket{3}{4}^4}{\ket{2}{3}^2\bra{2}{3}^2}  \frac{\bra{1}{2}^3 \ket{1}{4} \bra{1}{4}}{\ket{1}{2} \ket{1}{3} \bra{1}{3}}.
\ee
We now convert as many square bracket contractions into the round ones using the momentum conservation identities, e.g. $\bra{1}{4}/\bra{1}{3}=-\ket{2}{3}/\ket{2}{4}$. We finally get
\be\label{--++1}
{\cal M}^{--++} = 2\im \left(\frac{\kappa}{2}\right)^2 \ket{3}{4}^6 \frac{1}{\ket{1}{3}\ket{1}{4}\ket{2}{3}\ket{2}{4}} \frac{\bra{1}{2}}{\ket{1}{2}},
\ee
which is the usual GR result, see below. 

To rewrite (\ref{--++1}) in a more recognisable form, we evaluate the spinor contractions present in the centre of mass frame, and rewrite everything in terms of the Mandelstam variables. The relevant contractions are given in (\ref{contr}). We get
\be\label{--++}
{\cal M}^{--++} = \im \left(\frac{\kappa}{2}\right)^2 \frac{s^3}{tu},
\ee
which is the form one find this result in e.g. \cite{Bern:2002kj}, formula (17), modulo the factor of $\im$ that most likely has to do with different conventions, or in \cite{Grisaru:1975bx}, formula (40) with the coefficient $c$ given after formula (42). Our factors in (\ref{--++}) precisely match those in \cite{Grisaru:1975bx}. 

We emphasise that the result (\ref{--++}) holds for a general member of our family of theories. It is a particular case of a more general result that all amplitudes with just two positive helicity gravitons are the same for all members of our family, provided one identifies the Newton's constant as in (\ref{Mp}). The generality of the result (\ref{--++}) is important, because it immediately tells us important information about the types of higher derivative terms that are present in the Lagrangian of our theories if one interprets them as metric theories. Indeed, for a general counter-term corrected Einstein-Hilbert Lagrangian one expects (\ref{--++}) to get modified, see e.g. formula (61) of \cite{ArkaniHamed:2008gz}. The fact that this does not happen for our class of theories tells us that no such $R^4$ type modifications are present. Once again it illustrates a very tightly constrained character of the modifications present in our different from GR gravitational theories. 

\subsection{The $-+++$ amplitude}

We now compute the first example of an amplitude that is zero in GR, but non-zero in a general parity-violating theory (the $+++$ amplitude encountered in the previous section is zero once the momentum is taken to be real and the momentum conservation is imposed).

The first check that we need to do is that both for this, and the $++++$ amplitude only the 4-vertices with at least 4 derivatives can contribute, and so it is sufficient to restrict one's attention to (\ref{L4-M}). Let us first run the argument for the all plus case. Here we have 4 reference spinors $q^{A'}$, and these need to be contracted with something else than themselves. Therefore, one must have at least 4 derivatives. The same argument also works in the case $-+++$, if one chooses the positive helicity reference spinors to be equal to the momentum spinor for the negative helicity state. 

Let us now check that there cannot be any contribution to this amplitude from the 4-valent vertices. Since we have one negative helicity, only the vertices in (\ref{4vert-asd}) could contribute, with the negative helicity inserted into one of the ASD legs, i.e. where the contraction of the derivative with the connection is of the form $\partial^E_{M'} a_{EABN'}$. However, such a leg gets necessarily contracted with another ASD leg, where we will have a positive helicity graviton. As usual, we can choose the reference momenta of all positive helicity gravitons to be the same and equal to the momenta of the negative one. Let the negative helicity graviton be of momentum $k_1$, then we choose $p^{A'}=1^{A'}$. It is now easy to see that the primed index contraction will give that of $p^{A'}$ with itself, and so these 4-valent graph diagrams cannot contribute. Another way to phrase this argument is to note that there is always a dashed line connecting at least one pair of vertices in (\ref{4vert-asd}). Such dashed lines give a contraction of the reference spinors $p^{A'}$, or of these reference spinors with the momentum spinors $1^{A'}$, and this gives a zero result. 

It remains to compute contributions from the 3-valent graphs. There appears to be many possibilities, as we have 4 different types of 4-valent vertices to consider. Indeed, because we need just 4 derivatives to be present, it appears that even the one-derivative vertices can give some contribution. However, as usual, most of the possibilities give a zero result. In particular, the one-derivative vertices cannot contribute. Let us check this first. 

The way to check which diagrams can contribute is to follow the dashed lines. With our choice $p^{A'}=1^{A'}$ we know that we cannot put together states in a pair of legs that is connected by a dashed line. E.g., in the GR vertex (\ref{V-1}) we can only put the states one in an upper leg, one in a lower, but not two in the upper ASD legs, because this would result in a zero spinor contraction. If we do put two states into this vertex in the only allowed way, we will get the reference spinor $p^{A'}$ on the free dashed line, later to be contracted into some other 3-valent vertex. The same argument applies to (\ref{V-strange}) and (\ref{V-4}) where we again see that in order to avoid the contractions of the positive helicity reference spinors we need to use a very specific insertion. Thus, in the case of (\ref{V-strange}) we must necessarily insert the external states from the top, with again a reference spinor $p^{A'}$ appearing on the dashed line of the free leg (bottom in this case). With (\ref{V-4}) we can only insert the external states one into a top leg, one into a bottom, as these are not connected by a dashed line. Again, there is a $p^{A'}$ spinor appearing on the free leg. This immediately tells us that the vertices (\ref{V-1}), (\ref{V-strange}) and (\ref{V-4}) cannot be paired, i.e. any diagram involving only vertices of these types is zero (for this choice of helicities). This means that one of the vertices must be (\ref{V-2}). In particular, this argument explains why this amplitude is zero in GR (in our formulation). 

The easiest way to see that the diagrams involving vertices (\ref{V-strange}), (\ref{V-2}) and (\ref{V-4}), (\ref{V-2}) cannot contribute is to work out the $M$-dependence. The prefactor in (\ref{V-2}) goes as $M^3/M_p^5$. The helicity states give $M$ from the negative helicity times $1/M^3$ from the positive. Overall, this leaves a factor of $M$ to first positive power. Thus, in order for these diagrams to give a surviving in $M\to 0$ contribution the other 3-valent vertex must have a negative power of $M$ in front. However, this is not the case. The vertex (\ref{V-strange}) goes as $M/M_p$, and the vertex (\ref{V-3}) goes as $M^5/M_p^5$. Thus, no non-zero Minkowski limit contributions are produced in this case. 

The only possibly non-vanishing in the $M\to 0$ limit diagram is therefore one involving GR vertex (\ref{V-1}) and the vertex (\ref{V-2}). We must necessarily insert the negative graviton into the vertex (\ref{V-1}), in one of the upper legs. Also, some positive helicity graviton should go into the bottom leg of this vertex. Let us compute the corresponding contraction, including the propagator at the end. We take the positive graviton to be the one of momentum $k_2$, and get
\be
-\frac{\im\kappa}{M} M \frac{q_E q_F 1^{M'} 1^{N'}}{\ket{1}{q}^2} \frac{1}{M} 2^E 2^F 2^{(A} 2^B (k_1+k_2)^{C)}{}_{N'} \frac{1}{2\im \ket{1}{2}\bra{1}{2}} \\ \nonumber
= \frac{\kappa}{2M^2} \frac{\ket{2}{q}^2}{\ket{1}{2}\ket{1}{q}^2} 2^A 2^B 2^C 1^{M'}.
\ee
The free indices here are $ABCM'$. We have to contract this object with one obtained by combining the gravitons $3$ and $4$ via the vertex (\ref{V-2}). This gives
\be
\frac{\im \sqrt{2} (4g^{(3)}-3g^{(2)})}{M^2 (g^{(2)})^{3/2}} \frac{1}{M^2} 3^E 3^F 4_E 4_F 3_{(A} 3_B 4_C 4_{D)} (k_3+k_4)^D{}_{M'}.
\ee
The symmetrisation here, together with its contraction with $k_3+k_4$ can be written as
\be
3_{(A} 3_B 4_C 4_{D)} (k_3+k_4)^D{}_{M'} = \frac{1}{2}\left( 3_D 3_{(A} 4_B 4_{C)} + 4_D 4_{(A} 3_B 3_{C)}\right) (k_3+k_4)^D{}_{M'} \\ \nonumber
= \frac{1}{2}\left( \ket{4}{3} 3_{(A} 4_B 4_{C)} 4_{M'}  + \ket{3}{4} 4_{(A} 3_B 3_{C)} 3_{M'} \right) .
\ee
Overall, after contracting the above two quantities, we get for this contribution to the amplitude
\be\label{amp-3}
\frac{\im (4g^{(3)}-3g^{(2)})}{2M^2 M_p^4} \frac{\ket{2}{q}^2\ket{3}{4}^3\ket{2}{3}\ket{2}{4}}{\ket{1}{2}\ket{1}{q}^2}\left( \ket{2}{4} \bra{1}{4} -\ket{2}{3} \bra{1}{3}\right).
\ee 
This is $34$ symmetric, as it should be. Note that we have replaced $\kappa$ with $\sqrt{2}/M_p$ and also used the fact that $M^2g^{(2)}=M_p^2$. We should now add contributions from the 2 more diagrams in which $1$ first gets connected to $3$ and then to $1$ and $4$, and another one where $1$ first gets connected to $4$ and then to the rest. What one gets can be checked to be $q$-independent, as it should be of course. So, we shall make a choice and set $q^A=4^A$, so that there is just one more contribution to consider. It reads
\be\label{amp-4}
\frac{\im (4g^{(3)}-3g^{(2)})}{2M^2 M_p^4} \frac{\ket{3}{q}^2\ket{2}{4}^3\ket{3}{2}\ket{3}{4}}{\ket{1}{3}\ket{1}{q}^2}\left( \ket{3}{4} \bra{1}{4} -\ket{3}{2} \bra{1}{2}\right).
\ee 
We now set $q^A=4^A$ and add the above two quantities. We can also use the momentum conservation to note that the quantities in brackets in both (\ref{amp-3}) and (\ref{amp-4}) are equal, so that we can keep only the first one of them in each case, and double the result. We thus get
\be
\frac{\im (4g^{(3)}-3g^{(2)})}{M^2 M_p^4} \bra{1}{4} \frac{\ket{2}{4}^3\ket{3}{4}^3 \ket{2}{3}}{\ket{1}{4}^2} \left( \frac{\ket{2}{4}}{\ket{1}{2}} - \frac{\ket{3}{4}}{\ket{1}{3}}\right). 
\ee
Using the Schouten identity this finally gives
\be\label{-+++1}
{\cal M}^{-+++} = \im \frac{(4g^{(3)}-3g^{(2)})}{M^2 M_p^4}  \frac{\bra{1}{4}}{\ket{1}{2}\ket{1}{3}\ket{1}{4}} \ket{2}{4}^3\ket{3}{4}^3 \ket{2}{3}^2.
\ee
Using the momentum conservation this can be seen to be $234$ symmetric. 

It is instructive to rewrite the result (\ref{-+++1}) using the Mandelstam variables. The relevant spinor contractions in the centre of mass frame are given in the Appendix, see (\ref{contr}). One gets
\be
{\cal M}^{-+++} = \frac{(4g^{(3)}-3g^{(2)})}{8\im M^2 M_p^4}  stu.
\ee
Specializing to the one-parameter family of theories guaranteed to contain only Planckian modifications of gravity, we use (\ref{diff-g-1}) and get
\be\label{-+++}
{\cal M}^{-+++} = \im \frac{27\beta^2}{32 M_p^6}  stu.
\ee
Here $\beta$ is a parameter that controls the strength of deviations from GR, see (\ref{mod-action}).  We see that at high energies this amplitude goes as $E^6/M_p^6$. 

\subsection{The $++++$ amplitude}

We now study the final graviton-graviton amplitude, involving 4 incoming gravitons of positive helicity. This amplitude vanishes if there are just the GR vertices, but is non-vanishing in general, as we shall now compute. 

Many of the argument of the previous section apply, and we can see that the 4-vertices in (\ref{4vert-asd}) do not contribute. We also know that from the 3-valent diagrams, only ones involving the vertex (\ref{V-2}) can contribute. We will analyse the 3-valent diagrams containing just one copy of the vertex (\ref{V-2}) in the Appendix. Let us now consider the diagram containing two vertices of the type (\ref{V-2}). We know that the prefactor in this vertex goes as $M^3/M_p^5$. Given two such prefactors, and taking these together with the factor of $1/M^4$ coming from the positive helicity states, we see that the result goes as $M^2$, and so does not survive in the Minkowski limit. 

We thus have to consider only the contributions from the 4-valent graphs with the vertex (\ref{4vert-sd-2}), as well as 3-valent graphs involving just one copy of the vertex (\ref{V-2}). The latter are worked out in the Appendix. On the other hand, the computation of the 4-valent diagram is very easy, because after the derivatives get applied to the external states, one is just left with $(1/M) k^A k^B k^C k^D$ quantities, where $k^A$ is the corresponding momentum spinor, to contract as dictated by the black lines in figure (\ref{4vert-sd-2}). The result is
\be\nonumber
\frac{\im ( 8 g^{(4)} - 4g^{(3)} -(7/2) g^{(2)} )}{3 M^8 (g^{(2)})^2}  ( \ket{1}{3}^4 \ket{2}{4}^4+ \ket{1}{2}^4\ket{3}{4}^4 + \ket{1}{4}^4\ket{2}{3}^4).
\ee

We can rewrite the above results more compactly by going into the centre of mass frame and introducing the Mandelstam variables (\ref{mandel}). Using the results of spinor contractions given in (\ref{contr}) we get 
\be
{\cal M}^{++++}_{\rm 4-vert} = \frac{\im(2 g^{(4)} - g^{(3)} -(7/8) g^{(2)})}{12  M^4  M_p^4 }  (s^4+t^4+u^4).
\ee
For theories having only Planckian modifications, such as the family described in the Appendix, we have (\ref{diff-g-2}), and 
\be\label{++++4V}
{\cal M}^{++++}_{\rm 4-vert} = \im \frac{27\beta^3 }{48  M_p^8 }  (s^4+t^4+u^4).
\ee
Adding this to the contribution (\ref{app-3-valent}) from the 3-valent diagrams from the Appendix we get the following answer for this amplitude
\be\label{++++}
{\cal M}^{++++}= \im \frac{135\beta^2}{16M_p^6} stu+ \im \frac{27\beta^3 }{48  M_p^8 }  (s^4+t^4+u^4).
\ee
At sub-Planckian energies the first term dominates and the amplitude goes as $E^6/M_p^6$.

\chapter{Conclusions}

In this thesis we introduced and studied the quantised version of the Pure Connection Formulation (PCF) of gravity. In particular, in this thesis we  defined a framework to compute graviton scattering amplitudes in such formulation. Also, by computing the scattering amplitudes, we were able to verify the validity of the theory  comparing the results with the expressions for the GR amplitudes known in the literature \cite{Bern:2002kj}. The main 
conclusion is that the results are in complete agreement with the established amplitudes. Also, we have seen that the interaction vertices took a much smaller and simpler form in this novel formalism, especially when written in terms of spinors, with respect to the  corresponding Einstein-Hilbert vertices.

However, the most important outcome of this thesis is another. Indeed, the main result is that we managed to derive the graviton-graviton scattering amplitudes for a whole general class of $gravitational$ theories. Most importantly, this lead to the surprising find of $new$ amplitudes that are not present in GR. One important feature of these new amplitudes is that they are not preserving parity. 

We already knew that this formalism could be employed to describe a wider class of gravitational theories that includes GR as a special case \cite{Krasnov:2011pp} \cite{Krasnov:2011up}. Those theories were seen to share the same linearised dynamics with GR formulated in this language. Indeed, we have seen that all this theories represent interacting massless (after the limit) spin 2 particles. The key difference with GR is that a general theory from this class does not exhibit explicit parity invariance.
However, one can argue that there is no physical reason to restrict one's attention to only parity-invariant theories. For instance, we know that Nature does violate parity (in the Standard Model) in the strongest possible way.

Let us now discuss the new parity-violating amplitudes in more details. First, there is a set of amplitudes where parity is flipped just by one unit. Applying the crossing symmetry and converting two gravitons from incoming to outgoing states, we can draw these parity-violating processes as
\be\nonumber
 \lower0.6in\hbox{\includegraphics[height=1.2in]{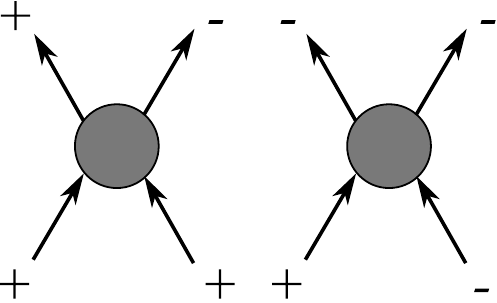}} 
\ee
In both cases one of the positive helicity gravitons is converted into a negative helicity one. We have seen \myref{-+++} that this amplitude goes as $stu/M_p^6$, where $s,t$ and $u$ are the usual Mandelstam variables. It is thus of equal importance as the usual parity-preserving amplitude $(1/M_p^2) s^3/tu$ at Planck energies, which is just an illustration of the fact that a generic member of our family of gravity theories is very different from GR at high energies. If one could extrapolate beyond the Planck barrier (which in reality one cannot, because the perturbation theory breaks down given that we are expanding in powers of $\f{\omega}{M_p}$), one could say that at higher energies the parity-violating processes scaling as $E^6$ become even more important than the parity-preserving ones scaling as $E^2$.

Another set of these processes is when the helicity is flipped by two units. This can be drawn as
\be\nonumber
 \lower0.6in\hbox{\includegraphics[height=1.2in]{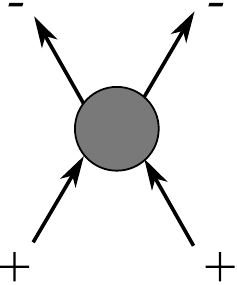}} 
 \ee
Two positive helicity gravitons are converted by this process into two negative helicity ones, with the amplitude computed in \myref{++++4V} going as $(s^4+t^4+u^4)/M_p^8$ and the amplitude in \myref{app-3-valent} going again as $stu/M_p^6$. 

One is then led to admittedly speculative, but thought-provoking picture of the dynamics of gravity at high energies, as predicted by our theories. Indeed, the parity-violating processes only go in one direction (the theories are not T-invariant). It is then clear that eventually all gravitons will get converted into gravitons of a single helicity (negative in our conventions). Whether this indeed has anything to do with what happens at the Planck scale remains to be seen, but it is clear that this picture is a valid, and to some extent unexpected outcome of our gauge-theoretic approach.

% 
% 
% The key feature of the PCF \cite{Krasnov:2011up} \cite{Krasnov:2011pp} is that it reduces the complexity of the field content to a minimum, only a connection $a^i_\m$. This connections takes values in the complexified algebra of $SO(3)$. The action itself takes a very simple form. Indeed, if expressed in terms of a ``defining function'' $f$, the action reduces to solely:
% 
% $$S = \int f(F\wedge F),$$ 
% 
% $F$ being the curvature of the connection. The defining function $f$ is uniquely defined for GR (with non-zero cosmological constant). However, one can relax some constraints and generalise $f$ to be any function homogeneous degree one, gauge-invariant and mapping $3 \times 3$ matrices to complex numbers. This more general class of theories can be shown to still propagate only two degrees of freedom and therefore can be regarded as a ``gravitational theory'', GR being a special case of it. 

Now let us summarise what has been done in the different chapters and the results that have been found. 

The general classical theory was reviewed at the beginning of chapter \myref{General_PCF_chapter}. Afterwards we started discussing problems connected with the quantisation of it, mostly following the paper of the present author $et$ $al.$ \cite{Delfino:2012zy}. We first went through the needed  Hamiltonian analysis in section \myref{Hamiltonian_analysis_section}. Here we took a novel approach introducing two differential operators $D$ and $\bar{D}$. Using this differential operators, in the subsequent section \myref{Reality_condition_section}, we discussed the problem of the reality condition that one has to apply before quantisation. One of the main results of this chapter, was that the correct reality condition to impose boils down to choosing that the ``usual'' metric perturbation $h$ is real. In other words, when we derived the canonical transformation that changes from the PCF back to the metric perturbation in EH gravity, in section \myref{Canonical_transformation_section}, we found that the quantity that gives $h_{ij}$ in terms of the connection $a_{ij}$ was the right field combination to be imposed to be real. This quantity was also extremely simple to define in terms of the previously introduced differential operators, and it read roughly as $h=\bar{D}a$. The chapter finally carries out the mode expansion of the connection field $a$ in section \myref{Mode_expansion_section}. Important considerations are made there on the choice of the relevant polarisation tensors. Indeed, this choice played a major role in the evaluation of the scattering amplitudes in the following chapter. Also important the conclusion that one of the polarisations of the connection must be intended as ``massive.'' Indeed, we saw that this asymmetry in the treatment of the two polarisations is a reflection of the fact that we are working in de Sitter instead of Minkowski. Finally, in section \myref{Discrete_symmetries_section}, we analysed the discrete symmetries CPT acting on the quantised connection field.

In chapter \myref{Amplitudes_chapter} we delved into the actual computations of the scattering amplitudes. All the content of this chapter is based on the work done by this author $et$ $al.$ in \cite{Delfino:2012aj}. Here we studied all the possible graviton-graviton scattering processes in the general theory. In GR the only non-trivial of these processes are those of the kind $++\rightarrow ++$, $--\to--$ and $+-\to+-$. However, in the more general theory, as mentioned, we encountered a wider range of non-trivial scattering amplitudes, such as the $++\rightarrow --$ and  $++\rightarrow +-$. These new amplitudes disappear in the GR limit as well as the interaction terms in  the Lagrangian that generate them. The chapter starts in section \myref{LSZ_section} with the discussion of the LSZ reduction formalism in the context of the theory previously introduced. Here we discussed the problem of how to take the Minkowski limit. In particular, we stressed how important it is to take the limit $after$ the amplitude is being computed, as doing so before would make us drop some terms which would have eventually produced a non-zero result. Then, in sections \myref{Gauge_fixing_propagator_section} and \myref{Interactions_section}, we went through the derivation of the Feynman rules for the propagator and for the interaction vertices (up to fourth order since we wanted to compute a four-graviton interaction). In the subsequent section \myref{spinor_technology_section}, instead, we finally introduced the $spinor$ formalism that is key in all our computations. This meant, in practice, introducing a map (the $soldering$ $form$) from the spacetime indices to a pair of $primed$-$unprimed$  $SL(2,\mathbb C)$ indices, and a map from the $SO(3)$ Latin indices to a pair of unprimed indices. The converted Lagrangian, using a convenient gauge, resulted greatly simplified. Remarkably we found that, in this formalism, the third order vertex comprises just a few terms and that even the fourth order vertex can be written in just a few lines. This is important given the fact that the usual fourth order GR vertex is composed by, at least, half a page of terms (see appendix in \cite{Goroff:1985th}). As a consequence of this simplicity, we showed that this formalism leads to very easy Feynman rules which we derived in \myref{Feynmans_rules_section}. Finally, in the last section \myref{graviton_graviton_scattering_section}, using such Feynman rules, we saw how to compute all possible graviton-graviton amplitudes of a general theory.% The results are encouraging in that they fully reproduce the known values for the amplitudes in the GR sector.

One of the final outcomes of this thesis, is that the PCF is a viable alternative to EH to tackle the issue of graviton scattering computation. Indeed, even if the theory itself introduced some technical challenges, as the need of a reality condition and the Minkowski limit issue, the final computations of the amplitudes were straightforward. % Indeed, this might have been expected as the premises of having a ``minimalistic'' approach with only one field in the Lagrangian drastically simplified the perturbative treatment. 
Furthermore, the promising results reported in this thesis, lead the present author, K.Krasnov and C.Scarinci to develop further the formalism. In fact, another paper will follow \cite{Delfino:2012zy} and \cite{Delfino:2012aj} dealing with general N-graviton scattering amplitudes given recursion relations which we can derive solely by looking at the Feynman rules of the theory.
 
Finally, the theory itself poses some interesting open issues to study. One example is to look into the ``unitarity'' of the theory at a non perturbative level (or prove it for every order in the perturbative expansion). In fact the PCF at linearised level, provided with a proper reality condition, is proved to be unitary by going to the metric variables (see section \myref{Canonical_transformation_section}); however,  the interactions could spoil this picture. Further study of this problem is needed. 

Another important problem is to study the renormalizability of this theory. In fact, although we have verified that the amplitudes of the GR sector agree with the expected results, we did so only at tree-level. Indeed, one can argue that if we include loops in the diagrams we could obtain a picture that is completely different from what happens in the usual approaches for GR. More research in this direction must be carried out.

Extending the theory is also possible in at least another direction, i.e. coupling to matter. The theory itself is formulated without the use of a metric tensor, therefore coupling to matter is problematic. %Still, there are two main ways one can circumvent this issue. First solution, and more obvious, is to use the Urbantke formula to produce a metric and use it to couple gravity to the energy-momentum tensor as usual. Another way 
One way to circumvent this issue is to enlarge the gauge group, especially to some $supersymmetric$ gauge group, having an $SU(2)$ embedded in it as symmetry of the gravity sector (see e.g. \cite{Krasnov:2011hi}). Adopting a SUSY group one could introduce fermionic and bosonic degrees of freedom into the theory, and have them couple to gravity. This also opens the door to possible Dark Matter models.

%%%%%%%%%%%%%%%%%%%%%%%%%%%%%%%%%%%%%%%%%%%%%%%%%%%%%%%%%%%
%%%%%%%%%%%%%%%%%%%%%%%%%%%%%%%%%%%%%%%%%%%%%%%%%%%%%%%%%%%
\appendix
%%%%%%%%%%%%%%%%%%%%%%%%%%%%%%%%%%%%%%%%%%%%%%%%%%%%%%%%%%%

%%%%%%%%%%%%%%%%%%%%%%%%%%%%%%%%%%%%%%%%%%%%%%%%%%%%%%%%%%%
\chapter{EH graviton-graviton scattering amplitude}\label{EH_Computation}
%%%%%%%%%%%%%%%%%%%%%%%%%%%%%%%%%%%%%%%%%%%%%%%%%%%%%%%%%%%
 
We work with the signature $(-,+,+,+)$ and our convention for the Riemann tensor is
$$R^\rho{}_{\sigma\mu\nu}=\partial_\mu\Gamma^\rho_{\nu\sigma}+\Gamma^\rho_{\mu\lambda}\Gamma^\lambda_{\nu\sigma}-(\mu\leftrightarrow\nu).$$ We consider small perturbations around the Minkowski metric $g_{\mu\nu}=\eta_{\mu\nu}+\kappa h_{\mu\nu}$. The constant $\kappa$ is introduced for dimensional reasons. Since $h_{\mu\nu}$ has dimensions of 
mass, we choose $\kappa^2=32\pi G$.

Note that
$$g^{\mu\nu}=\eta^{\mu\nu}-\kappa h^{\mu\nu}+\kappa^2 h^{\mu\rho}h_{\rho}{}^\nu-\kappa^3h^{\mu\rho}h_{\rho\sigma}h^{\sigma\nu}$$
so that
$$\delta g^{\mu\nu}-\kappa h^{\mu\nu},\quad \delta^2g^{\mu\nu}=2\kappa^2 h^{\mu\rho}h_{\rho}{}^\nu,\quad\delta^3g^{\mu\nu}=-3!\kappa^3h^{\mu\rho}h_{\rho\sigma}h^{\sigma\nu}.$$

We want to find Feynman rules for the graviton, therefore we need the variations of the Einstein-Hilbert action
$$S[g]=\frac{2}{\kappa^2}\int d^4x\sqrt{-g} R.$$ 

The actual computations of the variations up to the third order are rather lengthy and not particularly illuminating. We report here the results and refer to \cite{Goroff:1985th} for the details. For the second variation, including the gauge-fixing term, we have
\be
\nn S^{(2)} +  S_{gf} = \int d^4 x \bigg[ -\f{1}{2}(\p_\r h_{\m\n})^2+(\p^\m h_{\m\n})^2 - \p_\m h^{\m\n} \p_n h + \f{1}{2}(\p_\m h)^2 -\a \left( \p^\m h_{\m\n}-\f{1}{2}\p_\n h \right)^2 \bigg].
\\
\ee
We will choose $\a = 1$, the so called de Donder gauge. Then, employing the usual QFT machinery, we get the following propagator

\be
\Delta (p) =  \f{\eta^{\m \r} \eta^{\n\s}+\eta^{\m \s} \eta^{\n\r}- \eta^{\m \n} \eta^{\r\s} }{ 2p^2}.
\ee

For the third variation instead, we can already introduce an expedient that will simplify the interaction. Since we are interested in a tree-level process with four external legs, we can already set to be $on$-$shell$ two out of three fields in the vertex. We denote the off-shell field with an upper-case $H$, while the lower case fields are now to be understood as on-shell. Evaluated on Minkowski, the interaction reads:

\be
\nn S^{(3)}= \kappa \int d^4 x \bigg[ -H^{\m\n} h_{\r\s}\p_\m \p_\n h^{\r\s}-H_{\r\s}h^{\m\n}\p_\m\p_\n h^{\r\s}+2H^\n{}_\s h^\m{}_\r
\p_\m \p_\n h^{\r\s}+\f{3}{4}H \p^\l h_{\r\s}\p_\l h^{\r\s}
\\ \nn
-\f{1}{2}H^{\m\n}\p_\m h_{\r\s}\p_\n h^{\r\s}-H^\m{}_\r\p^\l h_{\m\s}\p_\l h^{\r\s}-\f{1}{2}H\p_\s h_{\m\r}\p^\m h^{\r\s}+H^\m{}_\n \p_\r h_{\m\r} \p^\r h^{\n\s}
\bigg].
\\
\ee

In momentum space this yields

\be
\nn S^{(3)}(p_1,p_2,p_3)= \f{\k}{2} \int \f{d^4 p_1}{(2\pi)^4}\f{d^4 p_2}{(2\pi)^4} h^{\m\n}(p_1)h^{\r\s}(p_2)H^{\a\b}(p_3) 
\\
\nn \bigg[
\eta_{\m\r}\eta_{\n\s}\big(p_{1\a}p_{1\b}+p_{2\a}p_{2\b}+p_{1\a}p_{2\b}-\f{3}{2}\eta_{\a\b}p_1^\l p_{2\l}\big)
\\
\nn -2 \eta_{\m\r}\big( \eta_{\a\n} p_{1\s}p_{1\b}+\eta_{\a\s} p_{2\n}p_{2\b}-\eta_{\n\a} \eta_{\s\b}  p_1^\l p_{2\l} -\f{1}{2} \eta_{\a\b} p_{2\n} p_{1\s} \big)
\\
  + \eta_{\a\m}\eta_{\b\n}p_{1\r}p_{1\s}+\eta_{\a\r}\eta_{\b\s}p_{2\m}p_{2\n}-2\eta_{\a\m}\eta_{\b\r}p_{2\n}p_{1\s}
\bigg].
\ee

Thus the vertex is

\be\label{V3_EH}
V^{(3)}_{\m\n\r\s\a\b}(p_1,p_2)= \f{\k}{2} \bigg[
\eta_{\m\r}\eta_{\n\s}\big(p_{1\a}p_{1\b}+p_{2\a}p_{2\b}+p_{1\a}p_{2\b}-\f{3}{2}\eta_{\a\b}p_1^\l p_{2\l}\big)
\\
\nn -2 \eta_{\m\r}\big( \eta_{\a\n} p_{1\s}p_{1\b}+\eta_{\a\s} p_{2\n}p_{2\b}-\eta_{\n\a} \eta_{\s\b}  p_1^\l p_{2\l} -\f{1}{2} \eta_{\a\b} p_{2\n} p_{1\s} \big)
\\
  + \eta_{\a\m}\eta_{\b\n}p_{1\r}p_{1\s}+\eta_{\a\r}\eta_{\b\s}p_{2\m}p_{2\n}-2\eta_{\a\m}\eta_{\b\r}p_{2\n}p_{1\s}
\bigg].
\ee

We now need to introduce the helicity tensors for the gravitons' external states. We refer to section \myref{spinor_technology_section} for a formal introduction to the technology of $SL(2,\mathbb C)$ spinors. In this formulation of GR, the helicity states can be defined as follows

\be
\nn\e^{\m\n}_{++}(p) =  \e^\m_+(p) \e^\n_+(p) , \quad \e^{\m \n}_{--}(p)  = \e^\m_-(p) \e^\n_-(p),
\\
\ee
where

\be
\e_{+}^{\m}(p) = \f{q^{M'} p^M \t^\m_{MM'} }{p_{A'}q^{A'}}, \quad \e_{-}^{\m}(p) = \f{q^M p^{M'} \t^\m_{MM'} }{p^{A}q_{A}}.
\ee

Here, $p$ is the momentum of the particle, $\t^\m_{MM'}$ is the soldering form (see section \myref{spinor_technology_section}) and the $q$'s are $reference$ spinors. These reference spinors can be chosen $arbitrarily$ for each external state. Also, a noteworthy consideration is that the helicities are essentially the complex conjugate of each other. This, indeed, is a reflection of the explicit parity invariance of the EH formulation. 

Now we have all the ingredients to compute a graviton-graviton scattering amplitude. Indeed, we do not need the four-vertex for our purposes. This is a great simplification since the fourth variation of the EH action, even with the fields on-shell and the background metric taken to be flat, occupies about half a page (see \cite{Goroff:1985th}, formula (A.6) of the Appendix). To understand why we do not need it let us examine the kind of terms we find in the fourth variation. We know that all the terms are given by contractions of four fields and two derivatives:

\be\label{S4_EH}
S^{(4)} \sim \sum h h (\p h)^2. 
\ee

We then can see what happens when we insert on the four external legs their respective helicity tensors. Let us compute, for instance, the amplitude of the process $--\rightarrow ++$. We can then choose out of convenience the following values for the reference momenta:

$$q_1=q_2=p_3,\quad q_3=q_4=p_2.$$
Therefore the $q$'s of the negative-helicity incoming gravitons are equal to the momentum of one of the positive-helicity outgoing gravitons. Similarly, the reference momenta of the outgoing gravitons are equal to the momentum of one of the incoming particles. Now we can look again at the vertex prototype in \myref{S4_EH}. We see that the two derivatives $\p$ will produce, in momentum space, two spinors of the kind $p^M_i$ and two $p^{M'}_i$, where $i$ is the label of the external leg the derivative acts on. Now, using the rules we defined for the reference spinors, we can write the four helicities we have at our disposal. Forgetting about the denominators and dropping the indices, we have

\be
\nn p_3 p_3 p_1'p_1', \quad  p_3 p_3 p_2'p_2';
\\
\nn  p_3 p_3 p_2'p_2',\quad  p_4 p_4 p_2'p_2'.
\ee
We notice that we have an excessive amount of primed spinors $p_2'$ and of unprimed spinors $p_3$. This is what we wanted. Now, reminding the anti-commutative nature of the contraction of the spinors ( $p_1^M p_{1M}=0$ ), we can quickly see why the contribution from the fourth vertex is automatically zero. We know that the vertex itself can only contract the external legs' helicities, in different ways, between themselves and with two derivatives. Then if the vertex contracts two identical spinors we know that the terms leads a zero contribution. Let us then start contracting the spinors assuming the most optimistic case possible. We then have, for instance, a contraction between the two unprimed $p_3$ coming from the second particle with the two unprimed $p_4$ from the fourth. Similarly, we can contract the primed spinors of the first particle and the third particle. We are left with four unprimed $p_3$ and four primed $p_2'$. We can at last assume that the derivatives drop the following convenient spinors $p_1 p_1' p_4 p_4'$. Then we see that we can successfully contract all of these with two $p_3$'s and two $p_2'$, but we are still left with the following four spinors: $p_3p_3p_2'p_2'.$

Given that we can only contract primed with primed and unprimed with unprimed, we have no choice but to contract two identical spinors. Therefore we can conclude that the fourth vertex contribution is always zero. Remarkably, we avoided computing half a page of terms  just by counting the number of derivatives and using a smart choice 
for the reference spinors.

	\begin{figure}[!htb] 
	\begin{center}
	\begin{tabular}{cccccccccccccccc}    %Adds several centered Columns

% 	
% 	
% 	\begin{fmffile}{four}
% 	  \fmfframe(1,7)(1,7){ 
% 	   \begin{fmfgraph*}(110,62)
% 	    \fmfleft{i1,i2}
% 	    \fmfright{o1,o2}
% 	    \fmflabel{$1$}{i1} %Labels one of the left sources
% 	    \fmflabel{$2$}{i2} %Labels one of the left sources
% 	    \fmflabel{$3$}{o1} %Labels one of the right outputs
% 	    \fmflabel{$4$}{o2} %Labels one of the right outputs
% 	    \fmf{photon}{i1,v,i2}
% 	    \fmf{photon}{o1,v,o2}
% 	   \end{fmfgraph*}
% 	  }
% 	\end{fmffile}
% 	
% 	&&&
	
	\begin{fmffile}{schannel} 	%one.mf will be created for this feynman diagram  
	  \fmfframe(1,7)(1,7){ 	%Sets dimension of Diagram
	   \begin{fmfgraph*}(110,62) %Sets size of Diagram
	    \fmfleft{i1,i2}	%Sets there to be 2 sources 
	    \fmfright{o1,o2}    %Sets there to be 2  outputs
	    \fmflabel{$1$}{i1} %Labels one of the left sources
	    \fmflabel{$2$}{i2} %Labels one of the left sources
	    \fmflabel{$3$}{o1} %Labels one of the right outputs
	    \fmflabel{$4$}{o2} %Labels one of the right outputs
	    \fmf{photon}{i1,v1,i2} %Connects the sources with a vertex.
	    \fmf{photon}{o1,v2,o2} %Connects the outputs with a vertex.
	    \fmf{photon,label=$ $}{v1,v2} %Labels the connecting line.
	   \end{fmfgraph*}
	  }
	\end{fmffile}
	%&&&&

	\begin{fmffile}{uchannel}
	  \fmfframe(1,7)(1,7){ 
	   \begin{fmfgraph*}(110,62)
	    \fmfleft{i1,i2}
	    \fmfright{o1,o2}
	    \fmflabel{$1$}{i1}
	    \fmflabel{$2$}{i2}
	    \fmflabel{$3$}{o1}
	    \fmflabel{$4$}{o2}
	    \fmf{photon}{i1,v1,o1}
	    \fmf{photon}{i2,v2,o2}
	    \fmf{photon,label=$ $}{v1,v2}
	   \end{fmfgraph*}
	  }
	\end{fmffile}

	\begin{fmffile}{tchannel}
	\fmfframe(1,7)(1,7){ 
		\begin{fmfgraph*}(110,62)
		\fmfleft{i1,i2}
		\fmfright{o1,o2}
		\fmflabel{$1$}{i1} %Labels one of the left sources
		\fmflabel{$2$}{i2} %Labels one of the left sources
		\fmflabel{$3$}{o1} %Labels one of the right outputs
		\fmflabel{$4$}{o2} %Labels one of the right outputs
		\fmf{photon}{i1,v1}
		\fmf{phantom}{v1,o1} % Invisible rubber band
		\fmf{photon}{i2,v2}
		\fmf{phantom}{v2,o2} % also invisible rubber band
		\fmf{photon,label=$ $}{v1,v2}
		% These are visible, but have no tension.
		\fmf{photon,tension=0}{v1,o2}
		\fmf{photon,tension=0}{v2,o1}
		%\fmfdot{v1,v2}
		\end{fmfgraph*}
	}
	\end{fmffile}

	\end{tabular}

	\caption{\label{Channels} S-Channel, T-Channel, U-Channel}\label{fey1}
	\end{center}
	\end{figure}
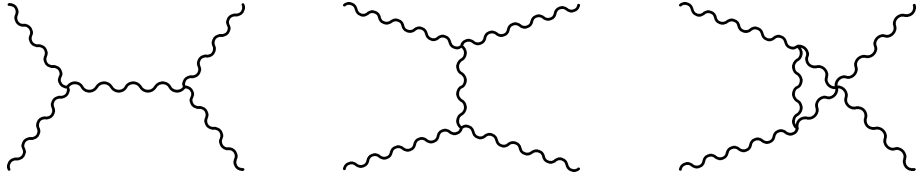

We can finally concentrate on the third vertex \myref{V3_EH}. Again a simple counting argument tells us that the non-vanishing contributions are all of the type
\be
\f{1}{p^2}(\e_1\cdot\e_4)^2(e_2\cdot p)^2(\e_3 \cdot p)^2,
\ee
where with the dot we indicate the complete contraction of the spinors. The contribution of the $s$ channel, with particles $1$ and $2$ in the same vertex is

\be
\nn \im V_3(1,2)\f{\Delta}{\im} \im V_3(3,4) = 2 \im \f{\k^2}{2}\f{(\e_1\cdot\e_4)^2(e_2\cdot p_1)^2(\e_3 \cdot p_4)^2}{(p_1+p_2)^2}
=\f{\im\k^2}{2}\f{\bra{1}{2}^4 \ket{3}{4}^4  }{\bra{1}{2}\ket{1}{2}\bra{3}{2}^2\ket{2}{3}^2},
\\
\ee

where we used the notation $p_{1M'}p_2^{M'}= \bra{1}{2}$ and $p_1^Mp_{2M}=\ket{1}{2}.$ Note that we added a symmetry factor of 2  in front of the amplitude. Similarly, the $t$ channel, with $1$ and $3$ sharing the same vertex

\be
\nn \im V_3(1,3) \f{\Delta}{\im} \im V_3(2,4) = 2\im \f{\k^2}{2}\f{(\e_1\cdot\e_4)^2(e_2\cdot p_4)^2(\e_3 \cdot p_1)^2}{(p_1+p_3)^2}
=\f{\im \k^2}{2}\f{\bra{1}{2}^4 \ket{3}{4}^4  }{\bra{1}{3}\ket{1}{3}\bra{3}{2}^2\ket{2}{3}^2}.
\\
\ee
 
While the remaining $u$ channel is zero since each term contain contractions of the kind $(\e_2\cdot p_3)$ or $(\e_3\cdot p_2)$ which vanish with our choice of reference spinors.
We already see that the two amplitudes coincide with the results obtained in \myref{amp-1} and \myref{amp-2}. If we sum them and use the usual momentum conservation identities, e.g. $\bra{1}{4}/\bra{1}{3}=-\ket{2}{3}/\ket{2}{4}$, we get

\be 
{\cal M}^{--++} = 2\im \left(\frac{\kappa}{2}\right)^2 \ket{3}{4}^6 \frac{1}{\ket{1}{3}\ket{1}{4}\ket{2}{3}\ket{2}{4}} \frac{\bra{1}{2}}{\ket{1}{2}},
\ee

Which is identical to what we obtained in \myref{--++1} computing the same amplitude in the PCF.

%%%%%%%%%%%%%%%%%%%%%%%%%%%%%%%%%%%%%%%%%%%%%%%%%%%%%%%%%%%
\chapter{Alternative derivation of the Pure Connection Action}\label{Appendix_AltDerivationOfAction}
%%%%%%%%%%%%%%%%%%%%%%%%%%%%%%%%%%%%%%%%%%%%%%%%%%%%%%%%%%%

We derive here the same action for the pure connection formulation of General Relativity that we obtain in section \myref{Section_From_PB_to_GR}. Here however we will follow a slightly different approach using only matrix manipulations. If we define the quantity:

\be \label{phi_def}
 \quad \Phi^{ij} = \left( \Psi^{ij} + \f{\Lambda}{3}\d^{ij} \right),
\ee

we see that we can re-write the action in (\ref{Pure_connection_with_LM3}) as following

\be\label{Pure_connection_with_LM2}
S (A, \Psi)=\frac{\im}{16\pi G}\int dx^4 \ \Phi^{-1}_{ij} \tilde{X}^{ij}.
\ee

Thus the new equation of motion for $\Psi$ is  

\be
%  B^i\wedge B^j = \f{\d^{ij}}{3}\d^{kl} B^k\wedge B^l=
%  \\
%  (\Phi^{-1})^i_m F^{m}\wedge (\Phi^{-1})^j_n F^{n} =  \f{\d^{ij}}{3}\d^{kl} (\Phi^{-1})^k_m F^{m}\wedge (\Phi^{-1})^l_n F^{n} =
%  \\
%   (\Phi^{-1})^i_m (\Phi^{-1})^j_n  F^{m}\wedge F^{n} =  \f{\d^{ij}}{3}  (\Phi^{-1})^k_m (\Phi^{-1})^k_n F^{m}\wedge  F^{n}  
%   \\
%     (\Phi^{-1})^i_m (\Phi^{-1})^j_n  d^4x *(F^{m}\wedge F^{n}) =  \f{\d^{ij}}{3}  (\Phi^{-1})^k_m (\Phi^{-1})^k_n d^4 x *(F^{m}\wedge F^{n})  
%      \\
%     (\Phi^{-1})^i_m (\Phi^{-1})^j_n  X^{mn} =  \f{\d^{ij}}{3}  (\Phi^{-1})^k_m (\Phi^{-1})^k_n X^{mn}
%     \\
%     (\Phi^{-1})^i{}_m X^{mn} (\Phi^{-1})^j{}_{n} = \f{\d^{ij}}{3}  (\Phi^{-1})^k{}_m (\Phi^{-1})^k{}_n X^{mn}
%      \\
    (\Phi^{-1})^i{}_k X^{kj}  = \f{1}{3} \Tr{(\Phi^{-1}) (\Phi^{-1}) X } \Phi^{ij},
\ee
or more simply in matrix form:

\be\label{EOM_PHI}
\Phi^{-1} X = \f{1}{3}\Tr{\Phi^{-1}\Phi^{-1} X} \Phi.
\ee

If we define the square root of a matrix as $\sqrt{M}^{ik}\sqrt{M}^{kj} = M^{ij}$, which we can for a positive definite matrix, then we can solve \myref{EOM_PHI} for $\Phi$

\be\label{PHI_sol}
\Phi = \sqrt{ \f{3X}{ \Tr{\Phi^{-1}\Phi^{-1} X} } }.
\ee

We can also trace the left and right hand sides of \myref{EOM_PHI} and use \myref{PHI_sol} to obtain

\be\label{TrPhiX}
\Tr{\Phi^{-1} X} = \f{1}{\sqrt{3}}\sqrt{\Tr{\Phi^{-1}\Phi^{-1} X}} \Tr{\sqrt{X}}.
\ee

Now we can use the definition of $\Phi$ \myref{phi_def} to see that $\Tr{\Phi} = \Lambda$. Thus from \myref{EOM_PHI} we also have

\be
\Tr{\Phi^{-1} X} = \f{1}{3}\Tr{\Phi^{-1}\Phi^{-1} X} \Lambda.
\ee

Substituting the latter in  \myref{TrPhiX} we finally obtain

\be 
\Tr{\Phi^{-1} X} = \f{1}{\sqrt{\Lambda}}\sqrt{\Tr{\Phi^{-1} X} } \Tr{\sqrt{X}},
\ee
which in turns implies

\be 
\Tr{\Phi^{-1} X} = \f{1}{ \Lambda }  \Tr{\sqrt{X}}^2.
\ee
 
 Therefore we can rewrite the action \myref{Pure_connection_with_LM2} as

\be 
S (A)=\frac{\im}{16 \Lambda \pi G}\int dx^4 \ \Tr{\sqrt{X}}^2.
\ee

%%%%%%%%%%%%%%%%%%%%%%%%%%%%%%%%%%%%%%%%%%%%%%%%%%%%%%%%%%%
\chapter{Modifications of GR and of defining coefficients}\label{Appendix_Mod_GR}
%%%%%%%%%%%%%%%%%%%%%%%%%%%%%%%%%%%%%%%%%%%%%%%%%%%%%%%%%%%

In this section we consider a simple one-parameter family of deformations of GR. This family has the advantage that the modifications one introduces are guaranteed to be ones relevant at the Planck scale. In contrast, as we have seen in the main text, a generic defining function would give rise to strong coupling phenomena at much lower energy scales. 

As in \cite{Krasnov:2011up}, consider the following Lagrangian
\be\label{mod-action}
\nn S[B,A,\Psi] = \frac{1}{8\pi \im G}\int \left[B^i\wedge F^i (A)- \frac{1}{2} \left( \Psi^{ij} - \frac{\Lambda}{3}\d^{ij}+ \frac{\b}{2M_p^2}  \Tr{\Psi^2} \d^{ij} \right) B^i\wedge B^j\right].
\\
\ee
When $\beta=0$ this is just the Lagrangian of GR. One obtains a pure connection Lagrangian for GR in the form (\ref{GEN_action}) by integrating out first the two-form field $B^i$, and then the Lagrange multiplier field $\Psi^{ij}$. On-shell, the Lagrange multiplier field $\Psi^{ij}$ receives the meaning of the self-dual part of the Weyl curvature. When $\beta\not=0$ we have added to the Lagrangian a term that becomes important when $(Weyl)/M_p^2$ is of order unity. Thus, (\ref{mod-action}) is guaranteed to produce only Planckian modifications of GR (for $\beta$ of the order unity).

Integrating out the two-form field in (\ref{mod-action}) one gets
\be
S[A,\Psi] = \frac{1}{16\pi \im G}\int\left( \Psi^{ij} - \frac{\Lambda}{3}\d^{ij}+ \frac{\beta}{2M_p^2}  \Tr{\Psi^2} \d^{ij} \right)^{-1} F^i\wedge F^j.
\ee
To prepare this functional for minimising with respect to $\Psi$ let us rescale $\tilde{\Psi}=\Psi/M^2$, where as before $M^2=\Lambda/3$. We can then rewrite the above action as
\be\label{S-A-Psi}
S[A,\tilde{\Psi}] = \frac{\im}{\alpha}\int\left( \delta^{ij} - \tilde{\Psi}^{ij} - \frac{\alpha\b}{2}  \Tr{\tilde{\Psi}^2} \d^{ij} \right)^{-1} F^i\wedge F^j.
\ee
The new Lagrange multiplier field $\tilde{\Psi}^{ij}$ is dimensionless, and 
\be
\alpha:= \frac{M^2}{M_p^2},
\ee
where as usual $M_p^2=1/16\pi G$. This way of writing the functional shows that deviations from GR are parametrised by $\beta$ always in the combination $\alpha\beta$. 

We can now extremise (\ref{S-A-Psi}) by passing to eigenvalues of both the $F^i\wedge F^j$ and $\tilde{\Psi}^{ij}$ matrices, as in \cite{Krasnov:2011pp}. Denoting the eigenvalues of $F^i\wedge F^j$ by $\lambda_{1,2,3}$, i.e. $F^i\wedge F^j = {\rm diag}(\lambda_1,\lambda_2,\lambda_3) \,d^4x$, we obtain the following expression for the extremum of (\ref{S-A-Psi}) as a function of the eigenvalues 
\be
S[A] = \im \int f[\lambda] \, d^4 x,
\ee
where 
\be\label{f-exp}
\nn f[\lambda] = \frac{ (\sum_i \sqrt{\lambda_i})^2 }{3\a} \left( 1 + \frac{3\a \b}{2}\right.
\\ 
\nn -\frac{9(\a \b)^2}{4(\sum_i \sqrt{\lambda_i})^3}
\bigg(4 \lambda _1^{3/2}-3 \lambda _1 \sqrt{\lambda _2}-3 \sqrt{\lambda _1} \lambda _2+4 \lambda _2^{3/2}-3 \lambda _1 \sqrt{\lambda _3}
\\
\nn +6 \sqrt{\lambda _1} \sqrt{\lambda _2} \sqrt{\lambda _3}-3 \lambda _2 \sqrt{\lambda _3}-3 \sqrt{\lambda _1} \lambda _3-3 \sqrt{\lambda _2} \lambda _3+4 \lambda _3^{3/2}\bigg)
\\ 
 \nn - \frac{27}{8}\frac{\a^3  \b^3 }{\left(\sqrt{\lambda _1}+\sqrt{\lambda _2}+\sqrt{\lambda _3}\right){}^4} \left(16 \lambda _1^2-17 \lambda _1^{3/2} \sqrt{\lambda _2}+6 \lambda _1 \lambda _2-17 \sqrt{\lambda _1} \lambda _2^{3/2} \right.
\\
\nn +16 \lambda _2^2-17 \lambda _1^{3/2} \sqrt{\lambda _3}+12 \lambda _1 \sqrt{\lambda _2} \sqrt{\lambda _3}
 +12 \sqrt{\lambda _1} \lambda _2 \sqrt{\lambda _3}-17 \lambda _2^{3/2} \sqrt{\lambda _3}
\\ \left. \left.
\nn +6 \lambda _1 \lambda _3+12 \sqrt{\lambda _1} \sqrt{\lambda _2} \lambda _3+6 \lambda _2 \lambda _3-17 \sqrt{\lambda _1} \lambda _3^{3/2}-17 \sqrt{\lambda _2} \lambda _3^{3/2}+16 \lambda _3^2\right)
+ O( (\a \b)^4 ) \right).
\\
\ee

This expression is sufficient to compute various coupling constants that appear in (\ref{f_second_variation})-(\ref{f_fourth_variation}). Thus, from (\ref{f_second_variation}) we have
\be
 - \frac{g^{(2)}}{3}= \frac{\partial^2 f}{\partial \lambda_1\partial \lambda_1}\bigg|_{\lambda_i = 1}.
\ee
This gives
\be
g^{(2)} = \frac{1}{ \alpha } \left(1+\frac{3 \alpha  \b}{2}+\frac{9 \alpha ^2 \b^2}{4}+\frac{27 \alpha ^3 \b^3}{8}+  O[\a \b]^4\right).
\ee
Similarly, from (\ref{f3}) we have
\be
\frac{\partial^3 f}{\partial \lambda_1\partial \lambda_1 \partial \lambda_1}\bigg|_{\lambda_i = 1} =  \, \frac{2g^{(3)}}{9}  + \frac{g^{(2)}}{3},
\ee
which gives
\be
g^{(3)} = \frac{3}{4\a}   \left( 1 +    \frac{3}{2} \a \b  - \frac{27}{4}\a^3 \b^3
+O[\alpha\b]^4\right).
\ee
We can now see how the coefficient in front of the vertex (\ref{V-2-app}) scales with $M$. We have
\be\label{diff-g-1}
g^{(3)} -\frac{3}{4}g^{(2)} =-\frac{27}{16} \a \b^2 + O(  \a^2 \b^3 ),
\ee   
and therefore this goes as $M^2/M_p^2$. 

Also, for $g^{(4)}$ in (\ref{f_fourth_variation}), the fourth derivative with respect to one of the eigenvalues gives
\be
\frac{\partial^4 f}{\partial \lambda_1 \partial \lambda_1\partial \lambda_1 \partial \lambda_1}\bigg|_{\lambda_i = 1} = -\frac{2 g^{(4)} }{9}-\frac{16  g^{(3)} }{27}-\frac{4  g^{(2)} }{9}, 
\ee
which in turn gives
\be
g^{(4)} = \frac{104+156  \alpha\b + 126 \alpha^2\b^2  +297  \alpha^3\b^3}{128 \alpha } +O(\alpha^3\beta^4).
\ee
The coefficient appearing in the 4-vertex and in the $++++$ amplitude then reads
\be\label{diff-g-2}
2g^{(4)}-g^{(3)}-\frac{7}{8}g^{(2)}  = \frac{27}{4} \a^2 \b^3 +O(\alpha^3\beta^4).
\ee
Note that this goes as a higher power of $M$ than the similar difference (\ref{diff-g-1}). 

Finally, from (\ref{f-exp}) we see that
\be
f(\delta)=\frac{3}{\alpha}\left( 1+ \frac{3\alpha\beta}{2} + O((\alpha\beta)^3)\right).
\ee
This implies that
\be\label{g2-f}
3g^{(2)}-f(\delta)= \frac{27\alpha\beta^2}{4} + O(\alpha^2\beta^3).
\ee

\chapter{Self dual Forms}\label{Self_dual_forms_Appendix}

Here we report some useful formulae and relations involving the Self dual forms \myref{self_dual_forms}. We can start by re-stating the (de Sitter) metric we are using for the reader convenience:

\be\label{de_sitter_metric_in_Appendix}
ds^2 = a^2 \left( -dt^2 + \sum_i (dx^i)^2\right).
\ee
   
The tetrad $\theta^I$, $I=0,1,2,3$ associated to the above metric reads:

\be
\theta^0 = a dt, \quad \theta^i = a d x^i,
\ee

so that $d s^2 = \theta^I \otimes \theta^J \eta_{IJ},$ where $\eta_{IJ} = diag(-1,1,1,1).$ Then it is convenient to rewrite the 2-forms \myref{self_dual_forms} as

\be\label{self_dual_forms_with_tetrads}
\S^i  = a^2 \left( \im \t^0 \wedge \t^i + \f{1}{2}\e^{ijk} \t^j \wedge \t^k \right)
\ee 

where $i=1,2,3.$ It is not hard to check that, indeed, the two forms \myref{self_dual_forms_with_tetrads} are self dual with respect to the Hodge star operation on two-forms defined by the metric in \myref{de_sitter_metric_in_Appendix}:

%[where does the factor 2 come from:] 
% 
% \be
% \S^i = \f{1}{2}\S^i_{\m\n} d x^\m \wedge d x^\n
% \ee
% but i can also change basis $dt = \t^0_\m d x^\m$
% \be
% \S^i = \left( \im \t^0_{[\m}  \t^i_{\n]} + \f{1}{2}\e^i{}_{jk} \t^j_{[\m}  \t^k_{\n]} \right)d x^\m \wedge d x^\n
% \ee
% 
% therefore
% 
% \be
% \S^i_{\m\n} = 2\left( \im \t^0_{[\m}  \t^i_{\n]} + \f{1}{2}\e^i{}_{jk} \t^j_{[\m}  \t^k_{\n]} \right)
% \ee

%there is a factor of 2 going to explicit forms \mu \nu
\be
\e_{\m\n}{}^{\r\s} \S^i_{\r\s}{}   %= \e_{\m\n}{}^{\r\s} 2 \left( \im \t^0_{[\r}  \t^i_{\s]} + \f{1}{2}\e^i{}_{jk} \t^j_{[\r}  \t^k_{\s]} \right)=
% \\
%  2\im \e_{\m\n}{}^{0 i} +  2 \f{1}{2}\e_{\m\n}{}^{\r\s}\e^{0i}_{\m\n} =
% \\
% 2\im \e_{\m\n}{}^{0 i} + 2\f{1}{2}(-4)\d^0_{[r} \d^i_{\s]} =
% \\
%  2\im\left( 2\f{1}{2} \e_{\m\n}{}^{0 i} + 2\im  \d^0_{[r} \d^i_{\s]} 
% \right)
% \\
%  2\im\left( 2\f{1}{2}\e^{oi}{}_{jk}\t^j_{[\m}\t^k_{\n]} + 2\im \t^0_{[\m}  \t^i_{\n]} 
% \right)
= 2\im \S^i_{\m\n}{} .
\ee

Here the object $\e_{\m\n}{}^{\r\s}$ is obtained from the volume form 
$\e_{\m\n\r\s}$ by rising two of its indices using the metric, and in our conventions $\e^{0123} = \e^{123} = +1$.

Dropping the conformal factor $a(t)$ we define the Minkowski counterpart of the above two-forms. We will denote such basis self-dual forms simply by $\S^i_{\m\n}{}_{M}$, $i =1,2,3$. 

\be
\S^i_M = \im dt \wedge dx +\f{1}{2}\e^{ijk} dx^j \wedge dx^k.
\ee

From the above definition, we can read off the components of the basis self-dual two forms under space$+$time decomposition. We have:
\be
\S^i_{0j}{}_M %= 2\im \t^0_{[0}\t^{i}_{j]} = 2\im  \d^0_{[0}\d^{i}_{j]}
 = \im \d^i_j, \quad
\S^i_{jk}{}_M = \e^i_{jk},
\ee 
where $\d^i_j$ is the Kronecker-delta. Then it is not hard to verify that the $\S^i$ as well as $\S^i_{M}$ satisfy the following relations

\be
\e^{\m\n\r\s}\S^i_{\m\n}\S^j_{\r\s} = 8 \im \d^{ij}
\ee
and 

\be\label{sigma_algebra}
\S^i_\m{}^\a \S^j_{\a\n} = -\d^{ij}g_{\m\n}-\e^{ijk}\S^k_{\m\n}.
\ee
Thus, the basic self-dual two-forms satisfy an algebra similar to that of Pauli matrices.

\chapter{Centre of mass frame momentum spinors and Mandelstam variables}

The centre of mass frame expressions for our momentum spinors can be obtained from (\ref{kA}). We take the particles $1,2$ to be moving in the direction of the $\vec{z}$ axes, positive and negative respectively. This gives
\be
1^A = 2^{1/4} \sqrt{\omega_k} o^A, \qquad 2^A = \im 2^{1/4} \sqrt{\omega_k} \iota^A.
\ee
The particles $3,4$ we take to be the scattered ones, moving at an angle $\theta$ to the $\vec{z}$ axes. For simplicity we put $\phi=0$. We get
\be
\nn 3^A = 2^{1/4}\sqrt{\omega_k} \left( \sin(\theta/2) \iota^A + \cos(\theta/2) o^A\right), 
\\ 
4^A = \im 2^{1/4}\sqrt{\omega_k} \left( \sin(\theta/2) o^A - \cos(\theta/2) \iota^A\right).
\ee
The non-zero contractions are
\be\label{contr}
\ket{1}{2} = -\im \sqrt{2}\omega_k, \quad \ket{1}{3} = - \sqrt{2}\omega_k \sin(\theta/2), \quad \ket{1}{4} = \im \sqrt{2}\omega_k \cos(\theta/2), \\ \nonumber
\ket{2}{3} = \im \sqrt{2}\omega_k \cos(\theta/2), \quad \ket{2}{4} = - \sqrt{2}\omega_k \sin(\theta/2), \quad \ket{3}{4} = \im \sqrt{2}\omega_k.
\ee
It is also customary to introduce the following Mandelstam variables
\be\label{mandel}
s = -4\omega_k^2, \quad t= 4\omega_k^2 \sin^2(\theta/2), \quad u=4\omega_k^2 \cos^2(\theta/2).
\ee

\chapter{Feynman rules}

In this Appendix we collect all the Feynman rules derived in the main text, directly in spinor notations that are most convenient for practical computations. 

In spinor notations the vertices involve only the factors of momentum spinors $k^{AA'}$ on the legs of the vertex, as well as factors of spinor metrics $\epsilon^{AB}, \epsilon^{A'B'}$. Writing down the corresponding expressions can quickly lead to horribly looking formulas. For this reason it is much more efficient to draw the vertices, indicating the factors of $\epsilon^{AB}, \epsilon^{A'B'}$ by lines. The only drawback of this procedure is that it is not easy to keep track of the signs (remember that raising-lowering a pair of spinor indices induces a minus sign). We have not tried to develop any convention for these signs, just going back to the corresponding term in the Lagrangian and seeing how the indices contract when there is a question. But it is possible that a more systematic sign convention can be developed. Here, in view of the fact that only simple computations are done, our rules are sufficient. 

With these remarks in mind, let us state the rules of the game. First, we only state here the rules of computing the Minkowski space amplitudes. The way these are obtained as a limit of more general de Sitter graviton amplitudes is explained in the main text. The field that propagates in our theory is an ${\rm SU}(2)$ connection, but after all the gauge-fixings and translation into the spinor notations, this is a field $a_{ABCC'}$ with 4 spinor indices, 3 unprimed and one primed. It is moreover symmetric in its 3 unprimed spinor indices, thus forming an object that takes values in an irreducible representation of the Lorentz group. Thus, only $4\times 2=8$ components of the field propagate, as compared to $10$ in the usual metric treatment. As in any textbook example, the scattering amplitudes are obtained from the field (connection) correlation functions by certain reduction formulas. These are most practical in the momentum space representation, where the correlation function is that of the Fourier coefficients of the field operator. The scattering amplitude then reads
\be\label{LSZ-app}
\langle k_- \ldots  k_+ \ldots  | p_- \ldots p_+ \ldots \rangle  = (-1)^{m+m'} (2\pi)^4 \delta^4\left(\sum k - \sum p\right)  
\\ \nonumber
\, \langle T \e^+(k_-) \cdot a(k_-) \ldots \e^-(k_+) \cdot a(k_+)\ldots  \e^-(p_-) \cdot a(p_-) \ldots  \e^+ (p_+) a(p_+)\rangle_{\rm amp},
\ee
where we denoted the contractions of the spinor indices involved by a dot, in other words $\e(k) \cdot a(k) \equiv \e_{ABCC'}(k) a^{ABCC'}(k)$, and $k_-, k_+$ are the momenta of the outgoing negative and positive helicity gravitons, $p_-, p_+$ are those of the incoming particles, and $m,m'$ are the numbers of positive helicity outgoing and incoming gravitons respectively. As usual, the momentum space amplitude is amputated from its external line propagators, and there is a factor of the total momentum conservation. 

The quantities $\e^\pm$ are the graviton helicity states that read
\be\label{helicity-app}
\e^{-}_{ABCA'}(k) = M \frac{ q_A q_B q_C k_{A'}}{\ket{q}{k}^3}, \qquad \e^{+}_{ABCA'}(k) = \frac{1}{M} \frac{k_A k_B k_C p_{A'}}{\bra{p}{k}},
\ee
where $M$ is the mass scale of the background de Sitter space, and $q^A, p^{A'}$ are the negative and positive helicity reference spinors, which we denote (for convenience) by different letters.

We have the usual statement of the crossing symmetry, which is that one can change an outgoing state into an incoming one, if one flips the direction of the arrow on the corresponding external leg, and flips the helicity. In addition, there is a factor of minus sign for any such flip, but this is a result of our convention choices. Because of the crossing symmetry, one can assume all particles to be e.g. incoming, which is what we do. 

Finally, the rule is that all positive helicity (incoming) particles are taken to be slightly massive, with the mass related to the mass scale $M$ in (\ref{helicity-app}). The meaning of $k_A$ in this formula is then explained by the following decomposition of the 4-momentum $k^{AA'}$
\be\label{mass-shell-app}
k^{AA'}= k^A k^{A'} + M^2 \frac{p^A p^{A'}}{\ket{p}{k}\bra{p}{k}}.
\ee
The convention is that the reference spinors $p^A, p^{A'}$ in this formula are the same as what is used in the positive helicity spinor in (\ref{helicity-app}). The negative helicity particles are all massless. 

The propagator of the theory is best represented as a drawing, consisting of a set of lines contracting the indices, and a black box denoting the symmetrisation of the unprimed spinor indices. Black lines represent unprimed indices, while the dashed line is for the single primed index. The propagator then reads
\be\nonumber
\frac{1}{\im k^2} \quad \lower0.07in\hbox{\includegraphics[height=0.3in]{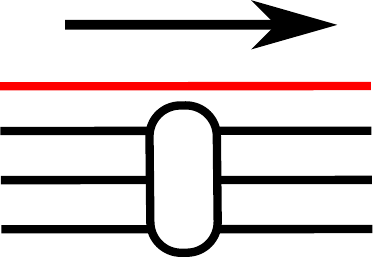}} 
\ee

In the same graphic notation, the 3-valent vertex that is relevant for the computation of the amplitudes with only two positive gravitons reads
\be\label{V-GR-app}
\frac{\im \kappa}{M} \quad \lower0.6in\hbox{\includegraphics[width=1.2in]{3vertexASD.pdf}} \, ,
\ee
where $\kappa^2=32\pi G$. For GR this is the only relevant vertex. The general member of our family of theories contains an additional 3-vertex, which reads
\be\label{V-2-app}
\im \kappa^3 M (2g^{(3)}-(3/2)g^{(2)}) \quad  \lower0.6in\hbox{\includegraphics[width=1.2in]{3vertexSD.pdf}} \, 
\ee
where $(2g^{(3)}-(3/2)g^{(2)})$ is a certain combination of the coupling constants that vanishes in the case of GR. For the family of theories considered in the next section, this combination goes as $M^2/M_p^2$, see (\ref{diff-g-1}). 

The other vertex we will need in this thesis is the single-derivative one present already in GR. It is not important for computations of the GR amplitudes, but contributes to some other amplitudes, e.g. the $++-$ amplitude (for complex momenta) and to the $++++$ amplitude. Represented pictorially, it reads
\be\label{V-3-app}
-\frac{\im}{3} \kappa^3 M^3 f(\delta) \quad \lower0.6in\hbox{\includegraphics[width=1.3in]{3vertex1derASD.pdf}} \,
\ee
Drawings for the 4-valent vertices can be found in the main text. 

\chapter{Contributions to the $++++$ amplitude from the 3-valent graphs}

The purpose of this section is to compute contributions to the all plus amplitude from two different 3-valent diagrams. The computation is more involved than any of those that we have done before, and so we decided to put it in the Appendix. 

There are two types of diagrams to consider. In one, we couple the GR-relevant vertex (\ref{V-GR-app}) to the vertex (\ref{V-2-app}), and in the other it is the 1-derivative vertex (\ref{V-3-app}) that is similarly coupled to (\ref{V-2-app}). Let us consider the 3-derivative vertex diagram first. We do the calculation in several steps. First, we will compute the diagram where the states $1,2$ are inserted into the 3-derivative vertex, and the states $3,4$ are inserted into (\ref{V-2-app}). Then we perform the sum over permutations. 

First, consider the 3-derivative vertex with the states $1,2$ inserted. We necessarily have to insert one of them into the SD leg, and another into an ASD one. Since the vertex is not symmetric in this leg, we will need to add the permutation of $1$ and $2$. So, let $1$ go into the ASD leg, and $2$ into the SD one. We change the orientation of the SD leg, so that the momentum on all the lines is now incoming, and add a minus sign resulting from this change. There is another minus sign coming from the fact that the SD leg insertion gives for $k^{(A}{}_{A'} a^{BCD)A'}$
\be
\frac{1}{M} 2^A 2_{A'} \frac{2^B 2^C 2^D p^{A'}}{\bra{p}{2}} = - \frac{1}{M} 2^A 2^B 2^C 2^D.
\ee
Insertion into the ASD leg gives, taking into account the fact that the positive helicity states are massive
\be
M \frac{1_A 1_B p_{M'} p_{N'}}{\bra{1}{p}^2}.
\ee
Combining everything, and adding a factor for the final momentum we get
\be\label{app4-1}
\frac{\im\kappa}{M} \frac{\ket{1}{2}^2}{\bra{1}{p}^2} \left( 2^{(A} 2^B 1^{C)} \bra{1}{p} + 2^A 2^B 2^C \bra{2}{p}\right) p^{A'}.
\ee
This will have to be multiplied by the propagator $1/2\im\ket{1}{2}\bra{1}{2}$, and then connected to the other vertex. However, before we do this, let us add all other contributions coming from connecting $1,2$ via the 3-derivative vertex. It is clear that we also need to add to (\ref{app4-1}) the same quantity with $1$ and $2$ exchanged. This gives
\be\label{app4-2}
\nn \frac{\kappa}{2M}\frac{\ket{1}{2}}{\bra{1}{2}}\left( 1^A 1^B 1^C \frac{\bra{1}{p}}{\bra{2}{p}^2} +2^A 2^B 2^C \frac{\bra{2}{p}}{\bra{1}{p}^2} + 1^{(A} 1^B 2^{C)} \frac{1}{\bra{2}{p}} + 2^{(A} 2^B 1^{C)} \frac{1}{\bra{1}{p}} \right) p^{A'},
\\
\ee
where we also multiplied by the propagator.

We now connect this to the result of insertion of states $3,4$ into the vertex (\ref{V-2-app}). Here we choose to represent the momentum on the internal line of the diagram, outgoing from the vertex (\ref{V-2-app}), as $-(1+2)^{AA'}$. We get for this vertex
\be\label{app4-4}
- \frac{\im\sqrt{2}(4g^{(3)}-3g^{(2)})}{M M_p^3} \ket{3}{4}^2 3_{(A} 3_B 4_C 4_{D)} (1^D 1_{A'} + 2^D 2_{A'}).
\ee
Contracting (\ref{app4-2}) and (\ref{app4-4}) we get, after simplifying the prefactor 
\be\nonumber
\frac{(4g^{(3)}-3g^{(2)})}{\im M^2 M_p^4} \frac{\ket{1}{2}\ket{3}{4}^2}{\bra{1}{2}\bra{1}{p}^2\bra{2}{p}^2} \left( 1^A 1^B  1^C 1^D \bra{1}{p}^4 + 2^A 2^B 2^C 2^D \bra{2}{p}^4 \right. \\ \nonumber
\left. + 2\cdot  2^A 1^B 1^C 1^D  \bra{1}{p}^3\bra{2}{p} + 2\cdot 1^A 2^B 2^C 2^D \bra{1}{p}\bra{2}{p}^3 + 2 \cdot 1^A 1^B 2^C 2^D \bra{1}{p}^2 \bra{2}{p}^2 \right)  3_{(A} 3_B 4_C 4_{D)} .
\ee
To perform the contraction, we rewrite the symmetrisation $3_{(A} 3_B 4_C 4_{D)}$ in several different ways, depending on the symmetries of the expression that it gets contracted to. For the contracting object that is $BCD$ symmetric we can write
\be\label{app-sym1}
3_{(A} 3_B 4_C 4_{D)} = \frac{1}{2}\left( 3_A 3_{(B} 4_C 4_{D)} + 4_A 4_{(B} 3_C 3_{D)}\right).
\ee
For the contracting object that is $AB$ and $CD$ symmetric, as well as symmetric under the exchange of pairs $AB$ and $CD$ we can write 
\be\label{app-sym2}
3_{(A} 3_B 4_C 4_{D)} = \frac{1}{6} \left( 3_A 3_B 4_C 4_D +4_A 4_B 3_C 3_D+ 4\cdot 3_{(A} 4_{B)} 3_{(C} 4_{D)}\right).
\ee
Performing the contractions we get the following amplitude
\be\label{app4-amp1}
\nn\frac{(4g^{(3)}-3g^{(2)})}{\im M^2 M_p^4} \frac{\ket{1}{2}\ket{3}{4}^2}{\bra{1}{2}} \Big[ \ket{1}{3}^2\ket{1}{4}^2 \frac{\bra{1}{p}^2}{\bra{2}{p}^2} + \ket{2}{3}^2\ket{2}{4}^2 \frac{\bra{2}{p}^2}{\bra{1}{p}^2}  \\ \nonumber 
+ \left( \ket{2}{3}\ket{1}{3}\ket{1}{4}^2 + \ket{2}{4}\ket{1}{4}\ket{1}{3}^2\right) \frac{\bra{1}{p}}{\bra{2}{p}}+ \left( \ket{1}{3}\ket{2}{3}\ket{2}{4}^2 + \ket{1}{4}\ket{2}{4}\ket{2}{3}^2\right) \frac{\bra{2}{p}}{\bra{1}{p}}  \\ \nonumber 
+\frac{1}{3} \left( \ket{2}{3}^2\ket{1}{4}^2 + \ket{1}{3}^2\ket{2}{4}^2 + 4 \ket{1}{3}\ket{1}{4}\ket{2}{3}\ket{2}{4}\right) \Big].
\\
\ee

The first four terms in (\ref{app4-amp1}) are $p$-dependent, while the ones in the last line are not. We expect the result to be $p$-independent, but this should only be true after all the permutations are added. Thus, the above result is $12$ and $34$ symmetric (by construction), but is not symmetric under the permutation of $12$ with $34$. This means that we have to consider overall 6 different permutations, and add them all up. We first need to show that the result of adding all these permutations is $p$-independent. To this end, let us consider all the terms that will have $\bra{1}{p}^2$ in the denominator. These are
\be\label{app4-5}
\frac{\ket{2}{3}^2\ket{2}{4}^2\ket{3}{4}^2}{\bra{1}{p}^2} \left( \bra{2}{p}^2 \frac{\ket{1}{2}}{\bra{1}{2}} + \bra{3}{p}^2 \frac{\ket{1}{3}}{\bra{1}{3}}+ \bra{4}{p}^2 \frac{\ket{1}{4}}{\bra{1}{4}} \right).
\ee
We can now use the Schouten identity to rewrite $\bra{2}{p}^2$ in terms of the square brackets $\bra{3}{p}$ and $\bra{4}{p}$. Thus, we replace
\be
\ket{1}{2}^2\bra{1}{2}^2 = \ket{1}{3}^2 \bra{3}{p}^2 + \ket{1}{4}^4 \bra{4}{p}^2 + 2\ket{1}{3}\ket{1}{4} \bra{3}{p}\bra{4}{p}.
\ee
Then, after some simplifications with the use of the Schouten identity we get for (\ref{app4-5})
\be\label{app4-6}
- \frac{\ket{2}{3}^2\ket{2}{4}^2\ket{3}{4}^2 \ket{1}{3}\ket{1}{4} \bra{3}{4}^2}{ \ket{1}{2}\bra{1}{2}\bra{1}{3}\bra{1}{4}},
\ee
where $\bra{1}{p}^2$ cancelled out. We can now rewrite this result in a bit more convenient form, by eliminating as many square brackets as possible using the momentum conservation. This gives for (\ref{app4-6})
\be\label{app4-8}
\ket{1}{2}\ket{1}{3}\ket{1}{4}\ket{2}{3}\ket{2}{4}\ket{3}{4} \frac{\ket{3}{4}}{\bra{1}{2}}.
\ee
The fact that this is $2,3,4$ symmetric follows easily from the momentum conservation. It can then be checked that the other terms in the sum of permutations of (\ref{app4-amp1}) containing squares of $p$-dependent square brackets in the denominator are all equal to (\ref{app4-8}). Thus, running the same argument for e.g. the terms proportional to $1/\bra{2}{p}^2$ gives precisely the same result as (\ref{app4-8}). This means that the sum of permutations of the first two terms in (\ref{app4-7}) is 4 times (\ref{app4-8}). 

Let us now consider permutations of terms containing just a single power of $p$-dependent square bracket. Let us concentrate on the terms proportional to $1/\bra{1}{p}$. These give rise to the sum
\be
\nn \frac{\ket{2}{3}\ket{2}{4}\ket{3}{4}}{\bra{1}{p}} \left( \left( \ket{1}{3}\ket{2}{4}+\ket{1}{4}\ket{2}{3}\right) \bra{2}{p} \frac{\ket{1}{2}\ket{3}{4}}{\bra{1}{2}} \right. \\ \nonumber \left. - \left( \ket{1}{2}\ket{3}{4}-\ket{1}{4}\ket{2}{3}\right) \bra{3}{p} \frac{\ket{1}{3}\ket{2}{4}}{\bra{1}{3}}-\left( \ket{1}{2}\ket{3}{4}+\ket{1}{3}\ket{2}{4}\right) \bra{4}{p} \frac{\ket{1}{4}\ket{2}{3}}{\bra{1}{4}}  \right).
\\
\ee
Using the momentum conservation to convert the determinants inside the brackets into multiples of $\bra{1}{2}$ we get
\be\label{app4-7}
\frac{\ket{2}{3}\ket{2}{4}\ket{3}{4}^2}{\bra{1}{p}\bra{1}{2}} \left( \left( \ket{1}{3}\ket{2}{4}+\ket{1}{4}\ket{2}{3}\right) \bra{2}{p} \ket{1}{2} \right. \\ \nonumber \left. + \left( \ket{1}{2}\ket{3}{4}-\ket{1}{4}\ket{2}{3}\right) \bra{3}{p} \ket{1}{3}-\left( \ket{1}{2}\ket{3}{4}+\ket{1}{3}\ket{2}{4}\right) \bra{4}{p} \ket{1}{4} \right).
\ee
We can now again use the momentum conservation, now in the form $\ket{1}{2}\bra{2}{p}=-\ket{1}{3}\bra{3}{p}-\ket{1}{4}\bra{4}{p}$. After some simple algebra involving the Schouten identity the expression in brackets becomes $-3\ket{1}{3}\ket{1}{4}( \ket{2}{3}\bra{3}{p}+\ket{2}{4}\bra{4}{p}) = - 3\ket{1}{2}\ket{1}{3}\ket{1}{4}\bra{1}{p}$, where in the last equality we used the momentum conservation. Overall, (\ref{app4-7}) becomes
\be
-3 \ket{1}{2}\ket{1}{3}\ket{1}{4}\ket{2}{3}\ket{2}{4}\ket{3}{4} \frac{\ket{3}{4}}{\bra{1}{2}}.
\ee
As previously the case with (\ref{app4-8}), there are in total a multiple of 4 of these, coming from applying the same analysis to different $p$-dependent denominators, e.g. to terms proportional to $1/\bra{2}{p}$, etc. 

It remains to consider terms in (\ref{app4-amp1}) that are $p$-independent, together with their permutations. For this, using Schouten identity it is convenient to rewrite the term in the last line of (\ref{app4-amp1}) as
\be\label{app4-9}
2 \ket{1}{3}\ket{1}{4}\ket{2}{3}\ket{2}{4} + \frac{1}{3} \ket{1}{2}^2\ket{3}{4}^2.
\ee
Then the first of these two terms, when multiplied by the prefactor in (\ref{app4-8}), is already of the form (\ref{app4-8}). Summing over 6 different permutations we thus get a multiple of 12 of (\ref{app4-8}) from the first term in (\ref{app4-9}). For the permutations of the second term in (\ref{app4-9}) we have
\be\label{app4-10}
\frac{2}{3}\left( \frac{\ket{1}{2}^4 \ket{3}{4}^4}{\bra{1}{2}\bra{1}{2}} + \frac{\ket{1}{3}^4 \ket{2}{4}^4}{\bra{1}{3}\bra{1}{3}}+\frac{\ket{1}{4}^4 \ket{2}{3}^4}{\bra{1}{4}\bra{1}{4}}\right).
\ee
Converting the square brackets in the denominators into multiples of $\bra{1}{2}$ using the momentum conservation, and then using the Schouten identity we get for (\ref{app4-10}) $- 2$ times (\ref{app4-8}). Thus, the sum of all permutations of (\ref{app4-amp1}) is equal $(1-3)4+12-2= 2$ multiples of  (\ref{app4-8}). Together with the prefactor, this gives the following result for this part of the amplitude
\be\label{app4-res1}
\frac{2(4g^{(3)}-3g^{(2)})}{\im M^2 M_p^4} \ket{1}{2}\ket{1}{3}\ket{1}{4}\ket{2}{3}\ket{2}{4}\ket{3}{4} \frac{\ket{3}{4}}{\bra{1}{2}}.
\ee

We now compute the other diagram, where the 1-derivative vertex (\ref{V-3-app}) is connected to (\ref{V-2-app}).  As before, we first consider contributions from connecting $1,2$ into the 1-derivative vertex  and $3,4$ into the vertex (\ref{V-2-app}), and then sum over the permutations. For the 1-derivative vertex, both states are now inserted into the legs with no derivatives in them. We also need to change the orientation of the ASD leg, so that all the momenta are incoming, with a resulting extra minus sign. Overall, we get
\be
 \frac{\im}{3} \kappa^3 M f(\delta) \frac{\ket{1}{2}^2}{\bra{1}{p}\bra{2}{p}} 1^{(A} 2^B \left( 1^{C)} \bra{1}{p} + 2^{C)} \bra{2}{p}\right) p^{A'}.
\ee
We now multiply this by the propagator, and connect to the other vertex (\ref{app4-4}). We get, after simplifying the prefactor
\be
\nn \frac{2f(\delta) (4g^{(3)}-3g^{(2)})}{3\im M_p^6} \frac{\ket{1}{2}\ket{3}{4}^2}{\bra{1}{2}} 1^A 2^B \left(   1^C 1^D \frac{\bra{1}{p}}{\bra{2}{p}} +  2^C 2^D \frac{\bra{2}{p}}{\bra{1}{p}} + 2\cdot 1^C 2^D \right) 3_{(A} 3_B 4_C 4_{D)}.
\\
\ee
We perform the contraction using (\ref{app-sym1}), (\ref{app-sym2}). We get for this contribution to the amplitude
\be
\nn \frac{2f(\delta) (4g^{(3)}-3g^{(2)})}{3\im M_p^6} \frac{\ket{1}{2}\ket{3}{4}^2}{\bra{1}{2}} \Big[ \frac{1}{2} \left( \ket{2}{3}\ket{1}{3}\ket{1}{4}^2 + \ket{2}{4}\ket{1}{4}\ket{1}{3}^2\right) \frac{\bra{1}{p}}{\bra{2}{p}}\\ \nonumber + \frac{1}{2} \left( \ket{1}{3}\ket{2}{3}\ket{2}{4}^2 + \ket{1}{4}\ket{2}{4}\ket{2}{3}^2\right) \frac{\bra{2}{p}}{\bra{1}{p}}   
+2 \ket{1}{3}\ket{1}{4}\ket{2}{3}\ket{2}{4}+ \frac{1}{3}\ket{1}{2}^2\ket{3}{4}^2 \Big],
\\
\ee
where we have used (\ref{app4-9}). We have already performed the sums over permutations required here, and so we can immediately write down the result. It is given by the prefactor, times a multiple $-(3/2)\cdot 4+ 12-2=4$ of (\ref{app4-8}), in other words this part of the amplitude equals
\be\label{app4-res2}
 \frac{8f(\delta) (4g^{(3)}-3g^{(2)})}{3\im M_p^6} \ket{1}{2}\ket{1}{3}\ket{1}{4}\ket{2}{3}\ket{2}{4}\ket{3}{4} \frac{\ket{3}{4}}{\bra{1}{2}}.
\ee

We now have to add the two results (\ref{app4-res1}) and (\ref{app4-res2}), and compute the $M\to 0$ limit. Since $4g^{(3)}-3g^{(2)}= - (27/4) \beta^2 M^2/M_p^2$ plus higher order in $M$ terms, and $f(\delta)=3M_P^2/M^2$ plus order unity terms, we only need to keep these leading orders. Thus, we get, overall, our sample one-parameter family of theories with Planckian modifications
\be\label{app-3-valent}
{\cal M}^{++++}_{\rm 3-vert} = \im \frac{135\beta^2}{2M_p^6} \ket{1}{2}\ket{1}{3}\ket{1}{4}\ket{2}{3}\ket{2}{4}\ket{3}{4} \frac{\ket{3}{4}}{\bra{1}{2}} = \im \frac{135\beta^2}{16M_p^6} stu.
\ee

	\phantomsection

\end{document}